\documentstyle[rotating,12pt]{article}
\newcommand{\be}{\begin{equation}}
\newcommand{\ee}{\end{equation}}
\newcommand{\bear}{\begin{eqnarray}}
\newcommand{\ear}{\end{eqnarray}}
\renewcommand{\theequation}{\arabic{section}.\arabic{equation}}
\newcommand{\hpv}{\mbox{$H_{PV}$}}
\newcommand{\hpve}{\mbox{$H_{PV}^{(1)}$}}
\newcommand{\hpvz}{\mbox{$H_{PV}^{(2)}$}}
\newcommand{\vecx}{\mbox{$\vec{x}$}}
\newcommand{\gaq}{\mbox{$G_A^{(q)}$}}
\newcommand{\gaqzn}{\mbox{$G_A^{(q)}(Z,N)$}}
\newcommand{\qwe}{\mbox{$Q_{W}^{(1)}$}}
\newcommand{\qwz}{\mbox{$Q_{W}^{(2)}$}}
\newcommand{\sint}{\mbox{$\sin^2 \theta_W$}}
\newcommand{\cale}{\mbox{$\cal E$}}
\newcommand{\calvece}{\mbox{$\vec{\cal E}$}}
\newcommand{\calf}{\mbox{$\cal F$}}
\newcommand{\rbz}{\mbox{$r_B(Z)$}}
\newcommand{\underm}{\mbox{$\underline{\cal M}$}}

\newcommand{\underg}{\mbox{$\underline{\Gamma}$}}
\newcommand{\undert}{\mbox{$\underline{\cal T}$}}
\newcommand{\underd}{\mbox{$\underline{D}$}}
\newcommand{\undere}{\mbox{$\underline{E}$}}
\newcommand{\underf}{\mbox{$\underline{F}$}}

\newcommand{\lefteil}{\mbox{$ ( \widetilde{\alpha,F_3 }|$}}

\newcommand{\lefteile}{\mbox{$ ( \widetilde{\alpha,F_3 ,\cale \vec{e}_3}|$}}
\newcommand{\righteire}{\mbox{$ | \alpha,F_3,\cale \vec{e}_3 )$}}

\newcommand{\eins}{\mbox{$1 \hspace{-1.0mm}  {\bf l}$}}

\newcommand{\eh}{\mbox{$\frac{1}{2}$}}
\newcommand{\dh}{\mbox{$\frac{3}{2}$}}
\newcommand{\ev}{\mbox{$\frac{1}{4}$}}

\newcommand{\dv}{\mbox{$\frac{3}{4}$}}
\newcommand{\da}{\mbox{$\frac{3}{8}$}}
\newcommand{\ih}{\mbox{$\frac{i}{2}$}}
\newcommand{\fh}{\mbox{$\frac{5}{2}$}}

\newcommand{\zsehnn}{\mbox{$ 2S, \frac{1}{2} ,0,0 $}}
\newcommand{\zsehen}{\mbox{$ 2S, \frac{1}{2} ,1,0 $}}
\newcommand{\zpehnn}{\mbox{$ 2P, \frac{1}{2} ,0,0 $}}
\newcommand{\zpehen}{\mbox{$ 2P, \frac{1}{2} ,1,0 $}}
\newcommand{\zpdhen}{\mbox{$ 2P, \frac{3}{2} ,1,0 $}}
\newcommand{\zpdhzn}{\mbox{$ 2P, \frac{3}{2} ,2,0 $}}
\newcommand{\zpdhzz}{\mbox{$ 2P, \frac{3}{2} ,2,2 $}}
\newcommand{\zpdhze}{\mbox{$ 2P, \frac{3}{2} ,2,1 $}}
\newcommand{\zpdhee}{\mbox{$ 2P, \frac{3}{2} ,1,1 $}}
\newcommand{\zpehee}{\mbox{$ 2P, \frac{1}{2} ,1,1 $}}
\newcommand{\zsehee}{\mbox{$ 2S, \frac{1}{2} ,1,1 $}}
\newcommand{\zpdhzmz}{\mbox{$ 2P, \frac{3}{2} ,2,-2 $}}
\newcommand{\zpdhzme}{\mbox{$ 2P, \frac{3}{2} ,2,-1 $}}
\newcommand{\zpdheme}{\mbox{$ 2P, \frac{3}{2} ,1,-1 $}}
\newcommand{\zpeheme}{\mbox{$ 2P, \frac{1}{2} ,1,-1 $}}
\newcommand{\zseheme}{\mbox{$ 2S, \frac{1}{2} ,1,-1 $}}
\newcommand{\ex}{\mbox{\boldmath$e$}_1}
\newcommand{\ey}{\mbox{\boldmath$e$}_2}
\newcommand{\ez}{\mbox{\boldmath$e$}_3}
\newcommand{\bra}[1]{\mathopen{\bigl\langle}_{\!\!\!\!\!\SSS #1\,\,\,\,\,}}
\newcommand{\ket}[1]{\mathclose{\bigr\rangle}_{\!\SSS #1}}
\def\vert     {\mathbin|}
\renewcommand{\tilde}{\widetilde}

\def\SS		{\scriptstyle}   
\def\SSS	{\scriptscriptstyle}
\def\nn		{\nonumber \\}
\newcommand{\Tr}{\mbox{Tr}}
\def\halb      {\textstyle \frac{1}{2} \displaystyle}
\def\viertel      {\textstyle \frac{1}{4} \displaystyle}
\def\shalb      {\SS \frac{1}{2} \displaystyle}

\def\ihalb      {\textstyle \frac{i}{2} \displaystyle}
\def\Im         {{\rm Im  } }
\def\Re         {{\rm Re   } }
\newcommand{\E}{{\cal E}}
\newcommand{\N}{{\cal N}}
\newcommand{\U}{{\cal U}}
\newcommand{\F}{\mbox{$\cal F$}}
\newcommand{\R}{\mbox{$\cal R$}}
\newcommand{\T}{\mbox{$\cal T$}}

\newcommand{\Gs}{\mbox{$\Gamma_S$}}
\newcommand{\Gp}{\mbox{$\Gamma_P$}}
\def\M		{\underline{{\cal M}}}
\def\calm   {{\cal M}}

\def\ssw		{\sin^2 \theta_W}

\newcommand{\pro}{\underline{{\mbox{\,P$\!\!\!\!\!${\rm I}$\,\,\,$}}}}
\newcommand{\Pro}{{\mbox{\,P$\!\!\!\!\!${\rm I}$\,\,\,$}}}   

\newcommand{\C}{\mbox{\large \bf C$\!\!\!\!${\normalsize \sf I}$\,\,\,$}}
\newcommand{\bm}[1]{\mbox{\boldmath$#1$}}  
\renewcommand{\vec}{\bm}
\newcommand{\Ehat}{\hat{\bm{\E}}}

\def\Qe		{\mbox{e}}   
\def\QWe	{Q_W^{(1)}}

\date{}
\begin{document}
\begin{titlepage}
\begin{flushright}
HD--THEP--94--14
\end{flushright}
\quad\\
\vspace{.8cm}
\begin{center}
{\bf\LARGE POLARIZATION ROTATION EFFECTS}\\
\bigskip
{\bf \LARGE DUE TO PARITY VIOLATION}\\
\bigskip
{\bf\LARGE IN ATOMS}\\
\vspace{1cm}
G. W. Botz\footnote{Partially supported by German National Scholarship
Foundation},
 D. Bru\ss\ \footnote{Partially supported by Landesgraduiertenf\"orderung,
Baden-W\"urttemberg}
and O. Nachtmann\\
\bigskip
Institut  f\"ur Theoretische Physik\\
Universit\"at Heidelberg\\
Philosophenweg 16, D-69120 Heidelberg, FRG\\
\bigskip
{\bf Abstract:}\\
\parbox[t]{\textwidth}{\small
We present a study of parity (P) violating polarization rotations
of atoms in external electric fields. Five different types of rotations
are identified and the consequences of time reversal invariance (T) are
discussed. The role played by Zeldovich's electric dipole moment of unstable
states is elucidated. To calculate the effects, we use the standard model
of elementary particle physics where P violation in atoms is due to the
exchange of the Z boson between the quarks in the nucleus and the atomic
electrons. We consider in detail hydrogen-like systems in $n=2$ states,
where $n$ is the principal quantum number, especially $^1_1$H,  $^4_2$He$^+$,
and $^{12}_6$C$^{5+}$. There one has the metastable $2S_{1/2}$ states which
are separated from the $2P_{1/2}$ states with opposite parity only by the small
Lamb shift leading to a relatively large P-violating state mixing. The
nominal order of magnitude of the polarization rotations is $10^{-11}-10^{-12}$
but we discuss ways to obtain in some cases large enhancement factors of
order $10^4$. We show that e.g. for $^1_1$H one could in principle observe
P-violating  polarization rotations as large as a few percent, where for a
statistically significant result one would need a total of $10^{15}$ polarized
atoms. We point out that some of our P-violating polarization rotations are
very sensitive to the nuclear spin-dependent part of the P-violating
Hamiltonian which receives a contribution from the polarized strange quark
density in polarized nuclei. Thus a measurement of P-violating polarization
rotations of
atoms would be of great interest in at least three respects: it would
demonstrate the existence of quantum mechanical effects in atoms where
P violation and T conservation interplay in a subtle way, it would allow
a measurement of polarized quark densities in
polarized nuclei, thus shedding light on problems associated with the
``spin crisis'' for the nucleons and it would allow precision determinations
of the electroweak parameters related to the $Z$ boson exchange at very low
energies. This shows again the interplay of atomic, nuclear, and particle
physics in P-violating effects in atoms.
\\[1cm]
Submitted to Ann. Phys. (N.Y.)
}
\end{center}
\end{titlepage}
\section{Introduction}
The advent of the standard model (SM) of elementary particle
physics \cite{1} and the experimental discovery of weak neutral
currents in neutrino interactions \cite{2} paved the way for detailed
predictions of parity(P)-violating effects in atoms \cite{3}, \cite{4}.
By now such effects are well studied theoretically and have been
verified experimentally for heavy atoms containing many electrons (for a
recent review cf. \cite{5}). On the other hand, P-violating effects have so far
not
been demonstrated experimentally for the simplest atomic systems: hydrogen-like
atoms, where theoretical calculations can be pushed to very high accuracy
\cite{6}.
Proposals for P-violating experiments in hydrogenic systems have been made
e.~g. in \cite{6a} -- \cite{6e}. These proposals as well as the experiments
actually performed for heavy atoms in general involve looking at P-violating
characteristics of light quanta emitted or absorbed by the atom. In the present
work we will discuss P-violationg effects in atoms which are of quite a
different nature. We propose to look at P-violating polarization rotations
of atoms when they are subjected to external electric fields. This has some
similarity to the proposals for regeneration-type experiments made in
\cite{6a}, \cite{6c}.
In the days of high precision physics in the framework of the SM at high
energies, e.g. at LEP (cf. \cite{7} for a review), we should discuss the
motivations for our study. They are as follows:

(1) P-violating effects in atoms can provide precision determinations of the
effective SM  couplings at low energies. Comparison with high energy precision
values for these
couplings allows to check their ``running'' as predicted by the renormalization
theory. In this way a very sensitive test of higher order effects in the SM
and a probing for new physics beyond the SM is possible.

(2) P-violating effects in atoms can give interesting information on properties
of
atoms and nuclei. They may, for instance, shed light on the so-called
``spin crisis'' for nucleons \cite{8}, \cite{9}.

(3) We will show in the following that P-violation  in atoms leads to
certain spin rotation effects for atoms in electric fields. Such effects
present interesting quantum mechanical  phenomena which should be worthwhile to
study for their own sake.

(4) There are many recent advances in experimental techniques of atomic
physics,
like the production and storage of ``cold'' heavy ion beams in storage rings
(cf. \cite{10} and references therein) and the observation of interference
effects with atomic beams (cf. \cite{11} for a review). This should also make
the tiny P-violating effects in atoms more easily accessible to experimental
studies.

Our paper is a continuation of the work done in \cite{12}, and much of the
formalism
will be taken over from there. In section 2 we briefly recall the P-violating
Hamiltonian. In section 3 we work out the general formalism for the angular
momentum precession effects for atoms in a constant external electric field.
We use the Wigner-Weisskopf approach for the description of unstable states.
In section 4 we discuss rotation effects for the case of an electric field
which is
piecewise constant in time, for instance the case where the electric field has
a value $\vec{\cale}^1$ for the time 0 to $t_1$, then $\vec{\cale}^2$ for the
time $t_1$ to $t_2$ and is then switched off.
In section 5 we present our results for hydrogen-like systems with spin-zero
nuclei.
Results for hydrogen-like systems with nuclei of spin $\eh $ are discussed in
section 6. Our conclusions are drawn in section 7.

All our formulae are written in natural units, $\hbar = c = 1$, if other units
are not explicitly indicated.
\section{The P-violating Hamiltonian}
\setcounter{equation}{0}
In the framework of the SM the effective P-violating Hamiltonian $H_{PV}$,
relevant for atomic physics, is due to $Z$ exchange between the atomic
electrons ---
or muons in the case of muonic atoms --- and the quarks in the nucleus. In
terms
of lepton and quark field operators we have
\bear
\label{2.1}
\hpv &=& \hpve + \hpvz, \\
 \label{2.2}
\hpve &=& - \frac{G}{\sqrt{2}} \int d^3 x \, 2  g_A^l \bar{l}(\vec{x})
          \gamma^{\lambda}
         \gamma_5 l(\vec{x}) \left( \sum_q g_V^q \bar{q}(\vec{x})
         \gamma_{\lambda} q(\vec{x})
         \right),   \\
\label{2.3}
\hpvz &=& - \frac{G}{\sqrt{2}} \int d^3 x \, 2  g_V^l \bar{l}(\vec{x})
            \gamma^{\lambda}
          l(\vec{x}) \left( \sum_q g_A^q \bar{q}(\vec{x})
           \gamma_{\lambda} \gamma_5 q(\vec{x})
         \right) .
\ear
Here $l=e$ or $\mu$ and $q=u,d,s$,  neglecting possible contributions
from the heavy quarks $c,b,t$. The notation is as in \cite{13} with $G$ Fermi's
constant and $g_{V,A}^{l,q}$ the neutral current coupling constants for $l,q$.

In the following we will neglect nuclear recoil and size effects, i.e. we will
consider an infinitely heavy point nucleus sitting at the origin
$\vec x=0$. P-violating effects in the nucleus like its anapole moment \cite{5}
will be neglected as they are higher-order effects. Let $\vec{I}, I,I_3$ denote
as usual the spin operator and
quantum numbers of the nucleus $Z,N$ ($Z$: number of protons, $N$: number of
neutrons) and let us normalize the nuclear states to
\be
<Z,N;I,I'_3 \, | \, Z,N;I,I_3 > \, = \, \delta_{I_3' I_3^{}}.
\label{2.4}\ee
{}From rotational invariance and conservation of parity, charge, strangeness
and baryon
number in strong interactions, we find now for the matrix elements of the
hadron
currents in (\ref{2.2}), (\ref{2.3}) the expressions:
\bear
\label{2.5}
<Z,N;I,I'_3 \, | \, \bar{q}(\vec{x}) \gamma^{0} q(\vec{x})\,| \,
          Z,N;I,I_3 > & =&
      \delta^3(\vec{x}) \, \delta_{I'_3 I_3^{}} \cdot
      \left\{   \begin{array}{ccc}
                 2Z+N & \mbox{ for}  & q=u , \\
                 Z+2N & \mbox{ for}  & q=d ,  \\
                 0 & \mbox{ for} & q=s  ,
 \end{array} \right.  \nonumber \\
<Z,N;I,I'_3 \, | \, \bar{q}(\vec{x}) \vec{\gamma} q(\vec{x})\,| \,
          Z,N;I,I_3 > & =& 0 ,\nonumber \\
       & & \\
\label{2.6}
<Z,N;I,I'_3 \, |\,  \bar{q}(\vec{x}) \gamma^{0}  \gamma_5 q(\vec{x})\,|\,
          Z,N;I,I_3 > & =& 0 ,\nonumber \\
<Z,N;I,I'_3 \, | \, \bar{q}(\vec{x}) \vec{\gamma} \gamma_5 q(\vec{x})\,| \,
          Z,N;I,I_3 > & =& \gaqzn\ \delta^3(\vecx )
           <I,I'_3  |  \vec{I}  | I,I_3 > .\nonumber \\
             & &
\ear
Here we have introduced the axial form factors $G_A^{(q)}(Z,N)$ of the nucleus
for
the various quark species $q$. For the proton and neutron these form factors
are related to the integral over the $g_1$ structure function of polarized
lepton-nucleon scattering (cf. \cite{14}--\cite{15d} for reviews). There one
defines $\Delta q$ as the ``total polarization'' carried by the quark species
$q$.
These $\Delta q$ are related to $G^{(q)}_A$ (\ref{2.6}) by
\be
(\Delta q )_{p,n} = \eh \gaq_{p,n}.
\label{2.7}\ee
There are some subtle renormalization problems concerning the axial flavour
singlet current in making the connection (\ref{2.7}). We will not enter into
this discussion here. For definiteness we choose a scheme where the axial
flavour singlet current is not renormalized. In other words, we consider
(\ref{2.7}) as \underline{definition} of $\Delta q$. The sumrule in deep
inelastic scattering related to $\Delta q$ will then have calculable
corrections due to gluonic effects.
(We thank U. Ellwanger for a discussion on this point).
We will generalize the definition of $\Delta q$ for nuclei of arbitrary spin
$I$ as
\be
\Delta q (Z,N) = I \,  \gaq (Z,N).
\label{2.8}\ee
In the naive parton model interpretation $\Delta q$ is then the number of
quarks $q$
and antiquarks $\bar q$ with spin parallel minus the number of quarks $q$ and
antiquarks
$\bar q$ with spin antiparallel to the direction of flight of the nucleus
moving
fast in the 3-direction and having $I_3=I$ as spin state.

What do we know about the values of $\Delta q$ for various nuclei?
By isospin invariance the quantity $\Delta u-\Delta d$ can in some cases
be related to the Gamow-Teller matrix element of nuclear $\beta$ decay. This is
discussed in Appendix~A.

For the proton we get for instance
\be
\eh (G_A^{(u)} - G_A^{(d)})_p = (\Delta u -\Delta d )_p = g_A,
\label{2.9}\ee
where $g_A=1.2573(28)$ is the axial decay constant of neutron $\beta$ decay
\cite{16}.
(In our convention \cite{13} we have $g_A>0$). The quantity $(\Delta s)_p$ is
derived
from the EMC measurement \cite{8} as
\be
\eh (\Delta s)_p = -0.095 \pm 0.016 \pm 0.023,
\label{2.10}\ee
where the first error is statistical, the second one systematic. The naive
expectation was $(\Delta s)_p=0$ and thus the value (\ref{2.10}) gave rise to
the ``spin crisis'' of the proton.
New -- as yet unpublished -- results from the SMC collaboration
(D. Adams et al of \cite{9}) seem to indicate a smaller value for
$(\Delta s)_p$ than (\ref{2.10}). In the following we will
always use for numerical estimates the central value of  $(\Delta s)_p$
of (\ref{2.10}). Other experimental results for $(\Delta s)_p$
are then easily taken into account by a suitable rescaling.
 An independent determination of $\Delta s$ by
an atomic P-violation experiment would clearly be a significant achievement.
As we will see, some of the effects discussed below
are indeed quite sensitive to $\Delta s$.

The next step is to use the matrix elements (\ref{2.5}), (\ref{2.6}) for the
hadronic
part and a nonrelativistic reduction for the leptonic part of $H_{PV}^{(1,2)}$
(cf. (\ref{2.2}), (\ref{2.3})).
The final result for the effective P-violating Hamiltonians reads as follows:
\bear
\label{2.11}
\hpve &=& \frac{G}{4\sqrt{2}} \frac{1}{m_l } \qwe  (Z,N) \cdot
                 \left\{ \delta^3(\vec{x}) (\vec{\sigma} \cdot \vec{p}) +
                 (\vec{\sigma} \cdot
              \vec{p})  \delta^3(\vec{x})   \right\}, \\
\hpvz &=& \frac{G}{4\sqrt{2}} \frac{1}{m_l } \qwz (Z,N) \cdot
                 \left\{ \delta^3(\vec{x}) (\vec{I} \cdot \vec{\sigma})
                 (\vec{\sigma} \cdot \vec{p}) + (\vec{\sigma} \cdot \vec{p})
                  (\vec{I} \cdot \vec{\sigma})
                   \delta^3(\vec{x})   \right\},
\label{2.12}
\ear
where $\vec p$ and $\vec\sigma$ are the momentum and spin operators of the
lepton $l$
and $Q_W^{(1,2)}$ are weak charges:
\bear
\qwe (Z,N) &=& -4 g_A^l \left\{ g_V^u (2Z+N) + g_V^d (Z+2N) \right\} , \\
\label{2.13}
\qwz (Z,N) &=& 4 g_V^l \cdot \sum_q g_A^q \gaqzn
\label{2.14}.\ear
In the framework of the SM we have
\bear
g_V^l & = & -\eh + 2\, \sint \nonumber ,\\
g_A^l & = & -\eh \nonumber ,\\
g_V^u & = & \eh -\mbox{$\frac{4}{3}$} \sint  ,\nonumber \\
g_V^{d,s} & = & -\eh + \mbox{$\frac{2}{3}$} \sint  ,\nonumber \\
g_A^u &=& \eh ,\nonumber \\
g_A^{d,s} & =  &  -\eh .
\label{2.15}\ear
\bear
\label{2.16}
\qwe (Z,N)  & = & \ (1-4 \,\sint ) Z-N  ,\\
&&\nn
\qwz (Z,N) & = & -\, (1-4\, \sint )\left[
                G_A^{(u)}(Z,N) -G_A^{(d)}(Z,N) -G_A^{(s)}(Z,N) \right]
            \nonumber \\
   & = & -\frac{1}{I} (1-4 \,\sint )\left[ \Delta u(Z,N) -
           \Delta d(Z,N) -\Delta s(Z,N) \right].
\label{2.17}\ear
The numerical values of these weak charges are collected for some nuclei
of interest in Table A2 of Appendix A.
\section{Rotation effects in a constant electric field}
\subsection{Hamiltonian and state vectors}
\setcounter{equation}{0}

Let us consider a hydrogen-like system, i.e. a nucleus with a single lepton
(electron or muon) around it. The energy levels in the Coulomb approximation
are $1S$, $2S$, $2P$ etc., where $2S$ and $2P$
are degenerate. These levels are split by the fine structure, the Lamb shift
and
further QED effects and the hyperfine structure if the nuclear spin $I\not=0$.
We will
consider P-violating effects for the states with principal quantum number $n=2$
since one can make beams of the metastable  $2S$ states which can readily
mix with the nearby $2P$ states through $H_{PV}$.

The Hamiltonian for the atom in a constant external electric field $\vec{\cale}
$
and in interaction with the radiation field can be written as
\bear
\label{3.1}
H & = & H_0 +\hpv + H_{ext} + H_{rad} , \\
H_{ext} & = & e \vecx \vec{\cale }.
\ear
Here $e>0$ is the positron charge, $H_0$ is the Hamiltonian for free photons
plus the atomic states before the inclusion of P-violating, radiative decay and
external electric field effects, $H_{PV}$ is given in (\ref{2.1}),
(\ref{2.11}),
(\ref{2.12}) and $H_{rad}$ is the term responsible for radiative transitions
between the atomic states. Compared to our work in \cite{12} we will include
now
nuclear spin effects and, in the following section, we will also consider
external
electric fields with a nontrivial time dependence. The spectrum of atomic
eigenstates of $H_0$ is shown schematically in Fig. 1 for $n=1,2$. By inclusion
of $H_{PV}$
and $H_{ext}$ the $S$ and $P$ states get mixed, by inclusion of $H_{rad}$ the
$n=2$ states become unstable. We neglect in the following radiative transitions
between levels with the same $n$, as such transitions have a much longer time
scale
than the effects we will be interested in. We also neglect the small mixing of
states of different $n$ due to $H_{PV}$ and $H_{ext}$.

The typical situation we will consider is that of a metastable $2S$ state which
is put into the external field $\vec{\cale} $ at time $t=0$ and
polarization-analysed
at some later time. The theoretical tool to deal with this situation is the
Wigner-Weisskopf method \cite{17}. The relevant states which we have to
consider
are the $n=2$ atomic states and their decay products: the $n=1$ atomic states
plus photon states. Let us choose as basis for the atomic states
the eigenstates of $H_0$ (\ref{3.1}) which we classify in the standard way:
\begin{displaymath}
|n,l,j,F,F_3\rangle.
\end{displaymath}
Here we consider a nucleus of spin $I$,
and $l, j, F$, and $ F_3$ are the quantum numbers
of orbital, total lepton, total, and third component of the total angular
momentum,
respectively. The corresponding operators of angular momenta are
\bear
\vec{L}, & & \nonumber \\
\vec{J} & = & \vec{L} + \eh \vec{\sigma }  ,\nonumber \\
\vec{F} & = & \vec{J} + \vec{I}.
\label{3.2}\ear
The phases of the above states are taken as in \cite{18}
except for an overall minus sign in all radial wave functions. To choose
some definite phase convention is important for our discussion of consequences
of time reversal (T) invariance below. For ease of notation we will denote
the indices $l, j, F, F_3$ collectively by letters $a,b$,...
in the following. The orthonormal eigenstates of $H_0$ which we take as basis
states
for our considerations are then the $n=2$ atomic states $ |2,a\rangle$
\be
H_0 |2,a\rangle=E_{2,a}^0 |2,a\rangle
\label{3.3}\ee
and the $n=1$ atomic states plus photon states $ |1,b;\gamma , k\rangle$
\be
H_0 |1,b; \gamma ,k\rangle=(E_{1,b} + E_{\gamma ,k}) |1,b; \gamma ,k\rangle.
\label{3.4}\ee
Here we consider the Fock-space of al photon states (with arbitrary number of
photons) and choose a convenient basis there, labelling by $\gamma, k,
(k=1,2,\dots)$ these basis states.
In the approximations stated above the $n=1$ levels are stable and stay
unperturbed. The state vector $|t\rangle$ at time $t$ for the situation
outlined
above can now be written as follows: (cf. \cite{12}):
\be
|t\ket{} \, =\,  \left(  \begin{array}{c}
                 \Psi (t) \\
                 \Phi_1(t)    \\
                 \Phi_2(t) \\ \stackrel{.}{\stackrel{.}{\stackrel{.}{}}} \\
                \end{array} \right)
\label{3.5}\ee
where
\be
\Psi(t) = \left( \bra{}2,a | t \ket{} \right) =
\left( \begin{array}{c}
\bra{}2,1 | t \ket{} \\
\bra{}2,2 | t \ket{} \\
\vdots
\end{array}
\right)
\label{3.6}\ee
is the projection  of the state vector onto the subspace of the $n=2$ atomic
states and the
\be
\Phi_k(t) = \left( \bra{}1,b;\gamma ,k | t\ket{} \right)
=
\left( \begin{array}{c}
\bra{}1,1;\gamma,k | t \ket{} \\
\bra{}1,2;\gamma,k | t \ket{} \\
\end{array}
\right)
\label{3.7}\ee
$(k=1,2,...)$ are the projections onto the subspaces corresponding to the $n=1$
atomic states times the photon state $\gamma,k$.

\subsection{The Wigner-Weisskopf solution}
The state vector $|t\rangle$ satisfies Schr\"odinger's equation:
\be
i \frac{\partial}{\partial t} \, |t\rangle = H \, |t\rangle
\label{3.8}.
\ee
We want to construct a solution of (\ref{3.8}) with the initial condition of
having
 some pure $n=2$ state at $t=0$:
\be
|t=0\ket{} \,\,\, = \, \left(  \begin{array}{c}
                 \Psi (0) \\
                 0    \\
                 0 \\ \stackrel{.}{\stackrel{.}{\stackrel{.}{}}} \\
                \end{array} \right).
\label{3.9}\ee
The Wigner-Weisskopf method gives this solution for $t\geq0$ as
\be
\Psi(t) = e^{-i\underline{\cal M} t} \Psi(0),
\label{3.10}\ee
\bear
\Phi_k(t) &=& \left( \langle1,b;\gamma ,k | t\rangle \right) ,\nonumber \\
\langle1,b;\gamma ,k | t\rangle & = & \sum_a \left( {\cal H}_k
          \frac{e^{-i\underline{\cal M} t}-
          e^{-i(E_{1,b}+E_{\gamma,k}) t}}{\underline{\cal M} -
          E_{1,b}-E_{\gamma,k}}
        \right)_{b,a} \Psi_a(0).
\label{3.11}\ear
Here $\M$ is the non-hermitian mass matrix:
\bear
\underm & = & \undere -\ih \underg \, , \nonumber \\
\undere & = & \left( E_{2,a}^0 \delta_{aa'} + \langle2,a | \hpv +H_{ext} |
2,a'\rangle
               \right) , \nonumber \\
\underg & = & 2 \pi \sum_k {\cal H}_k^{\dagger} \delta(E_1+E_{\gamma,k}
            -E_2) {\cal H}_k
\label{3.12}\ear
Matrices over the $n=2$ subspace will be denoted by a letter with a
bar underneath: $\underm, \underline E$, etc. $\underline E$ is the hermitian,
$\underline \Gamma$ is the antihermitian part of $\M$.
The radiative shift of the hermitian part of $\underm$ has to be thought of
being included in $E^0_{2,a}$. The ${\cal H}_k$ are matrices defined as
\be
{\cal H}_k = (\langle1,b;\gamma,k \, | \, H_{rad} \, | \, 2,a \rangle)
\label{3.13}\ee
and lead from the $n=2$ to the $n=1$ subspace.
The energies $E_1 (E_2)$ occuring in the definition of $\underline \Gamma$ can,
to a sufficient accuracy for our purpose, taken to be the mean energies of the
$n=1 (n=2)$ states.

If we also include  in $\M$  the rest energy of the nucleus and the lepton,
then $\M$ is indeed the physical mass matrix in
complete analogy to the mass matrix as defined for the neutral $K$-meson
system.
However, below we will choose as convenient zero point of the energy scale
the energy center of the $2P_{1/2}$ states, i.e. we will
subtract from $\M$ a piece proportional to the unit matrix which
is irrelevant for our discussion here.

{}From here on we restrict ourselves to atoms with a single electron and leave
the
muon case for further work. The mass matrices are given explicitly for a
nucleus $Z,N$ of spin $I=0,\halb$ and electric field $\vec{\cale}$ in Appendix
B,
where we also collect all relevant matrix elements of $H_{ext}, H_{PV}$, etc.
It is instructive to consider the $Z$-dependence of the relevant energy scales.
We define
\bear
\Delta(Z,N) & = & E_{2P_{3/2}} -E_{2P_{1/2}} , \nonumber \\
L(Z,N) & =&  E_{2S_{1/2}}-E_{2P_{1/2}} ,\nonumber \\
\Gamma_P(Z,N) &\equiv & (\tau_P(Z,N))^{-1} = \Gamma (2P \rightarrow 1S + \gamma
)\left( 1 + {\cal O} (\alpha)\right),
                 \nonumber \\
\Gamma_S(Z,N) &\equiv & (\tau_S(Z,N))^{-1} = \Gamma (2S \rightarrow 1S +
2\gamma )\left( 1 + {\cal O} (\alpha)\right).
\label{3.14}\ear
Here $E_{2P_{3/2}}(E_{2P_{1/2}}, E_{2S_{1/2}})$
is the energy center of the $2P_{3/2}(2P_{1/2},2S_{1/2})$ levels and
 $\Gamma_P(\Gamma_S)$ are the total decay
rates of the $2P(2S)$ states. For the hyperfine splitting and the external
field Hamiltonian $H_{ext}$ the relevant scales are
\bear
\Delta E_{hfs} (Z)& = & \frac{\alpha}{m_e m_p}(\rbz)^{-3} = \frac{\alpha^4
                     m_e^2 Z^3}{m_p},
  \nonumber \\
\calf  (Z,\cale ) & \equiv & e \,\rbz \cale
       =  \frac{e}{Z\alpha m_e} \cale, \label{3.15}\ear
where $r_B(Z)=(Z\alpha m_e)^{-1}$ is the Bohr radius for the atom with
nucleus $Z,N$ and $\E = |\vec \E |$. For the P-violating parts of the mass
matrix we introduce
dimensionless parameters, in essence the ratios of the matrix elements of
$H_{PV}^{(1,2)}$ divided by the Lamb shift:
\bear
\delta_i(Z,N) &=& -\frac{G\sqrt{3}}{64 \pi \sqrt{2}} \frac{Q_W^{(i)}(Z,N)}{m_e}
                    \frac{1}{(\rbz)^4 L(Z,N)} ,\nonumber \\
&&(i  = 1,2).
\label{3.16}\ear

Numerically, we have for ordinary hydrogen $Z=1, N=0$ from \cite{6}:
\bear
\Delta(1,0) & = & 45.364 \, \mu e\mbox{V} \, = h \cdot \, 10969 \, \mbox{MHz}
 = \hbar \cdot 6.892 \cdot 10^{10} \mbox{s}^{-1},             \nonumber \\
L(1,0) & = & 4.376 \, \mu e\mbox{V} \, = h \cdot  \, 1058 \, \mbox{MHz}
= \hbar \cdot 6.648 \cdot 10^{9} \mbox{s}^{-1}.\label{3.17}
\ear
In natural units $\hbar = c = 1$, energies, when measured in $\mbox{s}^{-1}$,
are to be set equal to the corresponding \underline{angular} frequencies. With
this we get
\bear
\Gamma_P(1,0) & = & (1.6 \, \mbox{ns} )^{-1} ,\nonumber \\
\Gamma_S(1,0) & = & (0.12 \, \mbox{s})^{-1} ,\nonumber \\
\frac{\Gamma_P(1,0)}{L(1,0)} & \cong  & 0.09, \nonumber \\
\frac{\Delta E_{hfs}(1)}{L(1,0)}& \cong  &0.18 ,\nonumber\\
\frac{\sqrt{3}\calf (1,|\vec \E |)}{L(1,0)} & = & 1 \hspace{5mm} \mbox{ for}
\hspace{5mm}
        |\vec \E |=477 \, \mbox{V/cm} . \label{3.18}
\ear
For the other hydrogen isotopes the quantities corresponding to (\ref{3.17})
and (\ref{3.18}) are only slightly different.
The P-odd parameters of (\ref{3.16}) for the hydrogen isotopes
\mbox{$Z = 1$},
$ N = 0,1,2$ are given by
\be
\delta_i(1,N)  = -Q_W^{(i)}(1,N) \cdot 6.14 \cdot 10^{-12}.
\label{3.19}\ee

For general nuclei $Z, N$ we can estimate the parameters of the mass matrix
from the
following approximate scaling laws:
\bear
\Delta(Z,N)& \cong  &Z^4 \Delta(1,0) ,\nonumber \\
L(Z,N)& \cong  &Z^4 L(1,0) ,\nonumber \\
\Gamma_P(Z,N)  & \cong  &Z^4  \Gamma_P(1,0)  ,\nonumber \\
\Gamma_S(Z,N)  & \cong  &Z^6  \Gamma_S(1,0) , \nonumber \\
\frac{\Delta E_{hfs}(Z)}{L(Z,N)}   & \cong  & Z^{-1}
           \frac{\Delta E_{hfs}(1)}{L(1,0)} , \nonumber \\
\calf (Z,|\vec \E |) & = &  Z^4 \calf (1,\frac{|\calvece |}{Z^5}) ,
                                \nonumber \\
\delta_i(Z,N) & \cong  & -Q_W^{(i)}(Z,N) \cdot 6.14 \cdot 10^{-12}.
\label{3.20}\ear
Thus, the mass matrix scales in essence as $Z^4$ if the external electric field
is scaled up by $Z^5$. Correspondingly we can scale down the time by $Z^4$ to
obtain a ``reduced $Z=1$'' problem from the general $Z$ case. The weak effects
in this ``reduced'' mass matrix scale like $N$ for $Q_W^{(1)}$ and are in
essence
$Z$-independent for $Q_W^{(2)}$ (cf. Table A2 of appendix A).

\subsection{Diagonalization of the mass matrix and the consequences of T
invariance}
The non-hermitian mass matrix $\underm $ (\ref{3.12}) has right and left
eigenvectors
which are in general different. In the following we will denote the vectors in
the $n=2$ subspace by $|\ )$
to distinguish
them from the complete state vectors $|\ \rangle$. For a constant electric
field $\vec
{\cale}$ in 3-direction, $\vec{\cale}={\cale}\vec e_3,\underm$ commutes
with the 3-component of the  total angular momentum operator:
\be
[\underm , \underf_3] =0.
\label{3.21}\ee
We can diagonalize $\underm$ and $\underline F_3$ simultaneously.
\newline
Right eigenvectors: $| \alpha,F_3 )$
\bear
 \underf_3 \, | \alpha,F_3 ) & = &  F_3 \, | \alpha,F_3 )  ,\nonumber \\
 \underm \, | \alpha,F_3 ) & = &  E(\alpha,F_3) \, | \alpha,F_3 );
\label{3.22}\ear
Left eigenvectors: $\lefteil $
\bear
\lefteil \underf_3 &=& \lefteil F_3   ,\nonumber \\
\lefteil \underm &=& \lefteil E(\alpha,F_3).
\label{3.23}\ear
Here $\alpha=1,2,...$ labels the eigenvectors to the same $F_3$. We will assume
that there is no degeneracy of the complex energy levels for fixed $F_3$:
\be\label{3.23a}
E(\alpha,F_3)\not=E(\alpha',F_3)\quad {\rm for}\quad \alpha\not=\alpha'.\ee
This
is true for all systems in vacuum or in external fields which we will consider.
The condition (\ref{3.23a})  guarantees that the
eigenvectors (\ref{3.22}), (\ref{3.23}) form a basis in the $n=2$
subspace where the normalization can be chosen
such that
\be
\lefteil \alpha',F_3') = \delta_{\alpha,\alpha'} \, \delta_{F_3,F_3'},
\label{3.24}\ee
and
\be \label{3.24a}
( \alpha, F_3 | \alpha, F_3) = 1
\ee
(no summation over $\alpha$ and $F_3$).
We introduce quasiprojectors
\be
\pro (\alpha,F_3) = | \alpha,F_3) \lefteil
\label{3.25}\ee
which satisfy
\bear
\pro(\alpha,F_3)  \pro(\alpha',F_3')  & = & \delta_{\alpha,\alpha'} \,
       \delta_{F_3,F_3'}  \pro(\alpha,F_3)  ,\nonumber \\
\Tr \,\pro(\alpha,F_3) & = & 1
\label{3.26}\ear
but are not necessarily hermitian.
We have
\bear
\underline{\eins} & = & \sum_{\alpha,F_3} \pro(\alpha,F_3) ,\\
\label{3.27}
\underm & = & \sum_{\alpha,F_3}   E(\alpha,F_3) \, \pro(\alpha,F_3) ,\\
\label{3.28}
\underf_3 & = & \sum_{\alpha,F_3}  F_3 \, \pro(\alpha,F_3) .
\label{3.29}\ear

Now we consider T-invariance. Let $V$(T) be the antiunitary operator realizing
the T-transformation in the state vector space of quantum field theory
(Q.F.T.). Choosing conventional
phases (in essence as in \cite{18}, cf. Appendix B) for the basis states
(\ref{3.3}) this induces a definite
matrix representation for T:
\be
V({\rm T}) |2,a\rangle= \sum_b |2,b\rangle \undert_{ba}.
\label{3.30}\ee
where the antiunitarity of $V$(T) implies
\be\label{3.31a}
{\undert}^\dagger{\undert}=\eins.\ee
The explicit form of $\undert $ is given in Appendix B.
The total angular momentum operator $\vec F$ in Q.F.T. satisfies
\be
\left( V({\rm T}) \, \vec{F} \, V^{-1}({\rm T}) \right) ^{\dagger} = -\vec{F} .
\label{3.32}\ee
Therefore we get for the corresponding matrix $\vec{\underline F}$ in the $n=2$
subspace
\be
\left( \undert^{\dagger} \, \vec{\underf} \, \undert \right) ^{T} =
           -\vec{\underf} .
\label{3.33}\ee
Similarly, the electric dipole moment operator
\be
\vec{D} = -e \vecx
\label{3.34}\ee
when restricted to the $n=2$ subspace (cf. Appendix B for the explicit form)
satisfies
\be
\left( \undert^{\dagger} \, \vec{\underd} \, \undert \right) ^{T} =
           \vec{\underd}.
\label{3.35}\ee
The Hamiltonian (\ref{3.1}), including the coupling to the external constant
electric
field, is thus T-invariant:
\be
\left( V({\rm T}) \, H \, V^{-1}({\rm T}) \right) ^{\dagger} = H .
\label{3.36}\ee
This implies
\be
\left( V({\rm T}) \, e^{-iHt} \, V^{-1}({\rm T}) \right) ^{\dagger} = e^{-iHt}
{}.
\label{3.37}\ee
We got the mass matrix $\underm$ as (approximate) solution of Schr\"odinger's
equation. This means that we have
\be
\langle2,a | e^{-iHt} | 2,b\rangle \approx (2,a | e^{-i\underline{\cal M}t} |
2,b).
\label{3.38}\ee
{}From this we get the T-invariance condition for $\underm$  as
\be
\left( \undert^{\dagger} \, \underm \, \undert \right) ^{T} = \underm .
\label{3.39}\ee

Consider now the following double resolvent:
\be
\frac{1}{(\underm - \xi)(\underf_3 -\eta)} = \sum_{\alpha,F_3}
     \frac{\pro(\alpha,F_3)}{(E(\alpha,F_3) - \xi)(F_3 -\eta)}
\label{3.40}\ee
$(\xi,\eta\in{\C})$.
Here we have used (\ref{3.21}) and the properties of the quasiprojectors
(\ref{3.25})
to exhibit the pole structure in $\xi,\eta$ on the r.h.s. of (\ref{3.40}).
Applying the T-transformation (\ref{3.33}), (\ref{3.39}) we get
\bear
\left\{  \undert^{\dagger} \frac{1}{(\underm - \xi)(\underf_3 -\eta)}
     \undert \right\} ^{T} &=& \frac{1}{(\underm - \xi)(-\underf_3 -\eta)} ,\\
\label{3.41}
\sum_{\alpha,F_3}\frac{(\undert^{\dagger}\pro(\alpha,F_3)\undert )^T}
         {(E(\alpha,F_3) - \xi)(F_3 -\eta)} & =&
 \sum_{\alpha,F_3} \frac{\pro(\alpha,F_3)}{(E(\alpha,F_3) - \xi)(-F_3 -\eta)} .
\label{3.42}\ear
{}From a comparison
of the poles in $\xi,\eta$ on the left- and right-hand
sides of (\ref{3.42}) we find that the spectrum of eigenvalues $\lbrace
F_3\rbrace$
must be symmetric around $0$ and that we can choose the numbering
$\alpha=1,2...$ of the complex energy eigenvalues such that
\bear
\label{3.43}
E(\alpha,F_3) & = & E(\alpha,-F_3) ,\\
\left( \undert^{\dagger}\pro(\alpha,F_3)\undert \right) ^T & = &
          \pro(\alpha,-F_3)  .
\label{3.44}\ear
\par These equations embody the consequences of T-invariance for the complex
energy eigenvalues and the eigenvectors of $\underm$.

\subsection{The absence of a linear Stark effect and of a linear shift of
the energy eigenvalues in the P-violating parameters}

It is well known that T invariance forbids a linear Stark effect \cite{19},
\cite{20}. To show that the spin rotations discussed below are not in
contradiction to this, we include here a proof of the absence of a linear Stark
effect.

Let us write the mass matrix in the external electric field $\vec{\cale}=\cale
\cdot
\vec e_3$ as
\bear
\underm (\calvece) & = & \underm_0 - \vec{\underd} \cdot \calvece
                               \nonumber \\
   & = & \underm_0 - \underd_3 \cdot \cale.
\label{3.45}\ear
Indicating the dependence on the electric field explicitly everywhere we get
from (\ref{3.43})
\be
E(\alpha,F_3,\cale \vec{e}_3) = E(\alpha,-F_3,\cale \vec{e}_3).
\label{3.46}\ee
{}From a rotation by $\pi$ around the 2-axis
we find that the numbering of eigenvalues for $\cale$ and $-\cale$
can be chosen such that
\be
E(\alpha,-F_3,\cale \vec{e}_3) = E(\alpha,F_3,-\cale \vec{e}_3).
\label{3.47}\ee
Together, (\ref{3.46}) and (\ref{3.47}) imply that $E(\alpha,F_3,\cale\vec
e_3)$
is an even function of $\cale$:
\be\label{3.47a}
E(\alpha,F_3,\cale\vec e_3)=E(\alpha,F_3,-\cale\vec e_3).\ee
However, by itself (\ref{3.47a}) is \underbar{not} sufficient to show the
absence of a linear Stark effect. We still need differentiability of $E$
with respect to $\cale$. But differentiability with respect to $\cale$
of $E(\alpha,F_3,\cale\vec e_3)$ and, with suitable phase conventions, of
the corresponding right and left eigenvectors, can easily be proven \underbar{
under the condition (\ref{3.23a})}, adapting the methods of ordinary
perturbation
theory of eigenvalues and eigenvectors to the case of non-hermitian matrices
(cf. (4.1)-(4.3) of \cite{12} and Appendix C).

With differentiability (\ref{3.47a}) leads, of course, directly to
\be\label{3.47b}
\left.\frac{\partial E(\alpha,F_3,\cale\vec e_3)}{\partial \cale}\right|
_{{\cal E}=0}=0,\ee
i.e. the absence of a linear Stark effect.

For comparison with later discussion we derive now an expression for the
derivative $\partial E/\partial\cale$ as matrix element of the dipole
operator. For the complex energy eigenvalues we have
  \be
E(\alpha,F_3,\cale \vec{e}_3) =(\widetilde{\alpha,F_3,\cale \vec{e}_3}
      | \underm(\cale \vec{e}_3) | \alpha,F_3,\cale \vec{e}_3).
\label{3.48}\ee
The differentiability with respect to $\cale$ discussed above allows us
to apply
Feynman's formula \cite{21}  which works also for our non-hermitian  mass
matrix:
\bear
\frac{\partial E(\alpha,F_3,\cale \vec{e}_3)}{\partial \cale} & = & \ \
     \left( \frac{\partial }{\partial \cale}
   \lefteile \right) \underm (\cale \vec{e}_3) \righteire  \nonumber \\
 & & +\ \lefteile  \underm (\cale \vec{e}_3)
     \left( \frac{\partial }{\partial \cale}   \righteire \right) \nonumber \\
  & & +\ \lefteile  \left(
     \frac{\partial \underm (\cale \vec{e}_3)}{\partial \cale} \right)
     \righteire  \nonumber \\
 & = & \ \ \left( \frac{\partial }{\partial \cale}
   \lefteile \right)  E(\alpha,F_3,\cale \vec{e}_3) \righteire \nonumber \\
  & & +\ \lefteile  E(\alpha,F_3,\cale \vec{e}_3)
    \left( \frac{\partial }{\partial \cale}   \righteire \right) \nonumber \\
  & & +\ \lefteile  \left(
     \frac{\partial \underm (\cale \vec{e}_3)}{\partial \cale} \right)
     \righteire  \nonumber \\
  & = & \ \ \lefteile  \left(
     \frac{\partial \underm (\cale \vec{e}_3)}{\partial \cale} \right)
     \righteire.
\label{3.49}\ear
Here we used (\ref{3.22})-(\ref{3.24}). With (\ref{3.45}) we get thus
\bear
\frac{\partial E(\alpha, F_3, \E \vec e_3)}{\partial \cale} & = &-\lefteile \,
\underd_3 \, \righteire
     \nonumber \\
 & = &-\Tr \left[ \, \underd_3 \, \pro(\alpha,F_3,\cale \vec{e}_3) \right] .
\label{3.50} \ear
and for $\E=0$   we find now from  (\ref{3.47b})
\be
\left.\frac{\partial E(\alpha, F_3, \E \vec e_3)}{\partial \cale}\right|_{\E=0}
=  -\left. \lefteile  \, \underd_3 \, \righteire  \right|_{{\cal E} =0} =  0.
\label{3.51}\ee

We will now prove a theorem related to the dependence of the energy
eigenvalues on the P-violating parameters $\delta_i$ (\ref{3.16}):
Let the external electric field and the mass matrix be as in (\ref{3.45}) and
let us assume that the nondegeneracy of complex energy eigenvalues
(\ref{3.23a}) holds also in the absence of P violation, i.e. for $\delta_i=0$
$(i=1,2)$, and in a whole interval around $\delta_i=0$ including the physical
values (\ref{3.16}) and their negatives. Then there is no shift of the
complex energy eigenvalues linear in $\delta_i$ $(i=1,2)$.

We show this as follows. For $\delta_i=0$, but $\cale\not=0$, the external
field breaks, of course, P invariance, but leaves the mass matrix still
invariant
under a reflection R on the 1-3 plane. R can be considered as product of a
parity operation followed by a rotation by $\pi$ around the 2-axis:
\bear\label{3.51a}
&&\mbox{R}:\left(\begin{array}{c}x_1\\ x_2\\
x_3\end{array}\right)\longrightarrow
\left(\begin{array}{r}x_1\\ -x_2\\ x_3\end{array}\right),\nonumber\\
&&\mbox{R}=e^{i\pi F_2}\cdot \mbox{P}.\ear
This induces a transformation $\underline{\R}$ in the space of $n=2$ states as
given
explicitly for $I=0$ and $I=1/2$ in appendix B.
\be\label{3.51b}
\mbox{R}: |2,a\bigr)
\to\underline{\R}|2,a\bigr)=\sum_b|2,b\bigr)\underline{\R}_{ba}.
\ee
Clearly, we have
\bear
\underline{\R}^\dagger\cdot\underline{\R}&=&{\eins},\nonumber\\
\underline{\R}^\dagger\underline
{F}_3\underline{\R}&=&-\underline{F}_3,\nonumber\\
\underline{\R}^\dagger{\M}(\cale\vec e_3,\delta_1,\delta_2)\underline{\R}
&=&{\M}(\cale\vec e_3,-\delta_1,-\delta_2).
\label{3.51c}
\ear
Here and in the following part of this subsection we indicate explicitly the
dependence of all quantities on $\delta_{1,2}$. From (\ref{3.51c})
we find easily that the numbering of eigenvalues for $\pm\delta_i$ can be
chosen such that
\bear
E(\alpha,F_3,\cale\vec e_3,\delta_1,\delta_2)&=&E(\alpha,-F_3,\cale\vec
e_3,-\delta_1,-\delta_2),\nonumber\\
\underline{\R}^\dagger{\pro}(\alpha,F_3,\cale\vec
e_3,\delta_1,\delta_2)\underline{\R}&=&{\pro}(\alpha,-F_3,\cale\vec
e_3,-\delta_1,-\delta_2).
\label{3.51d}
\ear
The T-invariance relation (\ref{3.43}) gives
\be\label{3.51e}
E(\alpha,-F_3,\cale\vec e_3,-\delta_1,-\delta_2)=E(\alpha,F_3,\cale
\vec e_3,-\delta_1,-\delta_2)\ee
and from (\ref{3.51d}), (\ref{3.51e}) we get
that the energy eigenvalues are symmetric functions of $\delta_i$:
\be\label{3.51f}
E(\alpha,F_3,\cale \vec e_3,\delta_1,\delta_2)=E(\alpha,F_3,
\cale\vec e_3,
-\delta_1,-\delta_2).\ee
Using now the assumption of non-degeneracy stated above and the results of
Appendix C we have that $E$
is differentiable with respect to $\delta_{1,2}$ and thus
\be\label{3.51g}
\left.\frac{\partial E(\alpha,F_3,\cale\vec e_3,\delta_1,\delta_2)}{\partial
\delta_i}\right|_{\delta_1=\delta_2=0}=0\quad{\rm for}\quad (i=1,2)\ee
q.e.d.
With the help of Feynman's formula we find also a relation analogous
to (\ref{3.50}):
\bear\label{3.51h}
\frac{\partial}{\partial\delta_i}E(\alpha,F_3,\cale\vec e_3,\delta_1,\delta_2)
&=&\widetilde{ (\alpha,F_3,\cale\vec e_3,\delta_1,\delta_2 }|\frac{\partial}
{\partial\delta_i}{\M}(\cale\vec e_3,\delta_1,\delta_2)|\alpha,F_3,\cale
\vec e_3,\delta_1,\delta_2)\nonumber\\
&=&{\rm Tr}\left[ \pro(\alpha,F_3,\cale\vec e_3,\delta_1,\delta_2)\,
\frac{\partial}{\partial \delta_i}{\M}(\cale\vec
e_3,\delta_1,\delta_2)\right],\nonumber\\
&&(i=1,2).\ear
In a situation which is not symmetric with respect to a suitable reflection the
above theorem, i.e. (\ref{3.51g}), will in general not hold. Then, linear
shifts of the complex energies in $\delta_{1,2}$ are in principle possible.
This will be elaborated on in a future publication.
\subsection{Polarization rotations of $n=2$ states}
We come now to the central question which we want to answer in this paper: Is
the
polarization of the atom rotated by the electric field? We will
always consider
a pure state $|\Psi(0))$ at $t=0$. The generalization to a mixed state is
straightforward.

The expectation value of the angular momentum operator for the
undecayed states in the electric field $\vec{\E}=\E \vec{e}_3$ at
time $t$ is (cf. (\ref{3.10}))
\bear
\label{3.52}
\vec{F}(t) &=& (\Psi(t) | \underline{\vec{F}}|\Psi(t)) \nonumber \\
&=& (\Psi(0)|e^{\textstyle i \M^\dagger(\vec{\cal E})t}\, \underline{\vec{F}}
\,e^{\textstyle -i\M( \vec{\cal E})t}|\Psi(0)).
\ear
The angular momentum per unit norm is
\be
\label{3.53}
\hat{\vec{F}}(t) = \vec{F}(t) / \N(t)
\ee
where
\be
\label{3.54}
\N(t) = (\Psi(t) | \Psi(t)).
\ee
The time derivatives are:
\bear
\label{3.55}
\stackrel{.}{\vec{F}}(t) &=& i (\Psi(t)|\M^\dagger(\vec{\E})
\underline{\vec{F}}
- \underline{\vec{F}}\,\M(\vec{\E}) |\Psi(t)) \nn
&=& i (\Psi(t) | \left( \left[ \underline{E},\underline{\vec{F}} \right] +
\ihalb \left\{ \underline{\Gamma}, \underline{\vec{F}} \right\}_+ \right) |
\Psi(t)) \nn
&=& ( \Psi(t) | \left( \underline{\vec{D}} \times \vec{\E} -
\halb \left\{ \underline{\Gamma}, \underline{\vec{F}} \right\}_+ \right) |
\Psi(t)), \\
\label{3.56}
\dot{\N}(t) &=& - (\Psi(t) | \underline{\Gamma} | \Psi(t)).
\ear
Here we used (\ref{3.12}) and (\ref{3.45}) and the fact that only the
coupling to the external electric field violates rotational invariance.

It is clear that in general both $\dot{\vec F}(t)$ and $\dot{\hat {\vec F}}(t)$
will be different from zero. An explicit example will be given in section 5.
In the following we will call this rotation of type I.

What is then the basis for the claim in \cite{19}, \cite{22} that the
angular momentum per unit norm stays constant? To analyse this we
consider an external field $\vec{\E}={\E}\vec e_3$ where (cf. (\ref{3.15}) ff)
\be
\label{3.57}
\frac{\sqrt{3} \F(Z, |\vec{\E}|)}{L(Z,N)} \ll 1.
\ee
The electric field is then a small perturbation compared to the Lamb shift
and we will, also in the presence of the electric field, have metastable states
which contain
mostly $2 S$ with only small $2 P$ admixtures due to $H_{PV}$ and $H_{ext}$.
Let
the corresponding eigenstates of $\M(\E\vec{e}_3)$ be
$$ | 2 \hat{S}, \alpha, F_3, \E \vec{e}_3) $$
where $\alpha$ denumbers the --- in general more than one --- linearly
independent
states to the same $F_3$. In addition, there are the eigenstates of
${\M}(\cale\vec e_3)$ containing mostly $2P$ with small admixtures of $2S$:
\begin{displaymath}
|2\hat P,\beta,F_3,\cale\vec e_3)\end{displaymath}
which have practically the same short lifetime $\tau_P$ as the unperturbed
$2P$ states.
We place now the atom in some $n=2$ state $|\Psi(0))$ into
the electric field at $t=0$. The state vector $|\Psi(t))$ can be expanded
in terms of the eigenstates of ${\M}(\cale\vec e_3)$, i.e. in terms of
the slightly perturbed $2S$ and $2P$ states discussed above:
\bear\label{3.57a}
&&|\Psi(t))=\sum_{\alpha,F_3}c_{\alpha,F_3}(t)|2\hat S,\alpha,F_3,\cale\vec
e_3)+\sum_{\beta,F_3}c_{\beta,F_3}'(t)|2\hat P,\beta,F_3,\cale\vec e_3),
\nonumber\\
(t\geq0).\ear
Here
\bear
c_{\alpha,F_3}(0)&=&(2\widetilde{\hat S,\alpha,F_3,}\cale\vec
e_3|\Psi(0)),\nonumber\\
c_{\beta,F_3}'(0)&=&(2\widetilde{\hat P,\beta,F_3,}\cale\vec e_3|\Psi(0)),
\label{3.57b}
\ear
and the time evolution of the coefficients $c$ is according to (\ref{3.22}),
(\ref{3.23})
\bear\label{3.58}
c_{\alpha,F_3}(t)&=&e^{-i E(2\hat S,\alpha,F_3,{\cal E}
\vec{e}_3)t}c_{\alpha,F_3}(0),
\nonumber\\
c_{\beta,F_3}'(t)&=&e^{-i E(2\hat P,\beta,F_3,{\cal E}
\vec{e}_3)t}c_{\beta,F_3}(0).
\ear

After a time $t$ much longer than the decay
time  $\approx \tau_P$ of the states $|2\hat P,\beta,F_3,\cale\vec e_3)$,
only the metastable part  of the state
vector will survive and we will have
\be
\label{3.59}
|\Psi(t)) = \sum_{\alpha, F_3} c_{\alpha,F_3}(t) | 2 \hat{S}, \alpha, F_3, \E
\vec{e}_3)
\ee
$$(t \gg \tau_P).$$
Consider now the case of a spin-0 nucleus, $I=0$. For $\E=0$ the total angular
momentum of the $2 S$ states is given by the electron spin and we have
$F=\halb$.
In the electric field only $F_3$ remains as a good quantum number. Under the
above assumptions we have no degeneracy of $2 S$ states for fixed $F_3$ and we
get only
\underbar{two} linearly independent metastable states also for $\E\not=0$
which we denote by
\be\label{3.60}
| 2\hat{S}, F_3, \E \vec{e}_3),\qquad F_3 = \pm \halb.
\ee
T-invariance requires these states to have the \underbar{same} complex energy
eigenvalue (cf. (\ref{3.43})):
\be
\label{3.61}
E(2\hat{S}, \halb, \E \vec{e}_3) = E(2\hat{S},-\halb, \E \vec{e}_3).
\ee
Thus, the state vector of the undecayed state (\ref{3.59}) evolves with time as
\be
\label{3.62}
|\Psi(t)) = e^{-i E(2\hat{S}, \shalb,\SS {\cal E} \vec{e}_3) t}
\sum_{F_3 = \pm \shalb} c_{F_3}(0) | 2 \hat{S}, F_3, \E \vec{e}_3)
\ee
$$(t \gg \tau_P)$$
i.e. it is only changed by multiplication with a complex number.

It is clear that in this case the angular momentum per unit norm
(\ref{3.53}) stays constant
\be
\label{3.63}
\hat{\vec{F}}(t) = const. \qquad \mbox{ for  } t \gg \tau_P.
\ee
Going through the arguments of \cite{19}, \cite{22}, we find that this is
the case considered there and thus we reproduce these results.

However, the case of only two linearly independent metastable states is clearly
a very
special one. In section 6 we will give explicit examples of polarization
rotations of metastable states for the case of nuclear spin $I\not=0$. We will
call this rotation of type II.

As a last situation we consider placing an atom which is in a metastable
$2 \hat S$ state corresponding to $\E=0$ for $t<0$ into the electric field
$\vec{\E}=\E\vec{e}_3$ at time $t=0$, leaving it there until a time
$t_1>0$ and then switching
off $\vec{\E}$ again. The undecayed part of the state vector will then
contain both $2 \hat S$ and $2 \hat P$ states (corresponding to $\vec{\E}=0$
again).
The $2 \hat P$ part will decay quickly, leaving us with the $2 \hat S$ part
(which we will construct, below, by a suitable projection of the state vector
at time $t_1$).  Is the polarization
of this $2 \hat S$ part
the same as for the original $2 \hat S$ state at $t=0$?

We will show that for nuclear spin $I=0$ where we have only 2 linearly
independent
$2 \hat S$ states the polarization stays the same whereas in the general case
it does
\underbar{not}.

To prove the first part of this statement consider thus nuclear spin $I=0$
and denote the $2 \hat{S}$ states as in (\ref{3.60}). We start at $t=0$ with a
state
\be
\label{3.64}
|\Psi(0)) = c_{\shalb} |2 \hat{S}, \halb,0) + c_{-\shalb} |2\hat{S}, -\halb,0)
\ee
where $c_{\pm\frac{1}{2}}$ are complex numbers such that
\be
\label{3.65}
( \Psi(0) | \Psi(0)) = 1.
\ee
After time evolution of this state until time $t_1$ in the electric field
$\vec{\E}=\E\vec{e}_3$ and subsequent projection onto the $2 \hat{S}$ states of
zero
field we get the state
\bear
\label{3.66}
| \Psi_{pr}(t_1)) &=& \sum_{F^{'}_3 = \pm \shalb} | 2 \hat{S}, F^{'}_3,0)
( \tilde{ 2 \hat{S},F^{'}_3, 0}| e^{-i \M(\E \vec{e}_3) t_1} | \Psi(0)) \nn
&=& \sum_{F'_3, F_3 = \pm \shalb}
| 2 \hat{S}, F^{'}_3,0)
( \tilde{ 2 \hat{S},F^{'}_3, 0}| e^{-i \M(\E \vec{e}_3) t_1} | 2 \hat S, F_3,0)
\, c_{F_3}.
\ear
Let us define
\be\label{3.67}
f_{F^{'}_3, F_3}(t_1) :=
( \tilde{ 2 \hat{S},F^{'}_3, 0}| e^{-i \M(\E \vec{e}_3) t_1} |
 2 \hat{S},F_3, 0).
\ee
Rotational invariance around the 3-axis gives
\be\label{3.68}
f_{F^{'}_3, F_3 }(t_1) = 0 \qquad \mbox{ for  }F^{'}_3 \neq F_3 .
\ee
The diagonal elements can be written as
\be\label{3.69}
f_{F_3, F_3}(t_1) = \Tr \left[ e^{-i \M(\E \vec{e}_3) t_1}  \pro ( 2 \hat{S},
F_3,0)\right],
\ee
where $\pro$ are the quasiprojectors (\ref{3.25}).
Now we can apply the T-invariance conditions (\ref{3.39}), (\ref{3.44})
to get
\bear\label{3.70}
f_{\shalb,\shalb}(t_1) &=&
\Tr \left[ e^{-i \M(\E \vec{e}_3) t_1}  \pro ( 2 \hat{S}, \halb,0)\right] \nn
&=& \Tr\left[ \pro^T(2 \hat{S}, \halb,0) e^{-i \M^T(\E\vec{e}_3) t_1}\right]
\nn
&=& \Tr\left[ \pro(2 \hat{S}, - \halb,0) e^{-i \M(\E\vec{e}_3) t_1}\right] \nn
&=& \Tr \left[ e^{-i \M(\E \vec{e}_3) t_1}  \pro ( 2 \hat{S}, - \halb,0)\right]
\nn
&=& f_{-\shalb,-\shalb}(t_1).
\ear
Inserting (\ref{3.68}) and (\ref{3.70}) in (\ref{3.66})
we get
\be\label{3.71}
|\Psi_{pr}(t_1)) = f_{\shalb,\shalb}(t_1) |\Psi(0)).
\ee
The projected state at time $t_1$ is a multiple of the initial state and no
polarization-rotation occurs, q.e.d..

This theorem does again not apply if there are more than 2 linearly independent
$2\hat S$ states, i.e. for nuclear spin $I>0$ (cf. section 5). We will call the
resulting rotation type III. Another way to circumvent in a certain sense
the above theorem even for $I=0$ will be discussed in section 4 below.
\section{Rotation effects in an electric field, piecewise constant in time}
\setcounter{equation}{0}

Consider a situation where we have an electric field which is piecewise
constant in
time. Let $t_k$ $(k=0,1,2,...,K)$ be a sequence of times such that
\be\label{4.1}
t_0 = 0, \quad t_{k-1} < t_k\ee
and let us assume that
\bear\label{4.2}
\vec{\E}&=& 0 \qquad\qquad\qquad \mbox{ for } t < t_0 = 0 \quad \mbox{ and } t
> t_K, \nn
\vec{\E}(t) &=& \vec{\E}^{k} = const. \qquad \mbox{for } t_{k-1} < t<t_k
\qquad(1\leq k\leq K).
\ear
The changes of the electric field at times $t_k$ should be such that they
are adiabatic with respect to transitions between states of different $n$
but sudden for the transitions between the $n=2$ sublevels. This implies
for the switching times $\Delta t$ the condition:
\be\label{4.3}
\left(E(n=3) - E(n=2)\right)^{-1} = \frac{36}{5 Z^2  Ry} \ll \Delta t \ll
\frac{1}{\Delta(Z,N)},
\ee
or in numbers:
\be\label{4.4}
Z^{-2}\, 3.5 \times 10^{-16} \mbox{s} \ll \Delta t \ll Z^{-4}  \, 1.5 \times
10^{-11} \mbox{s}.
\ee

The mass matrix will now be different for each time interval:
\be\label{4.5}
\M = \M(\vec{\E}^k) \qquad \mbox{ for } t_{k-1} < t < t_k.
\ee
Note that in general $\M(\vec{\E}^k)$ and $\M(\vec{\E}^{k'})$ will
\underbar{not}
commute for $k\neq k'$. This holds true even if $\vec{\E}^k$ and
$\vec{\E}^{k'}$
are parallel but of different length. For the evolution of the state vector
of the undecayed state we get now
\be
\label{4.6}
| \Psi(t)) = e^{\textstyle - i \M(\vec{\cal E}^k)(t - t_{k-1})} \ldots
e^{\textstyle -i \M(\vec{\cal E}^1)t_1}
| \Psi(0)),
\ee
$$(t_{k-1} < t < t_k).$$
Consider now the same type of situation as at the end of section 3. We start at
$t=0$ with some $2\hat S$ state, place it in electric fields $\vec{\E}^1,
\vec{\E}^2$, choosing $K=2$ for simplicity, and project onto the
$2\hat S$ states for $\vec{\E}=0$ at time $t_2$. Has the polarization changed?
In general it has changed even for $I=0$ nuclear spin. Indeed, consider this
case and
let $\vec{\E}^{1,2}$ be in 3-direction
\be\label{4.7}
\vec{\E}^{1,2} = \E^{1,2} \vec{e}_3.
\ee
Let the initial state be as in (\ref{3.64}). The projected state vector at time
$t_2$ is (cf. (\ref{3.66}), (\ref{3.67})):
\be
\label{4.8}
|\Psi_{pr}(t_2)) = \sum_{F_3^{'},F_3 = \pm \shalb} | 2 \hat{S}, F'_3,0)
f_{F'_3,F_3}(t_2,t_1) \, c_{F_3},
\ee
where now
\be
\label{4.9}
f_{F'_3,F_3}(t_2,t_1) = ( \tilde{ 2 \hat{S}, F'_3,0}|
e^{-i \M(\E^2\vec{e}_3)(t_2-t_1)}
e^{-i \M(\E^1\vec{e}_3) t_1}
| 2 \hat{S}, F_3,0).
\ee
Rotation invariance around the 3-axis gives as in (\ref{3.68})
\be\label{4.10}
f_{F'_3,F_3}(t_2,t_1) = 0\qquad \mbox{ for } F'_3\neq F_3.
\ee
The diagonal elements can be written as
\be\label{4.11}
f_{F_3,F_3}(t_2,t_1) = \Tr\left[
e^{-i \M(\E^2\vec{e}_3)(t_2-t_1)}
e^{-i \M(\E^1\vec{e}_3)t_1} \,\pro(2 \hat{S}, F_3,0)\right].
\ee
The T-invariance conditions (\ref{3.39}), (\ref{3.44}) give now
\be\label{4.11a}
f_{F_3,F_3}(t_2,t_1) = \Tr\left[
e^{-i \M(\E^1\vec{e}_3)t_1}
e^{-i \M(\E^2\vec{e}_3)(t_2-t_1)} \,\pro(2 \hat{S}, - F_3,0)\right].
\ee
Since in general $\M(\E^1\vec{e}_3)$ and $\M(\E^2\vec{e}_3)$ do not commute for
$\E^1\neq \E^2$, the r.h.s. of (\ref{4.11a})
is \underbar{not} equal to $f_{-F_3,-F_3}(t_2,t_1)$. Thus the state vector
$|\Psi_{pr}(t_2))$ (\ref{4.8})
is \underbar{not} a multiple of $|\Psi(0))$ and a rotation of the polarization
occurs. The relations (\ref{4.11}), (\ref{4.11a}) say that
\be\label{4.12}
f_{F_3,F_3}(t_2,t_1) = f_{-F_3,-F_3}(t_2,t_2 - t_1) \left|_{\E^1
\leftrightarrow\E^2}\right.
\ee
which can be used as check in numerical calculations.

To make sure that indeed a rotation which we will call type IV occurs, i.e.
that
\be\label{4.13}
\Delta f(t_2,t_1) := f_{\shalb,\shalb}(t_2,t_1) - f_{-\shalb,-\shalb}(t_2,t_1)
\neq 0,
\ee
we calculated the first few terms in the power series expansion of $\Delta f$
in
$t_1$ and
$(t_2-t_1)$:
\be\label{4.14}
\Delta f(t_2,t_1) = \sum_{r_2 = 0}^\infty \sum_{r_1 = 0}^\infty
\frac{(t_2 - t_1)^{r_2}}{r_2!}
\frac{(t_1)^{r_1}}{r_1!}(-i)^{r_2+r_1} \Delta f^{r_2,r_1}.
\ee
We find (cf. Appendix D)
\bear\label{4.15}
\Delta f^{r_2,0} &=& 0 \qquad \mbox{ for } r_2 = 0,1,2,\ldots, \nn
\Delta f^{0,r_1} &=& 0 \qquad \mbox{ for } r_1 = 0,1,2,\ldots, \\
\Delta f^{1,1} &=& 0,\nn
\Delta f^{1,2} &=&\Delta f^{2,1}= 0, \\
\Delta f^{1,3} &=&\Delta f^{3,1}= 0, \nn
\Delta f^{2,2} &=& i \E^1 \E^2 (\E^1 - \E^2) (e r_B(Z))^3 \, 12 \sqrt{3} \,
\delta_1(Z,N) \Delta(Z,N) \cdot \nn
\label{4.18}
&& \frac{L(Z,N)}{L(Z,N) - \frac{i}{2}(\Gs(Z,N) - \Gp(Z,N))}.
\ear
The dependence of $\Delta f^{2,2}$ on $(r_B(Z))^3$ reveals this quantity not to
be related to the atomic
dipole moment. It should be interpreted as a sort of octopole term.

For nuclear spin $I\not=0$ there are also terms \underbar{linear} in $r_B(Z)$
in quantities analogous to  $\Delta f$ leading to a polarization rotation even
for a single $\vec{\E}$-field,
i.e. for $t_2=t_1$ (cf. section 6, (\ref{6.32}), \ref{6.33})).

In sections 3 and 4 we have discussed 4 types of polarization rotations which
we summarize together with a fifth type, the angular momentum transfer process
studied in ref. \cite{12}, in Table 1.

Up to now we have always considered an atom at rest in an electric field. In
practice one will presumably deal with beams of neutral or charged atoms
entering and leaving electric fields. In this situation the atoms will
experience in their rest system also a magnetic field and ions will in addition
be accelerated. Can one hope to see then P-violating effects at all? In
principle yes. Let us discuss the simple example of an atomic beam traversing
the electric field of a plane plate capacitor (Fig. 2).
Let $\vec \E$ be in 3-direction and let the beam enter in 1-direction. Instead
of the parity operation P we consider now the reflection R on the 1-3 plane
(\ref{3.51a}).
 Let us choose, for instance, in the above situation the polarization state of
the atoms to be initially R symmetric. The initial state and the complete
set-up are then R symmetric. Observation of a R asymmetric polarization state
of the atoms at some later time would then be an unambiguous signal of R- and
thus P-violation.

In Fig. 3 we have schematically indicated the type I - V rotation effects
listed in Table 1 for the situation of an atomic beam entering plane plate
capacitors.
\section{Results for nuclear spin $I=0$}
\setcounter{equation}{0}

According to the theorems of subsection 3.5, the type II
and III rotations cannot occur for $I=0$. Thus we are left with types I, IV, V.
Possible
candidate atoms are $^4_2\mbox{He}^+$
or $^{12}_6\mbox{C}^{5+}$.
Some properties of these ions are collected in Table 2.
The mass matrix, T- and R-transformation matrices, energy eigenvalues
and right and left eigenvectors for the case $I=0$ are given in Tables B1--B4
in Appendix B.
A $2\hat S$ state in vacuum $(\vec{\E}=0)$ at $t=0$ is in general described by
a density matrix:
\be\label{5.1}
\underline{\varrho}(0) = \sum_{F'_3,F_3 = \pm \shalb} | 2 \hat{S}, F'_3,0)
\,\halb \left( \eins + \vec{s} \vec{\sigma} \right)_{F'_3,F_3} ( 2 \hat{S},
F_3,0|,
\ee
$$|\vec{s}| \leq 1,$$
where $\vec{s}$ is the polarization vector.
\subsection{Type I rotation}
Let us discuss first the type I rotation. (cf. subsection 3.5) and consider
a pure $2\hat S$ state (\ref{3.64}), (\ref{3.65}), at $t=0$, corresponding to
$|\vec s|=1$ in (\ref{5.1}). The expectation value of the
total angular momentum of the quantum state consists
of the part due to the undecayed state: $\vec F(t)$ (\ref{3.52}) and due to the
decay products: $\tilde{\vec{F}}(t)$,
\be\label{5.2}
\vec{F}_{tot}(t) = \vec{F}(t) + \tilde{\vec{F}}(t).
\ee
{}From (2.3), (2.4), (2.27) of \cite{12}, we get
\be\label{5.3}
\dot{\vec{F}}_{tot}(t) = ( \Psi(t) | \underline{\vec{D}} | \Psi(t)) \times
\vec{\E}
\ee
which, together with (\ref{3.55}) leads to
\be\label{5.4}
\dot{\vec{\tilde{F}}}(t) = \halb ( \Psi(t) | \left\{ \underline{\Gamma},
\underline{\vec{F}} \right\}_+ | \Psi(t)).
\ee
The initial condition is
\be\label{5.5}
\tilde{\vec{F}}(0) = 0,
\ee
and thus
\be\label{5.6}
\vec{\tilde{F}}(t) = \halb \int_0^t dt' (\Psi(t') | \left\{ \underline{\Gamma},
\underline{\vec{F}} \right\}_+ | \Psi(t')).
\ee
For the undecayed part of the state vector the norm is $\N(t)$ (\ref{3.54}),
for the part corresponding to the decay products the norm is $1-\N(t)$ and
the angular momentum  per unit norm
\be\label{5.7}
\hat{\tilde{\vec{F}}}(t) = \frac{\vec{\tilde{F}}}{1 - \N (t)} \ \ .
\ee
The linear superposition principle and rotation covariance allow us to write
general expansions for $\N(t)$ and $\vec F(t)$:
\bear\label{5.8}
\N (t) &=& a_1(t,|\vec{\E}|^2) + a_2(t,|\vec{\E}|^2) \vec{s} \cdot \Ehat, \\
\vec{F}(t) &=&
b_1(t,|\vec{\E}|^2) \vec{s} +
b_2(t,|\vec{\E}|^2) \left(\vec{s} \cdot\Ehat \right) \Ehat + \nn
&& b_3(t,|\vec{\E}|^2) \vec{s} \times \Ehat + b_4(t,|\vec{\E}|^2)  \Ehat,
\label{5.9}\ear
where $\hat{\vec\E}=\vec{\E}/|\vec{\E}|$ and the $a_i(t,|\vec{\E}|^2)
\ (i=1,2), \,b_i(t,|\vec{\E}|^2)\ (i=1,...,4)$
are scalar functions. Expansions similar to (\ref{5.9}) hold for $\vec
F_{tot}(t)$ and $\vec{\tilde F}(t)$.
The functions $a_1,b_1,b_2$ are P-conserving, $a_2,b_3,b_4$ P-violating.

For a constant field $\vec{\E}$ the component of the angular momentum in the
direction of $\vec{\E}$ is conserved and we have
\be\label{5.9a}
\left[\underline{\vec{F}}\cdot \hat{\vec{\E}},\,\M(\vec{\E})\right]
= 0.
\ee
This means that for an initial state with
\be\label{5.9b}
\vec{s} = \pm \hat{\vec{\E}}
\ee
the undecayed state at time $t$ must be an eigenstate of
$\underline{\vec{F}}\cdot \hat{\vec{\E}}$ with eigenvalue $\pm \frac{1}{2}$.
Thus we must have
\be\label{5.9.c}
\vec{F}\cdot \hat{\vec{\E}} = \pm \halb {\cal N}(t)
\ee
$$ \mbox{for } \vec{s} = \pm \hat{\vec{\E}}.$$
This implies
\bear \label{5.9d}
b_1(t,|\vec{\E}|^2) + b_2(t,|\vec{\E}|^2)  &=& \halb a_1(t,|\vec{\E}|^2),
\\
b_4(t,|\vec{\E}|^2) &=& \halb a_2(t,|\vec{\E}|^2)
\ear
and we have only 4 independent functions $a,b$ in (\ref{5.8}), (\ref{5.9}),
which we take to be $a_1$, $b_1$, $b_3$, $b_4$.

In the following calculations we will always expand in the P-violating
parameter $\delta_1(Z,N)$ (cf. (\ref{3.16}), (\ref{3.20})) and keep only the
leading term in this very small quantity.

For time $t=0$ we find from (\ref{3.65}), (\ref{3.56}):
\bear\label{5.10}
\N (0) &=& 1 ,\nn
\dot{\N}(0) &=& - ( \Psi(0) | \underline{\Gamma} | \Psi(0)) \nn
&=& i ( \Psi(0) | \M^\dagger(0) - \M(0) | \Psi(0)) \nn
&=& - \Gs(Z,N),
\ear
where $\M(0)$ is the mass matrix and $\Gs(Z,N)$ the decay rate
of the $2 \hat{S}$ states at zero field, $\vec{\E}=0$ (cf. (\ref{3.14})). From
(\ref{3.52}) and (\ref{3.55}) we get
\bear\label{5.11}
\vec{F}(0) &=& \halb \vec{s},  \\
\label{5.12}
\dot{\vec{F}}(0) &=& \Tr\left[ \underline{\varrho}(0)
\underline{\vec{D}}\right]
\times \vec{\E}  - \halb \Tr\left[ \underline{\varrho}(0) \left\{
\underline{\Gamma},\underline{\vec{F}}\right\}_+\right].
\ear
{}From rotational invariance we can write
\bear\label{5.13}
\Tr\left[ \underline{\varrho}(0) \underline{\vec{D}}\right] &=& (\Psi(0)
| \underline{\vec{D}} | \Psi(0)) \nn
&=& d_2 \vec{s},
\ear
where $d_2$ is the Zeldovich electric dipole moment \cite{23} of the unstable
$2 \hat{S}$
states at $\vec{\E}=0$.
We find (cf. (4.9) of \cite{12}, where we used
$\Gamma_S\ll\Gamma_P$):
\be\label{5.14}
d_2 = - \delta_1(Z,N) \sqrt{3} \Qe \, r_B(Z) \frac{L(Z,N) ( \Gp(Z,N) -
\Gs(Z,N))}{L^2(Z,N) + \frac{1}{4} (\Gp(Z,N) - \Gs(Z,N))^2} \ \ .
\ee
Note that (\ref{5.13}) implies for instance for $\vec{s}=\vec{e}_3$:
\be\label{5.15}
(2 \hat{S}, \halb,0| \underline{D}_3 | 2 \hat{S},\halb,0) = d_2 \neq 0,
\ee
where the notation for the state vectors is as in (\ref{3.60}).
This is \underbar{not} in contradiction with the absence of a linear Stark
effect which gave (\ref{3.51}), where a \underbar{left} eigenvector of $\M(0)$
stands as a bra-vector. The second term on the r.h.s. of (\ref{5.12}) gives:
\bear
- \halb \Tr\left[ \underline{\varrho}(0) \left\{
\underline{\Gamma},\underline{\vec{F}} \right\}_+ \right]&=&  i (\Psi(0) |
\left( \M^\dagger(0) \underline{\vec{F}} - \underline{\vec{F}} \, \M(0) \right)
| \Psi(0)) \nn
&=& - \halb \Gs(Z,N) \vec{s} .
\label{5.16}
\ear
{}From (\ref{5.12}), (\ref{5.10}), and (\ref{3.53}) we get now:
\bear
\dot{\vec{F}}(0) &=& d_2 \,\vec{s} \times \vec{\E} - \halb \Gs(Z,N) \vec{s},
\nn
\dot{\hat{\vec{F}}}(0) &=& d_2 \,\vec{s} \times \vec{\E}.
\label{5.17}
\ear
This shows explicitly that the angular momentum per unit norm of the undecayed
state \underbar{is} rotated at time $t=0$ with the amount of rotation being
determined
by the vector product of Zeldovich's dipole moment and the electric field.

Consider now weak electric fields $\vec{\E} = \E \vec{e}_3$, satisfaying
(\ref{3.57}).
For times $t\gg \tau_P(Z,N)$  we have then from the results
of subsection 3.5 (cf. (\ref{3.62})):
\be\label{5.18}
|\Psi(t)) = e^{- i E(2\hat{S},\frac{1}{2},
{\cal E}
\vec{e}_3
) t}
| \hat{\Psi}),
\ee
$$(t \gg \tau_P(Z,N)),$$
where
\be\label{5.19}
| \hat{\Psi}) = \sum_{F_3 = \pm \shalb} | 2 \hat{S}, F_3, \E \vec{e}_3)
(  2 \tilde{\hat{S}, F_3,} \E \vec{e}_3 | \Psi(0)).
\ee
This leads (always for $t\gg\tau_P(Z,N)$) to
\bear
\N (t) &=& e^{- \Gamma_S(Z,N,\E) t} (\hat{\Psi} | \hat{\Psi}), \nn
\vec{F}(t) &=& e^{- \Gamma_S(Z,N,\E) t}  (\hat{\Psi} | \underline{\vec{F}} |
\hat{\Psi}), \nn \hat{\vec{F}}(t) &=& \frac{(\hat{\Psi} | \underline{\vec{F}} |
\hat{\Psi})}{(\hat{\Psi} | \hat{\Psi})}
\label{5.20}= const.
\ear
where
\be\label{5.21}
\Gamma_S(Z,N,\E) = - 2 \Im E(2 \hat{S}, \halb,\E \vec{e}_3)
\ee
is the decay rate of the metastable $2\hat S$-type states in the electric
field.
We get then from (\ref{5.2}), (\ref{5.3}), and (\ref{5.20}):
\be\label{5.22}
\dot{\vec{\tilde{F}}}(t) = - \dot{\N}(t) \frac{(\hat{\Psi} |
\underline{\vec{F}}
| \hat{\Psi})}{(\hat{\Psi} | \hat{\Psi})} +
\N (t) \frac{(\hat{\Psi} | \underline{\vec{D}} | \hat{\Psi})}{(\hat{\Psi} |
\hat{\Psi})} \times \vec{\E},
\ee
$$(t \gg \tau_P(Z,N)).$$
In (\ref{5.20})-(\ref{5.22}) we have the situation discussed in \cite{22}. The
angular momentum per unit norm for the undecayed part of the state vector, i.e.
of the $n=2$ states, stays constant. The angular momentum increase of the decay
products is given by the loss of angular momentum of the $n=2$ states due
to their decay plus the torque exerted by the external field.

The initial polarization rotation (\ref{5.17}) of the
$n=2$ states is thus due to the reshuffling of metastable $2\hat S$ and
unstable $2 \hat P$ states when going from $\vec{\E}=0$ to $\vec{\E}\not=0$.
This takes place over a time scale proportional to  $\tau_P(Z,N)$, the decay
time of the
$2\hat P$ states. We verify this now for the examples of $^4_2\mbox{He}^+$ and
$^{12}_6\mbox{C}^{5+}$ in electric
fields  numerically.

In Fig. 4 we show the functions $a_1$, $b_1$, $b_3$, $b_4$ (cf. (\ref{5.8}),
(\ref{5.9})) for $|\vec{\E}| = 320$ V/cm for the $^4_2\mbox{He}^+$ ion as
function of time.
Here $\delta_1(Z,N)$ is given by (\ref{3.16}) and numerically we find
\bear\label{5.22a}
\delta_1(2,2) &=& - \QWe(2,2) \cdot 7.41 \cdot 10^{-12}\nonumber\\
&=& \ssw \cdot 5.93 \cdot 10^{-11}.
\ear
With $\ssw=0.23$ this gives
\be\label{5.29X}
\delta_1(2,2)=1.36\cdot 10^{-11}.\ee
The P-violating functions $b_3$, $b_4$ are directly proportional to
$\delta_1(2,2)$ and thus to
the sine squared of the weak mixing angle. The ratios $b_1/a_1$ and $b_3/(a_1
\delta_1)$, $b_4/(a_1 \delta_1)$ show damped oscillations where the frequency
is given in essence by the Lamb shift ($L/h \simeq (0.07 \mbox{ ns})^{-1}$) and
the damping time by $2 \tau_P(2,2) \simeq 0.2$ ns (cf. Table 2) corresponding
to $2S - 2P$ interference. After 1 ns essentially only the metastable $2\hat S$
states corresponding to the given electric field survive and an asymptotic
regime is reached.

The behaviour of $a_1$, $b_1/a_1$, $b_3/(a_1 \delta_1)$ and
$b_4/(a_1 \delta_1)$ at very short times is shown in Fig. 5. The limiting
behaviour deduced from (\ref{5.10}), (\ref{5.17}) is
\bear\label{5.22b}
a_1(t,|\vec{\E}|^2) &=& 1 - \Gs(Z,N)\, t + {\cal O}(t^2),\nn
a_2(t,|\vec{\E}|^2) &=& {\cal O}(t^2),\nn
b_1(t,|\vec{\E}|^2) &=&
\halb\left(1 - \Gs(Z,N)\, t\right) + {\cal O}(t^2),\nn
b_2(t,|\vec{\E}|^2) &=& {\cal O}(t^2),\nn
b_3(t,|\vec{\E}|^2) &=& d_2 \,|\vec{\E}|\, t + {\cal O}(t^2),\nn
b_4(t,|\vec{\E}|^2) &=& {\cal O}(t^2).
\ear
This is in accord with the numerical results.

In Fig. 6 we show the behaviour of the asymptotic values, i.e. the plateau
values of Fig. 4 for $b_3/(a_1 \delta_1)$ and $b_4/(a_1 \delta_1)$
as function of $|\vec{\E}|$. These plateau values depend linearly on
$|\vec{\E}|$
for small $|\vec{\E}|$, for larger values of $|\vec{\E}|$ one sees a quadratic
effect coming into play. For even larger values of $|\vec{\E}|$, corresponding
to (cf. (\ref{3.57})):
$$ \frac{\sqrt{3} {\cal F}(Z,|\vec{\E}|)}{L(Z,N)} \geq 1,$$
\be \label{5.22c}
\mbox{i.e. } |\vec{\E}| \geq 12.7 \mbox{ kV/cm}
\ee
the mixing of $2S$ and $2P$ states is practically 100\%. No metastable states
exist then any more in the electric field and no plateau as in Fig. 4
is any more observed.

In Fig. 7 we show results for the $^{12}_6\mbox{C}^{5+}$ ion for $a_1$ and the
ratios $b_1/a_1$, $b_3/(a_1 \delta_1)$ and $b_4/(a_1 \delta_1)$. Here we have
from (\ref{3.16}) and Table 2:
\bear\label{5.22d}
\delta_1(6,6) &=& - \QWe(6,6) \cdot 1.08 \cdot 10^{-11}\nonumber \\
&=& \ssw \cdot 2.59 \cdot 10^{-10}\nonumber\\
&=&5.96\cdot 10^{-11}\quad{\rm for}\quad \ssw=0.23.
\ear
Apart from the overall rescaling of $t$ with $Z^{-4}$ and $\vec{\E}$ with $Z^5$
(cf. (\ref{3.20})), the results are similar to those for $^4_2\mbox{He}^+$.
\subsection{Type IV rotation}
Let us discuss the type IV rotation and start again with a
pure 2$\hat{S}$ state (\ref{3.64}), corresponding to $\vert\vec s|=1$
in (\ref{5.1}) at $t=0$. Then we switch on electric fields $\vec
\E^1=\E^1\vec e_3,\vec\E^2=\E^2\vec e_3$
for times $t_1$ and  $t_2-t_1$, respectively, and project onto the $2\hat{S}$
states
at $t_2$ (cf. section 4). Using (\ref{4.8}) we define functions
$a(t_2,t_1),\vec b(t_2,t_1)$ through
\be
\label{5.23}
| \Psi_{pr}(t_2))(\Psi_{pr}(t_2)| = \sum_{F'_3,F_3 = \pm \shalb} |
2\hat{S},F'_3,0) \, \halb \left( a(t_2,t_1) + \vec{b}(t_2,t_1) \vec{\sigma}
\right)_{F'_3,F_3} ( 2 \hat{S}, F_3,0|.
\ee
The generalization to a mixed state (\ref{5.1}) is obvious.
{}From the linear superposition principle and rotation covariance we
can expand $a,\vec b$ as follows:
\bear\label{5.24}
a(t_2,t_1) &=& a_1(t_2,t_1;\E^2,\E^1) + a_2(t_2,t_1;\E^2,\E^1) \vec{s}
\vec{e}_3,\\
\vec{b}(t_2,t_1) &=& b_1(t_2,t_1;\E^2,\E^1) \vec{s}
+ b_2(t_2,t_1;\E^2,\E^1) \left(\vec{s} \vec{e}_3\right) \vec{e}_3 +\nn
\label{5.25}
&&b_3(t_2,t_1;\E^2,\E^1) \vec{s} \times \vec{e}_3 + b_4(t_2,t_1;\E^2,\E^1)
\vec{e}_3.
\ear
Here again $a_1,b_1,b_2$ are P-conserving, whereas $a_2,b_3,b_4$
are P-violating quantities. Under the procedure discussed in
sect. 4 the third component of the total angular momentum is conserved for the
undecayed
states and thus a state with $\vec s=\pm\vec e_3$ must go into a state
$|\Psi_{pr}(t_2))$ with
\be\label{5.25a}
\vec b = \pm a\,  \vec e_3.\ee
This implies
\begin{eqnarray}\label{5.25b}
b_1(t_2,t_1;\E^2,\E^1) +
b_2(t_2,t_1;\E^2,\E^1) &=&
a_1(t_2,t_1;\E^2,\E^1) ,\nn
b_4(t_2,t_1;\E^2,\E^1) &=& a_2(t_2,t_1;\E^2,\E^1)
\end{eqnarray}
Furthermore, an arbitrary pure state $|\Psi(0))$ must go into a
pure state $|\Psi_{pr}(t_2))$.
This implies
\be\label{5.26}
a(t_2,t_1;\E^2,\E^1) = | \vec b(t_2,t_1;\E^2,\E^1) |
\qquad \mbox{ for all } |\vec s | = 1
\ee
and leads to
\begin{eqnarray}
b_1^2 + b_3^2 + b_4^2 &=& a_1^2 ,\nn
(b_1 + b_2) b_4 &=& a_1 a_2 , \nn
 2 b_1 b_2 + b_2^2 - b_3^2 &=& a_2^2.
\label{5.27}
\end{eqnarray}
Neglecting here and in the following terms quadratic
in P-violating quantities -- which is a very good
approximation indeed -- we get, using also the continuity of the functions
$a,b$:
\begin{eqnarray}
b_1(t_2,t_1;\E^2,\E^1) &=& a_1(t_2,t_1;\E^2,\E^1) ,\nn
b_2(t_2,t_1;\E^2,\E^1) &=& 0,\nn
b_4(t_2,t_1;\E^2,\E^1) &=& a_2(t_2,t_1;\E^2,\E^1).
\label{5.28}
\end{eqnarray}
We note that these relations, (\ref{5.27}) or (\ref{5.28}), respectively,
could be used to check the superposition principle. In any case,
they reduce the number of independent functions $a_i,b_j$ to three:
$a_1,b_3,b_4$. In terms of the amplitudes $f_{F_3',F_3},\Delta f$,
defined in (\ref{4.9}), (\ref{4.13}) we get
\begin{eqnarray}
a_1 &=& \viertel \left[\vert f_{\frac{1}{2},\frac{1}{2}} +
f_{-\frac{1}{2},-\frac{1}{2}} \vert^2 +|\Delta f|^2 \right]\nn
b_3 &=&\halb \, \Im \left[ \Delta f \left(f_{\frac{1}{2},\frac{1}{2}}
+ f_{-\frac{1}{2},-\frac{1}{2}}
\right)^* \right], \nn
b_4 &=&\halb \, \Re \left[ \Delta f \left(f_{\frac{1}{2},\frac{1}{2}}
+ f_{-\frac{1}{2},-\frac{1}{2}}
\right)^* \right].
\label{5.29}
\end{eqnarray}

We have investigated the functions $a_i,b_j$ for various
values of $t_1,t_2$ and $\E^1,\E^2$. The
short-time behaviour $(t_1,t_2\to0)$ can be obtained from
(\ref{4.15})--(\ref{4.18}).
We find
\begin{eqnarray}
\label{5.30}
a_1(t_2,t_1;\E^2,\E^1) &=& 1 + ...\\
b_3(t_2,t_1;\E^2,\E^1) &=& (t_2 - t_1)^2 \, t_1^2 \E^1 \E^2 (\E^1-\E^2) (\Qe
r_B(Z))^3 \, 3 \sqrt{3} \delta_1(Z,N) \Delta(Z,N) \cdot \nn
&& \frac{L^2(Z,N)}{L^2(Z,N) + \frac{1}{4}\left[\Gamma_P(Z,N) -
\Gamma_S(Z,N)\right]^2}+...,
\label{5.31} \\
b_4(t_2,t_1;\E^2,\E^1) &=& (t_2 - t_1)^2 \, t_1^2 \E^1 \E^2 (\E^1-\E^2) (\Qe
r_B(Z))^3 \, \frac{3 \sqrt{3}}{2} \delta_1(Z,N) \Delta(Z,N) \cdot \nn
&& \frac{L(Z,N) \left[\Gamma_P(Z,N) - \Gamma_S(Z,N)\right]}{L^2(Z,N) +
\frac{1}{4}\left[\Gamma_P(Z,N) - \Gamma_S(Z,N)\right]^2}+...,
\label{5.32}
\end{eqnarray}
where the dots stand for terms with higher powers in $t_1$, $(t_2 - t_1)$.
{}From (\ref{5.31}) we see that here a P-violating rotation remains
also if we set the widths $\Gamma_P$ and $\Gamma_S$ to zero, i.e. for a
\underline{stable} atom. Type IV P-violating rotations for stable
atoms will be investigated in a future publication.

In Fig.~8 we show numerical results for the functions $a_1,b_3/(a_1\cdot
\delta_1)$ and $b_4/(a_1\cdot\delta_1)$ for $^4_2\mbox{He}^+$. The
parameters are chosen as follows:
\begin{eqnarray}
t_1 &=& 1 \mbox{ ns}, \nn
\E^1 &=& 4000 \mbox{ V/cm}, \nn
\E^2 &=& -4000 \mbox{ V/cm}.
\label{5.33}
\end{eqnarray}
Note that we plot in Fig. 8b-d the functions $a_1$ etc. as function
of the time $t_2$ where the electric field is switched off and the
$2\hat S$ content of the undecayed state is analysed. The
interpretation of Fig. 8 is as follows: For $t_1=1$ns the undecayed
state in the electric field $\E^1$ has reached a metastable state
(cf. Fig. 4). The sudden switch to $\E^2$ and then to
electric field zero produces for short times $t_2-t_1\ll 2\tau_P
(2,2)$ final polarizations which vary in an oscillatory manner with
the time $t_2$. The frequencies of these oscillations are given
by the energy differences of the $2S$ and $2P$ levels in the
field $\vec\E^2$:
$L(2,2,\E^2)/h,\ \Delta(2,2,\E^2)/h$ and $(\Delta(2,2,\E^2)-L(2,2,\E^2))/h$.
For $t_2-t_1  \stackrel{\scriptstyle>}{\sim} 1 $ns the oscillatory behaviour
dies
out and a plateau value for $b_3/(a_1\cdot\delta_1)$ and
$b_4/(a_1\cdot\delta_1)$ is reached.
This comes about, since then a metastable state in the field $\E^2$
has been reached prior to the switch off of the field and the projection onto
the $2\hat S$ states of zero field. The variation of the height of the plateau
values with the electric field $\E^1$ is shown in Fig.~9 where we choose
\bear
t_1 &=& 1 \mbox{ ns}, \nn
t_2 &=& 3 \mbox{ ns}, \nn
\E^1 &=& - \E^2.
\label{5.34}\ear
We see that $b_{3,4}/(a_1\delta_2)$ have roughly a cubic dependence on
$\cale^1$ as we expect from (\ref{5.31}), (\ref{5.32}).

Finally we note that the type IV rotations discussed in this subsection are
intimately related to interference effects for split beams of atoms.  To see
this
consider  a beam of polarized $2\hat S$-atoms entering an electric field $\E^1$
for a time $t_1$, then being split in 2 beams with amplitudes
$\chi_1$ and $\chi_2$ where
\be\label{5.34a}
\vert \chi_1\vert^2 + \vert \chi_2\vert^2 = 1.
\ee
Let the two beams traverse electric fields $\E^1$ and $\E^2$,
respectively, for a time $t_2-t_1$. Then, let the two beams interfere and be
split
again in two beams $(\pm)$ where the amplitudes add and substract,
respectively.
Suppose that finally one analyses their  $2\hat S$ content (Fig. 10). Let the
initial
state be as in (\ref{3.64}) with density matrix as in (\ref{5.1}). The
undecayed state
at time $t_2$ in the beams $(\pm)$ is then:
\be\label{5.35}
\vert \Psi^\pm(t_2)) = \frac{1}{\sqrt{2}}
\biggl[ \chi_1 e^{-i \M(\E^1 \vec e_3) t_2} \pm \chi_2 e^{- i \M(\E^2 \vec
e_3)(t_2 - t_1)} e^{-i\M(\E^1 \vec e_3)t_1} \biggr]
\vert \Psi(0)).
\ee
The projected state is
\be
\vert \Psi_{pr}^\pm(t_2)) = \frac{1}{\sqrt{2}} \sum_{F_3',F_3 = \pm
\frac{1}{2}} \vert 2 \hat S, F_3',0)
\biggl[ \chi_1 f_{F_3',F_3}(t_2,t_2) \pm \chi_2 f_{F_3',F_3}(t_2,t_1)\biggr]
c_{F_3},
\label{5.36}\ee
where $f_{F'_3,F_3}(t_2,t_2)$ and $f_{F'_3,F_3}(t_2,t_1)$ are as defined in
(\ref{4.9}). The unnormalized density matrix for the projected states
(\ref{5.36})
is again of the form (\ref{5.23}) where we get now for the functions
$a_1,b_3,b_4$ at
time $t_2$ the following:
\bear
a_1^\pm &=& \frac{1}{8} \biggl| \chi_1 \biggl(
f_{\frac{1}{2},\frac{1}{2}}(t_2,t_2) +
f_{-\frac{1}{2},-\frac{1}{2}}(t_2,t_2)\biggr)
\pm \chi_2 \biggl(
f_{\frac{1}{2},\frac{1}{2}}(t_2,t_1) +
f_{-\frac{1}{2},-\frac{1}{2}}(t_2,t_1)\biggr)
\biggr|^2,
\nn
b_3^\pm &=& \pm \frac{1}{4} \, \Im \left[ \chi_2 \Delta f \left\{ \chi_1
\biggl(
f_{\frac{1}{2},\frac{1}{2}}(t_2,t_2) +
f_{-\frac{1}{2},-\frac{1}{2}}(t_2,t_2)\biggr)
\pm \chi_2 \biggl(
f_{\frac{1}{2},\frac{1}{2}}(t_2,t_1) +
f_{-\frac{1}{2},-\frac{1}{2}}(t_2,t_1)\biggr)
\right\}^* \right],\nn
b_4^\pm &=& \pm \frac{1}{4} \, \Re \left[ \chi_2 \Delta f \left\{ \chi_1
\biggl(
f_{\frac{1}{2},\frac{1}{2}}(t_2,t_2) +
f_{-\frac{1}{2},-\frac{1}{2}}(t_2,t_2)\biggr)
\pm \chi_2 \biggl(
f_{\frac{1}{2},\frac{1}{2}}(t_2,t_1) +
f_{-\frac{1}{2},-\frac{1}{2}}(t_2,t_1)\biggr)
\right\}^* \right].\nn
\label{5.37}\ear
Here $\Delta f$ is as in (\ref{4.13}). The polarization rotation of the
interfering
beams is therefore governed by $\Delta f$ and qualitatively of the same type as
shown
in Figs.~8,~9. By a suitable choice of the splitting parameters $\chi_{1,2}$
one has
now the possibility to obtain \underline{large} rotations of the polarization
but in a
low intensity beam. Consider as an example the choice
\bear
\chi_1 &=& n_\chi (1+\epsilon) \biggl(
f_{\frac{1}{2},\frac{1}{2}}(t_2,t_1) +
f_{-\frac{1}{2},-\frac{1}{2}}(t_2,t_1)\biggr), \nn
\chi_2 &=& n_\chi (1-\epsilon) \biggl(
f_{\frac{1}{2},\frac{1}{2}}(t_2,t_2) +
f_{-\frac{1}{2},-\frac{1}{2}}(t_2,t_2)\biggr), \nn
n_\chi &=&  \left\lbrace\left|
f_{\frac{1}{2},\frac{1}{2}}(t_2,t_1) + f_{-\frac{1}{2},-\frac{1}{2}}(t_2,t_1)
\right|^2
+
\left|
f_{\frac{1}{2},\frac{1}{2}}(t_2,t_2) + f_{-\frac{1}{2},-\frac{1}{2}}(t_2,t_2)
\right|^2\right\rbrace^{-1/2}\!\!\!\!,
\label{5.37a}\ear
where $0 < \epsilon\ll 1$. We get then to leading order in $\epsilon$\, :
\bear
a_1^- &=& \frac{\epsilon^2 \left| \chi_1 \chi_2\right|^2}{2 n_\chi^2},
\nn
b_3^- &=& - \frac{\epsilon}{2} \frac{|\chi_2|^2}{n_\chi} \Im \biggl( \Delta f
\cdot \chi_1^*\biggr), \nn
b_4^- &=& - \frac{\epsilon}{2} \frac{|\chi_2|^2}{n_\chi} \Re \biggl( \Delta f
\cdot \chi_1^*\biggr).
\label{5.37b}\ear

Suppose we start originally with $N$ atoms. In order to observe the P-violating
rotation in the ($-$) beam at the $1\sigma$ level, we should have
\be\label{5.53a}
\frac{\sqrt{(b^-_3)^2+(b_4^-)^2}}{a^-_1}\geq\frac{1}{\sqrt{Na^-_1}},\ee
i.e.
\be\label{5.53b}
N\geq\frac{a^-_1}{(b_3^-)^2+(b_4^-)^2}.\ee
Inserting here (\ref{5.37b}) we see that the required $N$ is independent
of $\epsilon$. One does not gain in statistical accuracy choosing a
small $\epsilon$. But things should be different with regard to systematic
errors.

For a numerical example we choose (cf. (\ref{5.33}), (\ref{5.34})):
\bear\label{5.53c}
t_1&=&1 \mbox{ ns},\nonumber\\
t_2&=&3 \mbox{ ns},\nonumber\\
\cale^1&=&-\cale^2=4000 \mbox{ V/cm}.\ear
We get then
\bear
a^-_1&=&\epsilon^2\cdot6.1\cdot10^{-2},\nonumber\\
b^-_3&=&\epsilon\cdot 6.9\cdot 10^{-4}\,\delta_1(2,2),\nonumber\\
b^-_4&=&\epsilon\cdot1.2\cdot10^{-4}\,\delta_1(2,2),\nonumber\\
b_3^-/a^-_1&=&\epsilon^{-1}\cdot1.1\cdot10^{-2}\,\delta_1(2,2)\nonumber\\
&=&\epsilon^{-1}\cdot1.5\cdot10^{-13},\nonumber\\
b_4^-/a^-_1&=&\epsilon^{-1}\cdot2\cdot10^{-3}\,\delta_1(2,2)\nonumber\\
&=&\epsilon^{-1}\cdot2.7\cdot10^{-14},
\label{5.53d}
\ear
where we used (\ref{5.29X}). To get sizeable rotation effects $\epsilon$
should be very small indeed, of order $10^{-10}-10^{-13}$! The number $N$
(\ref{5.53b}) of required atoms is here
\be
\label{5.53e}
N\stackrel{\scriptstyle>}{\sim}10^{27}.
\ee
Thus, such a type of experiment requires beams of macroscopic
intensities with microscopic ``side beams''. It may be that the techniques
of accelerator mass spectroscopy \cite{24} could be used in this connection.
We also note that we have disscussed her only the analogue of a double slit
interference experiment. One could envisage analogues of multi slit
interference devices where the sensitivity to our P-violating effects should be
much bigger and thus the number of required atoms greatly reduced.
\subsection{Type V rotation}

This is the type of rotation effect studied extensively in~\cite{12}. Consider
an atom
in a $2\hat S$ state, described by the density  matrix (\ref{5.1}) being placed
in an
electric field $\vec\E$ at $t=0$ and left there to decay to the ground state.
The total transfer of angular momentum $\bra{}\Delta\vec J\,\ket{}$ from the
external field device
to the atomic system is then (cf. (2.7) of~\cite{12}):
\be
\bra{} \Delta \vec J \, \ket{} = \Delta J_1 ( \vec s \times \Ehat) \times \Ehat
+ \Delta J_2 \, \vec s \times \Ehat
\label{5.38}\ee
where $\Ehat=\vec\E/|\vec\E|$.

The quantity $\Delta J_1$ is P-conserving, $\Delta J_2$ is P-violating. The
norm of the
undecayed state as function of time is given by (\ref{5.8}), the lifetime of
the state
in the electric field is given by
\bear
\tau_S^{-1}(Z,N,\E) &=& \int_0^\infty dt \, t \biggl( - \frac{d}{dt}
\N(t)\biggr) \nn
 &=& \int _0 ^\infty dt \biggl[ a_1(t,|\vec\E|^2) + a_2(t,|\vec\E|^2) \vec s
\cdot \Ehat \biggr].
\label{5.39}\ear
The $a_2$-term is P-violating and of order $\delta_1(Z,N)$, thus very small
compared
to the $a_1$-term. Neglecting the $a_2$-term leads to a lifetime independent of
the
initial polarization  $\vec s$ as shown in Fig. 11 for $^4_2\mbox{He}^+$ and in
Fig. 12
for $^{12}_6\mbox{C}^{5+}$ as function of $\cale=|\vec\E|$.

These results can be understood as follows. The electric field produces a $2P$
admixture
to the $2S$ state of order (cf.(\ref{3.15}) ff.) $\sqrt 3{\cal
F}(Z,\E)/L(Z,N)$.

Schematically we have for $\sqrt 3{\cal F}/L\ll 1$:
\be\label{5.40}
|2 \hat S(\E) ) = | 2 \hat S ) + \kappa \frac{\sqrt{3} \F(Z,\E)}{L(Z,N)} |2 P )
\ee
where $\kappa$ is a constant of order 1. The decay rate of the state $2\hat
S(\E)$
is then estimated as follows:
\be\label{5.41}
\tau_S^{-1}(Z,N,\E) \cong \Gamma_S(Z,N) + |\kappa|^2 \left|\frac{\sqrt{3}
\F(Z,\E)}{L(Z,N)}\right|^2 \Gamma_P(Z,N).
\ee

{}From Figs. 11~b and 12~b we find both for $^4_2\mbox{He}^+$ and
$^{12}_6\mbox{C}^{5+}$
\be\label{5.42}
|\kappa|^2 =1.0.\ee

In Figs. 13, 14 we show results for $\Delta J_1$ and $\Delta J_2$ as function
of
$\cale $ for $^4_2\rm He^+$ and $^{12}_6\rm C^{5+}$.

The occurrence of the maxima in $-\Delta J_2$ can easily be understood. For
small times
the P-violating torque  is given by the vector product of the Zeldovich dipole
moment
and the electric field (cf. (2.4), (4.9) of~\cite{12}). We estimate then
$|\Delta J_2 |\cong $ torque $\times$ lifetime and thus from (\ref{5.13}),
(\ref{5.14})
\bear
|\Delta J_2| &\cong& | d_2 \, \vec s \times \vec\E | \, \tau_S(Z,N,\cale)
\nn
&\cong& |\delta_1(Z,N)| \frac{\Gamma_P(Z,N) \sqrt{3} \F(Z,\cale ) \,
L^{-1}(Z,N)}{\Gamma_S(Z,N) + |\kappa|^2 \left[\sqrt{3} \F(Z,\cale ) \,
L^{-1}(Z,N)\right]^2 \Gamma_P(Z,N)}
\label{5.43}\ear
where we used $\Gamma_S\ll\Gamma_P\ll L$ (cf. (\ref{3.18}), (\ref{3.20})).

This leads to a maximum in $|\Delta J_2|$ expected at
\be\label{5.44}
\frac{\sqrt{3} \F(Z,\cale )}{L(Z,N)} = \frac{1}{|\kappa|}
\sqrt{\frac{\Gamma_S(Z,N)}{\Gamma_P(Z,N)}},
\ee
corresponding roughly to
\be\label{5.45}
\cale  \cong 0.05 \cdot  Z^6 \mbox{ V/cm},\ee
where we used the scaling laws (\ref{3.20}).
The height of the maximum is then expected from (\ref{5.43}) to be
\bear
\left| \Delta J_2 \right|_{max} &\cong& |\delta_1(Z,N)| \, \frac{1}{2 |\kappa|}
\sqrt{\frac{\Gamma_P(Z,N)}{\Gamma_S(Z,N)}} \nn
&\cong& N \, \frac{1}{2} \frac{1}{Z} \sqrt{\frac{\Gamma_P(1,0)}{\Gamma_S(1,0)}}
\cdot 6.14 \cdot 10^{-12}
\label{5.46}\\
&\cong& \frac{N}{Z} \, 2.7 \times 10^{-8}.\nonumber
\ear
The numerical results of Figs. 13, 14 confirm these estimates. Note the large
enhancement factor $[\Gamma_P/\Gamma_S]^{1/2}$ in (\ref{5.46}) which is
responsible
for the ``large'' numerical result $\approx 10^{-8}$ compared to the nominal
order of
magnitude given by $\delta_1(Z,N)$ (cf. (\ref{3.19}), (\ref{3.20})).

To estimate also the P-conserving term $\Delta J_1$ in (\ref{5.38}) we note
that the
induced dipole moment $d_1(\vec\E^2,t)$ in (2.5) of~\cite{12} should
be of order
\be\label{5.47}
d_1(|\vec\E|^2, t \approx \tau_P(Z,N)) \cong \kappa' \frac{\sqrt{3}
   \F(Z,\cale )}{L(Z,N)} \, \Qe r_B(Z)
\ee
with $\kappa'$ of order 1.

Estimating again $\Delta J_1$ by the torque  times the lifetime we get
\bear
\left| \Delta J_1 \right| &\cong&  |\kappa'| \, \frac{\sqrt{3}
   \F(Z,\cale )}{L(Z,N)} \, \Qe r_B(Z) \E \tau_S(Z,N,\E) \nn
&\cong& |\kappa'| \frac{1}{\sqrt{3}} \left|\frac{\sqrt{3}
\F(Z,\cale)}{L(Z,N)}\right|^2 \frac{L(Z,N)}{\Gamma_S(Z,N) + |\kappa|^2
\left|\sqrt{3} {\cal F}(Z,\cale )\,L^{-1}(Z,N)\right|^2 \Gamma_P(Z,N)}
\nn
\label{5.48}\ear
{}From Figs. 13~b and 14~b we get
\be\label{5.49}
|\kappa'| \simeq
\renewcommand{\arraystretch}{1.5}
\left\{
\begin{array}{ll}
0.2 & \mbox{ for } ^4_2 \mbox{He}^+,\\
0.3 & \mbox{ for } ^{12}_6 \mbox{C}^{5+}.
\end{array}\right.
\renewcommand{\arraystretch}{1.0}
\ee
Thus at the maximum of $|\Delta J_2|$ given by  (\ref{5.44}), (\ref{5.46}), we
get
for the P-conserving ``background'' term $\Delta J_1$:
\be\label{5.50}
|\Delta J_1| \cong \left|\frac{\kappa'}{\kappa}\right| \frac{1}{2 \sqrt{3}}
\frac{L(Z,N)}{\Gamma_P(Z,N)}\cong 0.5.
\ee
{}From (\ref{5.46}) and (\ref{5.50}) we see that -- quite surprisingly --  both
the
maximal P-violating effect $|\Delta J_2|_{max}$ and the background term
$\Delta J_1$
at the corresponding value of the electric field (\ref{5.45}) are roughly
independent
of $Z$.
\section{Results for nuclear spin $I=1/2$}
\setcounter{equation}{0}
In this section we discuss polarization rotations for atoms with $I=\halb$.
Here all
types of rotations listed in Table 1 occur. In the present paper we will only
discuss
rotations of type II and III for $I=\halb $ since these are absent for the
$I=0$ case,
whereas the other  types of rotations also occurred there.

A schematic drawing  of the (zero field) energy levels for  the $n=2$ states of
hydrogenic atoms with $I=\halb $ is given in Fig. 15. The hyperfine splitting
of the
$2S_{1/2},\ 2P_{1/2}$ and $2P_{3/2}$  levels is denoted by $A_1,\ A_2$ and
$A_3$,
respectively:
\bear
A_1(Z,N) &=& E(2 S_{\frac{1}{2}},F=1) - E(2 S_{\frac{1}{2}},F=0), \nn
A_2(Z,N) &=& E(2 P_{\frac{1}{2}},F=1) - E(2 P_{\frac{1}{2}},F=0), \nn
A_3(Z,N) &=& E(2 P_{\frac{3}{2}},F=2) - E(2 P_{\frac{3}{2}},F=1).
\label{6.1}\ear
The hyperfine splittings scale with $Z$ roughly as $Z^3$ (cf.(\ref{3.15})).

Interesting examples of atoms with $I=\halb$ are ordinary hydrogen $^1_1$H
and the $^3_2\rm
He^+$ ion. The energy levels and other quantities of interest for the $^1_1$H
atom are collected in Table 3.

The weak charges $Q^{(1,2)}_W(Z,N)$  for these atoms  are given in Table A2 of
appendix A. Note
that $Q_W^{(2)}$ is rather sensitive to $\Delta  s$, the difference between
$\Delta
s=0$ and the EMC value for $\Delta s$ being $\approx 15 \%$.

The mass matrix $\M$ for the $n=2$ states of an atom with $I=\halb$ is given
in Table B5 of appendix B where we also list in Tables B6, B7  the matrices
$\underline{\cal T}$  and $\underline{\R}$
representing the time reversal and reflection
operations and $\underline{\vec D}$
for the dipole operator. The right and left eigenvectors of $\M$ for zero
external electric field are given in Table B8.
Note that due to the hyperfine interaction the $2\hat S$ states with $F=1$
contain now P-violating contributions from both states $2P_{1/2}$
and $2 P_{3/2}$ with $F=1$. But the amplitude for the $2P_{3/2},\ F=1$ state
is suppressed by $A_3/\Delta$ relative to the one for the $2P_{1/2},\ F=1$
state. For $^1_1$H we have (cf. Table 3)
\be
\label{6.1a}
A_3/\Delta=2.16\cdot10^{-3}\ee
and for higher $Z$ ions the ratio scales roughly as $Z^{-1}$ (cf.
(\ref{3.20})),
thus getting even smaller. Neglecting terms proportional to this small ratio,
we find for the $2\hat S$ eigenstates $|2 \hat{S},F,F_3) $ of $\M$ at zero
external field
from Table B8:
\bear
|2\hat S,0,0)&=&|2S,0,0)+i(\delta_1\eta_0-\dh \delta_2\eta_0)
|2P_{1/2},0,0),\nonumber\\
|2\hat S,1,F_3)&\cong&|2S,1,F_3)+i(\delta_1\eta_1+\eh \delta_2\eta_1)
|2P_{1/2},1,F_3),
\label{6.1b}
\ear
where $F_3 = 0, \pm 1$ and
\bear
\eta_0=\frac{L}{L-\frac{3}{4}(A_1-A_2)-\frac{i}{2}(\Gamma_S-\Gamma_P)},
\nonumber\\
\eta_1=\frac{L}{L+\frac{1}{4}(A_1-A_2)-\frac{i}{2}(\Gamma_S-\Gamma_P)}.
\label{6.1c}
\ear
Note that $|A_1-A_2|\ll L$ (for all $Z$). Thus $\eta_0\cong\eta_1$ and the
P-violating admixtures of the $2P$ states in (\ref{6.1b}) proportional
to $\delta_1$ which are due to the nuclear spin-independent Hamiltonian
$H^{(1)}_{PV}$ (\ref{2.2}) are practically the same for $F=0$ and $F=1$.
However,
the P-violating  admixtures proportional to $\delta_2$ which are due
to $H^{(2)}_{PV}$ (\ref{2.3}) differ  by roughly a factor ($-3$) for $F=0$ and
$F=1$. This will have drastic consequences in the following.

Consider now an atom at zero external field in a pure $2\hat S$ state  at time
$t=0$:
\be\label{6.2}
|\Psi(0)) = \sum_{F,F_3} | 2 \hat S, F, F_3) \, c_{F,F_3}
\ee
where $F=1,0;\, F_3=-F,...,F$.

As normalization condition we get
\be\label{6.3}
( \Psi(0) | \Psi(0)) = \sum_{F,F_3} | c_{F,F_3}|^2 = 1.
\ee
Here we use the normalization condition (\ref{3.24a}) and the fact that the
basis
states on the r.h.s. of (\ref{6.2}) are orthogonal in the usual sense, being
eigenstates to different eigenvalues of the \underline{hermitian} operators
$\underline{F}_3$ and $\underline{\vec F}^2$.

It is also convenient to introduce vector notation for describing the $F=1$
part   of
the state (\ref{6.2}) by setting
\be\label{6.4}
\vec c = \sum_{F_3} c_{1,F_3} \, \vec e_{F_3}
\ee
where
\bear
\vec e_\pm &=& \mp \frac{1}{\sqrt{2}} \left( \vec e_1 \pm i \vec e_2\right),
\nn
\vec e_0 &=& \vec e_3
\label{6.5}\ear
are the usual spherical basis vectors.

A general, pure or mixed, initial state is described by a density matrix
\be\label{6.6}
\underline{\varrho}(0) = \sum_{F',F_3',F,F_3} | 2 \hat S,F',F_3')
\,\varrho_{F',F_3';F,F_3}(0) \,( 2 \hat S,F,F_3| \ .
\ee
This general density matrix can be parametrized by one scalar $S_0$, one real
vector
$\vec S$, one complex vector $\vec K$ and one real symmetric  and traceless
second
rank tensor $S_{ij}$ as shown in Table 4.

These parameters must, of course, be such that $\underline{\varrho}$ is
positive
semidefinite and of trace 1:
\bear
\underline{\varrho}(0) &\geq& 0, \nn
\Tr \underline{\varrho}(0) &=& 1 .
\label{6.7}\ear
For our pure state (\ref{6.2})--(\ref{6.4}) we have
\bear
S_0 &=& 1 - |c_{0,0}|^2 = |\vec c|^2, \nn
\vec S &=& i \vec c \times \vec c^*, \nn
 \vec K &=& \vec c \, c_{0,0}^*, \nn
S_{ij} &=& - \frac{1}{2} \left( c_i c_j^* + c_j c_i^* \right) + \frac{|\vec
c|^2}{3} \delta_{ij}.
\label{6.8}\ear

If P-violation is disregarded, the state (\ref{6.2}) transforms as follows
under a reflection R on the 1-3 plane (cf. (\ref{3.51a})):
\bear\label{6.8a}
{\rm R}:|\Psi(0)\, )&\longrightarrow&|\Psi'(0)\, ),\nonumber\\
c_{0,0}&\longrightarrow& c_{0,0}'=c_{0,0},\nonumber\\
c_{1,0}&\longrightarrow& c_{1,0}'= - c_{1,0},\nonumber\\
c_{1,\pm1}&\longrightarrow& c_{1,\pm1}'=c_{1,\mp 1},\nonumber\\
\left(\begin{array}{c} c_1\\ c_2\\c_3
\end{array}\right)&\longrightarrow&\left(\begin{array}{c} c_1'\\
c_2'\\c_3'\end{array}\right)=\left(\begin{array}{r} -c_1\\ c_2\\
-c_3\end{array}\right);\ear

\bear\label{6.8b}
S_0&\longrightarrow&S_0'=S_0,\nonumber\\
\vec S &\longrightarrow&\vec S'=\left(\begin{array}{r}
-S_1\\ S_2\\-S_3\end{array}\right),\nonumber\\
\vec K &\longrightarrow&\vec K'=\left(\begin{array}{r}
-K_1\\ K_2\\-K_3\end{array}\right),\nonumber\\
S_{ij}&\longrightarrow&S_{ij}'=\left\lbrace\begin{array}{ccc}
-S_{ij},&\quad{\rm for}\quad (i,j)=(1,2),(2,3),(2,1),(3,2),\\
S_{ij}&\quad {\rm all\  other}\quad (i,j).\end{array}\right.
\ear
Thus, if P violation is neglected, $(\delta_1=\delta_2=0)$, the states with
$F=0,F_3=0$ and $F=1,F_3=0$ are R-eigenstates with opposite eigenvalues
\bear
\label{6.8c}
{\rm R}:\quad
|2\hat S,0,0)&\longrightarrow&|2\hat S,0,0),\nonumber\\
|2\hat S,1,0)&\longrightarrow& - |2\hat S,1,0),\nonumber\\
(2\widetilde{\hat S,0, }0|&\longrightarrow&
(2\widetilde{\hat S,0, }0|,\nonumber\\
(2\widetilde{\hat S,1, }0|&\longrightarrow&
-(2\widetilde{\hat S,1, }0|.
\ear

\subsection{Type II rotations}

Let us place the atom in the initial state (\ref{6.2}) in an electric field
$\vec\E=\E\vec e_3$ at time $t=0$ and observe the total angular
momentum of the undecayed state in the field for $t>0$. Let the electric field
be
weak in the sense that (cf. (\ref{3.15}), (\ref{3.18}))
\be\label{6.9}
\frac{\sqrt{3} \F(Z,\E)}{L(Z,N)} \ll 1.
\ee
Then  metastable eigenstates of
$\M(\vec\E)$ exist  in the presence of  the electric field, too, and  we denote
them by (cf. (\ref{3.57}) ff.)
\be\label{6.9a}
|2\hat S,\ F,F_3,\ \E\vec e_3),\qquad(F=0,1;\ F_3=-F,...,F).\ee
{}From our discussion of T-invariance in sect. 3 we know that the energies of
these
states satisfy (cf. (\ref{3.43})):
\be\label{6.10}
E(2 \hat S, F, F_3, \E \vec e_3) = E(2 \hat S, F,-F_3, \E \vec e_3).
\ee
Here, of course, only $F_3$ is still a good quantum number, $F$ is just a label
indicating the eigenvalue of the total angular momentum of the corresponding
state at zero electric field. The complex energies are shown as function
of $(\sqrt3 er_B{\E}/L)^2$ for $^1_1$H in Fig. 16. Their imaginary parts give
the decay rates
\be\label{6.10a}
\Gamma(2\hat S,F,F_3,{\E}\vec e_3):=-2\Im E(2\hat S,F,F_3,{\E}\vec e_3).\ee

Note that the states with $F=0, \ F_3=0$ and $F=1,\ F_3=0$ are not orthogonal
for
$\E\not= 0$ as they are for $\E=0$.

The time evolution of the undecayed part of the state vector with initial
condition (\ref{6.2}) is given by
\bear
|\Psi(t)) &=& e^{-i \M(\E \vec e_3) t} |\Psi(0))
\label{6.11} \\
&& (t > 0).\nonumber
\ear
For time $t\gg \tau_P(Z,N)$ only the new metastable states survive, and we have
(cf.
(\ref{3.59})):
\bear
|\Psi(t)) &=& \sum_{F',F_3'} e^{-i E(2 \hat S,F',F_3',{\cal E} \vec e_3) t} |2
\hat S, F',F_3',\E \vec e_3) \, c'_{F',F_3'}\\
\label{6.12}
&& ( t \gg \tau_P(Z,N)),\nonumber
\ear
where
\bear
c'_{F',F_3'} &=& ( 2 \tilde{ \hat S, F',} F_3', \E \vec e_3 | \Psi(0)) \nn
&=& \sum_{F,F_3} ( 2 \tilde{ \hat S, F',} F_3', \E \vec e_3 | 2 \hat S, F, F_3)
\, c_{F,F_3}.
\label{6.13}
\ear
Due to (\ref{6.10}) we get
\bear
|\Psi(t)) &=&
e^{-i E(2 \hat S,0,0,{\cal E} \vec e_3) t} |2 \hat S, 0,0,\E \vec e_3) \,
c'_{0,0} \nn
&&
\!\!\!\!\!+ e^{-i E(2 \hat S,1,0,{\cal E} \vec e_3) t} |2 \hat S, 1,0,\E \vec
e_3) \, c'_{1,0}
\nn
&&\!\!\!\!\!+ e^{-i E(2 \hat S,1,1,{\cal E} \vec e_3) t}
\left\{ |2 \hat S, 1,1,\E \vec e_3) \, c'_{1,1} + |2 \hat S, 1,-1,\E \vec e_3)
\, c'_{1,-1}
\right\}
\label{6.14}
\\
&&(t \gg \tau_P(Z,N)). \nonumber
\ear
In contrast to the case $I=0,F=\halb$, where the corresponding state vector
(\ref{3.62})
evolves by multiplication with a single complex number the evolution in
(\ref{6.14})
is non-trivial and can lead to rotations of the expectation value of the
angular momentum of the undecayed state. The oscillation frequencies are given
by the hyperfine splittings in the electric field.

We demonstrate this now  by  examples. Consider ordinary hydrogen
$^1_1$H and choose as initial state a $2\hat S$ state (\ref{6.2})
with polarization in 2-direction:
\bear
c_{0,0} = 0 , \quad c_{1,0} &=& -\frac{1}{\sqrt{2}}, \quad c_{1,\pm1} =
\pm\frac{i}{2}, \nn
\mbox{i.e.}\qquad \qquad
\vec c &=&- \frac{1}{\sqrt{2}} \left( \vec e_3 +i \vec e_1\right).
\label{6.15}\ear
This corresponds to an eigenstate of the angular momentum component $\vec
e_2\cdot\vec{\underline{F}}$
with eigenvalue +1 and to density matrix parameters (\ref{6.8}):
\bear
S_0 &=& 1, \nn
\vec S &=& \vec e_2, \nn
\vec K &=& 0, \nn
S_{ij} &=& diag \left( -\frac{1}{6}, \frac{1}{3},-\frac{1}{6}\right).
\label{6.16}\ear
The initial state and the set-up are then R-symmetric (cf. (\ref{6.8b})).
Therefore, if any of the components $\vec e_1\cdot \vec F(t)$ or $\vec e_3\cdot
\vec F(t)$ is different from zero for $t>0$, this is  a clear
indication of R-symmetry- and thus also parity-violation. Here $\vec F(t)$ is
the expectation value of the  angular momentum of the $n=2$ states
(cf. (\ref{3.52})).

Inserting the initial values (\ref{6.15}) in (\ref{6.13}), we get
\bear\label{6.18}
c_{0,0}'&=&-(\widetilde{2\hat S,0, }0,{\E}\vec e_3|2\hat
S,1,0)\frac{1}{\sqrt2},
\nonumber\\
c_{1,0}'&=&-(\widetilde{2\hat S,1, }0,{\E}\vec e_3|2\hat
S,1,0)\frac{1}{\sqrt2},
\nonumber\\
c_{1,\pm1}'&=&\pm(\widetilde{2\hat S,1, }\pm1 ,{\E}\vec e_3|2\hat
S,1,\pm1)\frac{i}{2}.\ear
We consider now again the reflection R (\ref{3.51a}) on the 1-3 plane.
Neglecting P-violation, i.e. for $\delta_1=\delta_2=0$, the eigenstates
of ${\M}({\E}\vec e_3)$ in the electric field have the same definite R
transformation properties as we discussed for ${\E}=0$ in
(\ref{6.8a})--(\ref{6.8c}). This implies, together with the differentiability
with
respect ot $\delta_{1,2}$ as discussed in sect. 3.4 and Appendix C,
that $c_{0,0}'$ in (\ref{6.18}) is R-violating and thus linear in $\delta_1,
\delta_2$ whereas $c_{1,0}'$ has a R-conserving contribution plus terms
quadratic in $\delta_{1,2}$. Similarly we find that the overlap of the
two states with $F_3=0$ in the electric field is linear in $\delta_{1,2}$:
\be
\label{6.19}
(2\hat S,0,0,{\E}\vec e_3|2\hat S,1,0,{\E}\vec e_3)={\cal
O}(\delta_1,\delta_2).
\ee
Neglecting terms quadratic in $\delta_{1,2}$, we get now for $t\gg\tau_P$
for the R-even quantities $\N(t)$ and $\vec e_2\cdot \vec F(t)$:
\bear
{\N}(t)&=&(\Psi(t)|\Psi(t))\nonumber\\
&=&e^{-\Gamma(2\hat S,1,0,{\E}\vec e_3)t}|c_{1,0}'|^2\nonumber\\
&&+e^{-\Gamma(2\hat S,1,1,{\E}\vec e_3)t}\left[|c_{1,1}'|^2+|c_{1,-1}'|^2
\right],
\label{6.20}
\ear
\bear
\vec e_2\cdot \vec F(t)&=&e^{i[E(2\hat S,1,1,{\E}\vec e_3)^*-E(2\hat S,1,0,{\E}
\vec e_3)]t} \cdot \nonumber\\
&&[{c'}_{1,1}^*(2\hat S,1,1,{\E}\vec e_3|+{c'}_{1,-1}^*(2\hat S,1,-1,{\E}\vec
e_3|] \cdot \nonumber\\
&&\vec e_2\cdot\underline{\vec F} \
   |2\hat S,1,0,{\E}\vec e_3)c_{1,0}'\nonumber\\
&&+c.c.
\label{6.21}
\ear
and for the R-odd ones $\vec e_1\cdot{\vec F}(t),\vec e_3\cdot\vec F(t)$:
\bear
\vec e_1\cdot\vec F(t)&=&e^{i[E(2\hat{S},1,1,{\E}\vec e_3)^*-E(2\hat
S,0,0,{\E}\vec e_3)]t} \cdot
\nonumber\\
&&\left[{c'}_{1,1}^*(2\hat S,1,1,{\E}\vec e_3|+{c'}^*_{1,-1}(2\hat
S,1,-1,{\E}e_3|\right]\cdot
\nonumber\\
&&\vec e_1\cdot \underline{\vec F}\,|2\hat S,0,0,{\E}\vec
e_3)c_{0,0}'\nonumber\\
&&+e^{i[E(2\hat{S},1,1,{\E}\vec e_3)^*-E(2\hat S,1,0,{\E}\vec e_3)]t}\cdot
\nonumber\\
&&\left[{c'}_{1,1}^*(2\hat S,1,1,{\E}\vec e_3|+{c'}^*_{1,-1}(2\hat
S,1,-1,{\E}e_3|\right]\cdot
\nonumber\\
&&\vec e_1\cdot \underline{\vec F}\,|2\hat S,1,0,{\E}\vec
e_3)c_{1,0}'\nonumber\\
&&+c.c.,
\label{6.22}
\ear
\be
\label{6.23}
\vec e_3\cdot\vec F(t)=e^{-\Gamma(2\hat S,1,1,{\E}\vec
e_3)t}\left[|c_{1,1}'|^2-|c_{1,-1}'|^2\right].
\ee

To investigate all this numerically we choose as electric field
${\E}=100$ V/cm which is small enough for metastable states to
exist (cf. (\ref{3.18}) and Table 3). We get then for the complex energies:
\bear
\label{6.24}
E(2\hat S,0,0,{\E}\vec e_3)=(6.078\cdot 10^9-i\,1.640\cdot
10^7)\mbox{s}^{-1},\nonumber\\
E(2\hat S,1,0,{\E}\vec e_3)=(7.131\cdot 10^9-i\,1.103\cdot
10^7)\mbox{s}^{-1},\nonumber\\
E(2\hat S,1,1,{\E}\vec e_3)=(7.145\cdot 10^9-i\,1.211\cdot 10^7)\mbox{s}^{-1}.
\ear
Remember that the real parts correspond to the \underbar{angular} frequencies.
In Fig. 17 we show results for $\E=$ 100 V/cm for $\N(t)$ (\ref{3.54}) and
$\hat{\vec F}(t)$ (\ref{3.53}). For small times
\be\label{6.25}
0 \leq t \stackrel{\scriptstyle<}{\sim} 6\tau_P=9.6 \, \mbox{ ns}
\ee
there are damped oscillations with a frequency of order $L/h$ and a damping
time of order $2\tau_P$. This is due to the reshuffling of $2S$ and $2P$ states
in the electric field and
completely analogous to the $I=0$ case in Figs.~4,~7. On larger time
scales one sees the behaviour expected from (\ref{6.20})--(\ref{6.23}).
The norm ${\N}(t)$ and the P-violating component $\vec e_3\cdot
\hat{\vec{F}}(t)=\vec e_3\cdot\vec F(t)/\N(t)$ show no oscillations any more.
(cf. (\ref{6.20}), (\ref{6.23})).
The oscillation and damping times seen for $\vec e_2\cdot \hat{\vec{F}}(t)$ and
$\vec e_1\cdot\hat{\vec{F}}(t)$ are easily identified as the differences in the
real and imaginary
parts of the complex energies (\ref{6.24}) as it should be, dividing
(\ref{6.21}) by (\ref{6.20}) and (\ref{6.22}) by (\ref{6.20}).


We have thus demonstrated that the metastable $2 \hat{S}$ hydrogen atoms in the
presence of the elecric field show both P-conserving and P-violating
polarization precessions. This is in striking contrast to the case of $2
\hat{S}$ atoms with nuclei of spin $0$ where T-invariance forbids such
precessions (cf. (\ref{3.63})).
\subsection{Type III rotations}
Consider again an initial state (\ref{6.2}). Let it evolve in an electric field
$\vec\E=\E\vec e_3$ from time 0 to $t_1$, then switch off $\vec\E$
and project onto the $2\hat S$ states of zero field. We get
\bear
|\Psi_{pr}(t_1)) &=& \sum_{F'F_3'} | 2 \hat S, F' , F_3') \, ( 2
\widetilde{\hat S, F'}, F_3' | e^{-i \M({\cal E} \vec e_3) t_1 } | \Psi(0)) \nn
&=& \sum_{F',F_3';F,F_3} |2 \hat S, F' F_3')\,  f_{F',F_3';F,F_3}(t_1) \,
c_{F,F_3},
\label{6.27}\ear
where we define
\be\label{6.28}
f_{F',F_3';F,F_3}(t_1) := ( 2 \widetilde{\hat S, F'}, F_3' |
e^{-i \M({\cal E} \vec e_3) t_1 } | 2 \hat S, F, F_3).
\ee
Using the conservation of the third component of the total angular momentum
and the consequences of T-invariance (\ref{3.39}),
(\ref{3.44}) we get with the same
reasoning as for $I=0$ in section 3.5 (cf. (\ref{3.68})--(\ref{3.70})):
\bear\label{6.29}
f_{F',F_3';F,F_3}(t_1) &=& 0 \qquad \mbox{     for   }\qquad F_3' \neq F_3,
\nonumber\\
f_{F,F_3;F,F_3}(t_1) &=& f_{F,-F_3;F,-F_3}(t_1).
\ear
With the T-transformation properties of the states given in (\ref{B.10}) we
find
in addition
\be
\label{6.30}
f_{1,0;0,0}(t_1)=-f_{0,0;1,0}(t_1).
\ee
Clearly, the relations (\ref{6.29}), (\ref{6.30})
are not sufficient to guarantee that the state $|\Psi_{pr}
(t_1))$ in (\ref{6.27}) is a multiple of $|\Psi(0))$, as was the case for $I=0$
in (\ref{3.71}).

The amplitudes $f$ are, of course, not directly observable. Observable
quantities involve $f^*f$. However, it is  convenient to work with
the amplitudes.

We get further relations for the functions $f$ (\ref{6.28}) by considering the
reflection R of (\ref{3.51a}). Indicating explicitly the dependence on the
P-violation parameters $\delta_i(i=1,2)$ of (\ref{3.16}) we get from
(\ref{6.28}), (\ref{3.51c}) and (\ref{3.51d})
\bear
f_{F,F_3;F,F_3}(t_1,\delta_1,\delta_2)
&=&\Tr\left[ e^{-i\M({\cal E}\vec e_3,\delta_1,\delta_2)t_1}\pro(
2\hat S,F,F_3,\delta_1,\delta_2)\right]\nonumber\\
&=&\Tr\left[ e^{-i\M({\cal E}\vec e_3,-\delta_1,-\delta_2)t_1}\pro(
2\hat S,F,-F_3,-\delta_1,-\delta_2)\right]\nonumber\\
&=&f_{F,-F_3;F,-F_3}(t_1,-\delta_1,-\delta_2).
\label{6.31}
\ear
Together with (\ref{6.29}) this shows that the diagonal elements of
the functions $f$ are even functions of $\delta_{1,2}$ and thus, neglecting
terms quadratic in $\delta_{1,2}$, they get no contribution from the
P-violating
Hamiltonian (\ref{2.1}). The off diagonal elements $f_{0,0;1,0}$ and $f_{1,0;
0,0}$ are according to (\ref{6.8c}) R-symmetry violating and thus linear
in $\delta_{1,2}$.

Let us first investigate the behaviour of the functions $f$ (\ref{6.28}) for
$t_1\to 0$. Expanding up to terms linear in $t_1$ we get with the notation
as in (\ref{3.45}):
\bear
f_{F',F_3';F,F_3}(t_1)&=&(2\widetilde{\hat S,F',}F_3'|\left\lbrace\eins
-i[\M_0-\underline D_3\E]t_1+...\right\rbrace|2\hat S,F,F_3)\nonumber\\
&=&\delta_{F',F_3';F,F_3}[1-iE(2\hat S,F,F_3)t_1]\nonumber\\
&&+i\E\,t_1(2\widetilde{\hat S,F'},F_3'|\underline D_3|2\hat S,F,F_3)+...
\label{6.32}
\ear
We know from (\ref{3.51f}), (\ref{3.51g}) that the complex energy eigenvalues
have no linear term in $\delta_{1,2}$ and from (\ref{3.51}) that the
\underline{diagonal} elements of the dipole operator between a left and right
eigenvector of ${\M}_0$ vanish. Thus we verify in the zero and first order
terms in $t_1$ that the diagonal functions $f$ are even in $\delta_{1,2}$.
But there is no reason for
 the off diagonal matrix elements of $\underline D_3$  to
vanish and indeed, using the approximate forms (\ref{6.1b}) for the
right eigenvectors and the corresponding forms for the left eigenvectors we
find:
\be
(2\tilde{\hat S,1},0|\underline D_3|2\hat S,0,0)
\cong-i\sqrt{3} e\, r_B\left[\delta_1(\eta_0-\eta_1)-\eh\delta_2(3\eta_0
+\eta_1)\right].
\label{6.33}
\ee
We have $\eta_0\cong\eta_1$ (cf. (\ref{6.1c})) and thus in (\ref{6.33}) the
nuclear spin-independent P-violating contribution proportional to $\delta_1$
is suppressed relative to the spin-dependent contribution proportional
to $\delta_2$. For hydrogen $_1^1$H we have a particularly strong enhancement
of the nuclear spin-dependent term since $\delta_1$ is here proportional
to $(1-4\ssw)$ (cf. (\ref{2.16}) and Table A2). We get, using $\ssw=0.23$:
\bear
\delta_1(\eta_0-\eta_1)\Bigr|_{^1_1H}&=&(5.87-i\cdot 0.57)\cdot
10^{-14},\nonumber\\
\eh\delta_2(3\eta_0+\eta_1)\Bigr|_{^1_1H}&=&(-2.61+i\cdot 0.31)
\cdot10^{-12}\cdot(1-g^{-1}_A \Delta s_p).
\label{6.34}
\ear
Setting $\Delta s_p=0$, the naive expectation, or $\Delta s_p=-0.19$, the
EMC value (\ref{2.10}), makes a difference of $\approx 15\%$ in the
second term.

To discuss the behaviour of the functions $f$ for general times we insert
in (\ref{6.28}) the decomposition of the mass matrix (\ref{3.27}) using the
quasiprojectors appropriate for the electric field:
\be
\label{6.35}
f_{F',F_3;F,F_3}(t_1)=\sum_\alpha e^{-iE(\alpha,F_3,\E\vec e_3)t_1}
(2\widetilde{\hat S,F'},F_3|\pro(\alpha,F_3,\E\vec e_3)|2\hat S,F,F_3).
\ee
Here $\alpha$ denumbers a complete set of states in the electric field to the
given $F_3$.
For small electric fields $\sqrt 3\F/L\ll1$ (cf. (\ref{3.15}), (\ref{3.18}))
and for $t\gg\tau_P$ only the contribution from the metastable $2\hat S$ states
in the electric field survives in (\ref{6.35}) and we get:
\bear
f_{F',F_3;F,F_3}(t_1)&=&\sum_{F''}e^{-iE(2\hat S,F'',F_3,\E\vec e_3)t_1}
\cdot\nonumber\\
&&(2\widetilde{\hat S,F'},F_3|2\hat S,F'', F_3,\E\vec e_3)(2\widetilde{\hat S,
F''},F_3,\E\vec e_3|2\hat S,F,F_3)\nonumber\\
&&(t\gg\tau_P)
\label{6.36}
\ear
where the notation is as in (\ref{6.9a}), (\ref{6.10}). Neglecting here terms
of order
$\delta^2_{1,2}$ we obtain
\bear
f_{F,F_3;F,F_3}(t_1)&=&e^{-iE(2\hat S,F,F_3,\E\vec e_3)t_1}
\Tr\left[\pro(2\hat S,F,F_3,\E\vec e_3)\pro(2\hat S,F,F_3)\right],\nonumber\\
f_{1,0;0,0}(t_1)&=&\sum_{F''=0,1}e^{-iE(2\hat S,F'',0,\E\vec e_3)t_1}
(2\widetilde{\hat S,1},0|\pro(2\hat S,F'',0,\E\vec e_3)|2\hat S,0,0)\nonumber\\
&&(t\gg\tau_P).
\label{6.37}
\ear
This shows that the diagonal elements have here single oscillation frequencies
and damping times whereas the P-violating function $f_{1,0;0,0}$ has two such
terms.

In Fig. 18 we show numerical results for $\E=100$ V/cm for the functions
$f$ for $^1_1$H. At short times, $0\leq t\stackrel{\scriptstyle<}{\sim}
6\tau_p\approx 10$ ns we expect to see oscillations and dampings due to
$S-P$ interference as given by the general formula (\ref{6.35}). But this
turns out to be a very small effect, hardly visible at the resolution of the
figures. Thus in essence we see already from very short times the behaviour
from (\ref{6.36}), (\ref{6.37}): the diagonal elements of $f$ show single
oscillation frequencies and damping times, the P-violating term $f_{1,0;0,0}$
shows a clear beat pattern of two frequencies.

We conclude that an atomic system with $I=1/2$ can have a polarization
rotation of type III. If we start at $t=0$ for instance with a pure $|2\hat
S,0,0)$ state
\be
\label{6.38}
|\Psi(0))=|2\hat S,0,0)  \ ,\ee
switch on an electric field for a time $0\leq t\leq t_1$, switch it off and
project onto the $2\hat S$ states we will find a state
\be
\label{6.39}
|\Psi_{pr}(t_1))=|2\hat S,0,0)f_{0,0;0,0}(t_1)
+|2\hat S,1,0)f_{1,0;0,0}(t_1).
\ee
The $|2\hat S,1,0)$ admixture is R symmetry violating (cf.~(\ref{6.8c}))
and represents a P-violating polarization rotation. Of course, the admixture
of the ``wrong'' R parity in (\ref{6.39}) is tiny, of order $10^{-12}$.
Can we enhance this tiny amplitude?

Consider a repeated pattern  of switching on and off the electric field, where
with $0<t_1<t_2$ we take
\bear\label{6.40}
&& \vec\E (t)=\left\{  \begin{array}{ll}
0&\mbox{for }  t<0,\\
{}& \ \ \ \ (k-1) t_2+t_1\leq t<kt_2,\\
{}& \ \ \ \ K t_2\leq t,\\
    & \\
\E \vec e_3&\mbox{for } (k-1)t_2\leq t<(k-1)t_2+t_1
\end{array} \right. \nonumber\\
&&(k=1,...,K).
\ear
We choose $t_1=1$ ns and $t_2-t_1\geq 10$ ns such that the $2P$-admixture
present after each swith off of  the field has decayed  and the state is
practically equal to the projected state. Let us start at $t=0$ with the state
(\ref{6.2}). We get then at $t=K t_2$
\bear
|\Psi_{pr}(K t_2))&=&\sum_{F',F'_3}|2\hat S,F',F'_3)\,c'_{F',F'_3},\nonumber\\
c'_{F',F'_3}&=&\sum_{F,F_3} f^{(K)}_{F',F'_3;F,F_3}\, c_{F,F_3}
\label{6.41}
\ear
where in matrix notation:
\be\label{6.42}
\underline{f}^{(K)}=(\underline{g})^K,\ee
with
\be\label{6.43}
g_{F',F'_3;F,F_3}=e^{-iE(2\hat S,F')(t_2-t_1)}f_{F',F'_3;F,F_3}(t_1).\ee
We have from (\ref{6.29})
\be\label{6.44}
f^{(K)}_{F',F'_3;F,F_3}=0\quad\mbox{for } F'_3\not= F_3.\ee
and keeping only terms of zeroth and first order in $\delta_{1,2}$:
\bear
f^{(K)}_{F,F_3;F,F_3}&=&(g_{F,F_3;F,F_3})^K,\nonumber\\
f^{(K)}_{0,0;1,0}&=&g_{0,0;1,0}\sum^{K-1}_{k=0}g^k_{0,0;0,0}\,
g^{K-k-1}_{1,0;1,0},\nonumber\\
f^{(K)}_{1,0;0,0}&=&g_{1,0;0,0}\sum^{K-1}_{k=0}g^k_{0,0;0,0}\,
g^{K-k-1}_{1,0;1,0}.
\label{6.45}
\ear
Let us define with $t_1=1$ ns
\bear
\tilde L_{F,F_3} &:=&\Re \left[\frac{i}{t_1}\ln
f_{F,F_3;F,F_3}(t_1)\right],\nonumber\\
\tilde\Gamma_{F,F_3} &:=&-2\Im \left[\frac{i}{t_1}\ln
f_{F,F_3;F,F_3}(t_1)\right].
\label{6.46}
\ear
Thus we have:
\be\label{6.47}
f_{F,F_3;F,F_3}(t_1)=\exp\left\{-i\tilde
L_{F,F_3}t_1-\eh\tilde\Gamma_{F,F_3}t_1\right\}.\ee

{}From numerical analysis we find for small electric fields,  $\sqrt3 {\cal
F}/L\ll 1:$
\bear\label{6.48}
\tilde L_{F,F_3}&=&\Re E(2\hat S,F)+\tilde
\kappa_{F,F_3}\cdot\left(\frac{\sqrt3 {\cal F}}{L}\right)^2L,\nonumber\\
\tilde\Gamma_{F,F_3}&=&\Gamma_S+\kappa_{F,F_3}\left(\frac{\sqrt3{\cal
F}}{L}\right)^2\Gamma_P,
\ear
where $\tilde \kappa,\kappa$ are numerical factors of order 1 which are listed
in Table 5. For the P-violating amplitude $f_{1,0;0,0}(t_1)$ we get
\bear
f_{1,0;0,0}(t_1)\left|_{t_1=1 ns}\right.&=&\sqrt3 {\cal F}
t_1\left[\xi_1\cdot\delta_1+\xi_2\cdot\delta_2\right],\nonumber\\
\xi_1&=&0.11-i\cdot 0.021,\nonumber\\
\xi_2&=&-2.0+i\cdot  0.27.
\label{6.49}
\ear
The results, (\ref{6.48}), (\ref{6.49}) are easily understood using arguments
as in subsect. 5.3, (\ref{5.40}) ff. From the experience gained there we expect
to get large effects of P-violation for small electric fields and high values
of $K$.

Consider thus small electric fields $\sqrt{3} {\cal F}/L\ll1$. We get then from
(\ref{6.43})--(\ref{6.48}):
\be
\label{6.50}
g_{F,F_3;F,F_3}=\exp \left\{-i\left[ E (2\hat S,F)t_2
+\tilde \kappa_{F,F_3}\left(\frac{\sqrt3{\cal F}}{L}\right)^2 Lt_1
-\frac{i}{2}\kappa_{F,F_3}\left(\frac{\sqrt3{\cal F}}{L}\right)^2\Gamma_P
t_1\right]\right\},
\ee
\be
\label{6.51}
\sum^{K-1}_{k=0} g^k_{0,0;0,0}
\,g^{K-k-1}_{1,0;1,0}=g^{K-1}_{1,0;1,0}\left(1-\frac{g_{0,0;0,0}}{g_{1,0;1,0}}
\right)^{-1}\left[1-\left(\frac{g_{0,0;0,0}}{g_{1,0;1,0}}\right)^K\right],
\ee
\bear
\frac{g_{0,0;0,0}}{g_{1,0;1,0}}&=&\exp\left\{-i\left[-A_1
t_2+(\tilde{\kappa}_{0,0}-
\tilde \kappa_{1,0})\left(\frac{\sqrt 3{\cal F}}{L}\right)^2Lt_1\right]\right.
\nonumber\\
&&\left.-\frac{1}{2}(\kappa_{0,0}-\kappa_{1,0})\left(\frac{\sqrt3{\cal
F}}{L}\right)^2\Gamma_P t_1\right\}.
\label{6.52}
\ear
Let us now choose $t_2$ such that all terms in the sum
on the l.h.s. of  (\ref{6.51}) add coherently, i.e. we choose
\be
\label{6.53}
A_1 t_2-(\tilde \kappa_{0,0}-\tilde \kappa_{1,0})\left(\frac{\sqrt3{\cal
F}}{L}\right)^2 Lt_1=2\pi n
\ee
with $n$  an integer.
We have then
\be\label{6.54}
\frac{g_{0,0;0,0}}{g_{1,0;1,0}}=\exp\left[-\frac{1}{2}
(\kappa_{0,0}-\kappa_{1,0})\left(
\frac{\sqrt3{\cal F}}{L}\right)^2\Gamma_Pt_1\right].
\ee
\par
Let us discuss now the following set-up for a ``theoretical'' experiment. We
start at $t=0$ with an atom where the initial state is as in (\ref{6.2}) with
\bear
c_{0,0}&=&1,\ c_{1,0}=0,\ c_{1,1}=c_{1,-1}=\epsilon,\nn &&
1\gg\epsilon\gg|\delta_{1,2}|.
\label{6.56}
\ear
Neglecting P-violation, this initial state is R-symmetric (cf. (\ref{6.8a})).
We let the atom pass $K$ electric fields as indicated in (\ref{6.40}) with
$\sqrt{3}{\cal F}/L\ll1$. The state at time $K t_2$ is then as in (\ref{6.41})
with
\bear
c'_{0,0}&=& g^K_{0,0;0,0}\ \ ,\nonumber\\
c'_{1,0}&=& g_{1,0;0,0}\,
g^{K-1}_{1,0;1,0}\left(1-\frac{g_{0,0;0,0}}{g_{1,0;1,0}}\right)^{-1}\cdot\left[
1-\left(\frac{g_{0,0;0,0}}{g_{1,0;1,0}}\right)^K\right],\nonumber\\
c'_{1,\pm 1}&=& g^K_{1,1;1,1}\,\epsilon.
\label{6.57}
\ear
Suppose we separate now the $F=1$ part of the atomic state  vector and  measure
its angular momentum component in 1-direction, where a nonzero value means
R-symmetry violation. We get for the $F=1$ part of the atomic state, its norm
\mbox{$\N(K t_2,F=1)$}  and for the angular momentum per unit norm in
1-direction:
\be
\label{6.58}
|\Psi(Kt_2,F=1))=\sum_{F_3}|2\hat S,1,F_3)\,c'_{1,F_3},\ee
\bear
\N(K t_2,F=1)&=&(\Psi(K t_2,F=1)|\Psi(K t_2,F=1))=\left[ |c'_{1,1}|^2
+|c'_{1,0}|^2+|c'_{1,-1}|^2\right]\nonumber\\
&=&2\epsilon^2|g_{1,1;1,1}|^{2K},
\label{6.59}
\ear
\bear
\vec{e}_1 \cdot \hat{\vec{F}}(K t_2)&=&(\Psi(K t_2,F=1)|\,\vec e_1\cdot
\underline{\vec{F}}\,
|\Psi(Kt_2,F=1))\N^{-1}(Kt_2,F=1)\nonumber\\
&=&\frac{\sqrt{2} \Re\left[ c'_{1,0}(c'_{1,1}+c'_{1,-1})^*\right]}{|c'_{1,1}
|^2+|c'_{1,0}|^2+|c'_{1,-1}|^2}\nonumber\\
&=&\sqrt2 \epsilon^{-1}\left(1-\frac{g_{0,0;0,0}}{g_{1,0;1,0}}\right)^{-1}
\left[1-\left(\frac{g_{0,0;0,0}}{g_{1,0;1,0}}\right)^K\right]\cdot\Re
\left[g_{1,0;0,0}\cdot g^{K-1}_{1,0;1,0}\cdot g^{-K}_{1,1;1,1}\right],
\nn
&&
\label{6.60}
\ear
\bear
(\Psi(K t_2,F=1)|(\vec e_1\cdot \underline{\vec{F}})^2|\Psi(K
t_2,F=1))\N^{-1}(K t_2,F=1)
& = &
\frac{[|c_{1,1}'+c_{1,-1}'|^2+2|c_{1,0}'|^2]}{2[|c_{1,1}'|^2+
|c_{1,0}'|^2+|c_{1,-1}'|^2]}\nn
&=&1.
\label{6.61}
\ear
Here we always use $1\gg\epsilon\gg|\delta_{1,2}|$.

Suppose we start with $N_0$ atoms at time $t=0$. In order to obtain
a $1\sigma$ R-violating effect at time $K t_2$ for the $F=1$ part
of the atomic state we must have
\be
\label{6.62}
|\vec e_1\cdot \hat{\vec{F}}(K t_2)|>\sqrt{\frac{1}{N_0 \N(K t_2,F=1)}},
\ee
i.e. $N_0 > \bar{N}_0$, where
\be
\label{6.63}
\bar{N}_0=\frac{1}{4}\left(1-\frac{g_{0,0;0,0}}{g_{1,0;1,0}}\right)^2
\left[1-\left(\frac{g_{0,0;0,0}}{g_{1,0;1,0}}\right)^K\right]^{-2}
\left[\Re\left(g_{1,0;0,0}\cdot g^{K-1}_{1,0;1,0}\cdot g^{-K}_{1,1;1,1}\cdot
|g_{1,1;1,1}|^K
\right)\right]^{-2}.
\ee
We want to minimize $\bar N_0$. The important factor in $\bar N_0$ is
\bear
Q &:=& \frac{1}{4}\left(1-\frac{g_{0,0;0,0}}{g_{1,0;1,0}}\right)^2
\left[1-\left(\frac{g_{0,0;0,0}}{g_{1,0;1,0}}\right)^K\right]^{-2}
(\sqrt3{\cal F} t_1)^{-2}|g_{1,0;1,0}|^{-2K}\nonumber\\
&=&\frac{(\kappa_{0,0}-\kappa_{1,0})\Gamma_P K}{8L^2 t_1}
\frac{1}{x}
(1-e^{-x/K})^2(1-e^{-x})^{-2}
\exp\left[\frac{2\kappa_{1,0}}{\kappa_{0,0}-\kappa_{1,0}}x\right]
e^{K\Gamma_St_2},  \nonumber \\
& &
\label{6.64}
\ear
where
\be\label{6.65}
x:=\frac{1}{2}(\kappa_{0,0}-\kappa_{1,0})\left(\frac{\sqrt3{\F}}{L}\right)^2
\Gamma_Pt_1K.\ee
We have then
\be
\label{6.66}
\bar N_0 =Q \left\{ \Re\left[(\xi_1\delta_1+\xi_2\delta_2)
\exp(-iE(2\hat S,1)(t_2-t_1))\,g^{-1}_{1,0;1,0}
\left(\frac{g_{1,0;1,0}\cdot|g_{1,1;1,1}|}{|g_{1,0;1,0}|\cdot
g_{1,1;1,1}}\right)^K
\right]\right\}^{-2}
\ee
and the term in curly brackets is of order $\delta_2$. For $K\gg1$ we get
\be\label{6.67}
Q\cong\frac{(\kappa_{0,0}-\kappa_{1,0})\Gamma_P}{8L^2t_1}x(1-e^{-x})^{-2}
\exp\left[\frac{2\kappa_{1,0}}{\kappa_{0,0}-\kappa_{1,0}}x\right]
 K^{-1}e^{K\Gamma_St_2}.\ee
Given the values for $\kappa_{0,0}$, $\kappa_{1,0}$ of Table 5, the minimum of
the
quantity $Q$ is well approximated by $Q$ at the values
\bear\label{6.68}
x&=&\frac{\kappa_{0,0}-\kappa_{1,0}}{2\kappa_{1,0}},\nonumber\\
K&=&\frac{1}{\Gamma_S t_2},\ear
leading to
\be\label{6.69}
Q_{min}\cong\frac{(\kappa_{0,0}-\kappa_{1,0})^2\Gamma_P\Gamma_St_2}{16
\kappa_{1,0} L^2t_1}\,e^2
\cdot\left[1-\exp\left(-\frac{\kappa_{0,0}-\kappa_{1,0}}{2\kappa_{1,0}}\right)
\right]^{-2}.
\ee
The optimal value for the electric field is then from (\ref{6.65})
\be\label{6.70}
\left(\frac{\sqrt3{\F}}{L}\right)^2=\frac{\Gamma_S
t_2}{\kappa_{1,0}\Gamma_Pt_1}
\ee
and the magic condition for $t_2$ (\ref{6.53}) reads now:
\be\label{6.71}
\left[A_1-\frac{\tilde\kappa_{0,0}-\tilde\kappa_{1,0}}{\kappa_{1,0}}
\frac{L}{\Gamma_P}\Gamma_S\right]t_2=2\pi n\ee
which is conveniently satisfied for $n=2$ with
\be\label{6.72}
t_2=11.3\mbox{ ns}.\ee
This leads to
\bear
K&=&1.08\cdot10^7,\nonumber\\
Kt_2&=&\Gamma_S^{-1}=0.12\mbox{ s}\nonumber\\
\left(\frac{\sqrt3\F}{L}\right)^2&=&7.00\cdot10^{-8},\nonumber\\
\E &=&0.13\mbox{ V/cm},\nonumber\\
Q_{min}&=&6.6\cdot10^{-9},\nonumber\\
\bar N_0&\cong&1.1\cdot10^{15},
\label{6.73}
\ear
\bear\label{6.74}
\N(Kt_2,F=1)&=&2\epsilon^2\exp\left[-\frac{\kappa_{1,0}
+\kappa_{1,1}}{\kappa_{1,0}}
\right]\nonumber\\
&=&\epsilon^2\cdot 0.30\ear
and to a signal of R-symmetry violation:
\be\label{6.75}
\vec e_1\cdot\hat{\vec{F}}(K t_2)=Q'\cdot \Re\left\{
(\xi_1\delta_1+\xi_2\delta_2)
\exp[-iE(2\hat S,1)(t_2-t_1)]
\,g^{-1}_{1,0;1,0}\cdot\left(\frac{g_{1,0;1,0}\cdot
|g_{1,1;1,1}|}{|g_{1,0;1,0}|\cdot g_{1,1;1,1}}\right)^K\right\}
\ee
where $Q'$ is an enhancement factor:
\bear
Q'&=&\epsilon^{-1}\sqrt2(1-e^{-x/K})^{-1}(1-e^{-x})\sqrt{3}\F t_1
\Bigl|\frac{g_{1,0;1,0}}{g_{1,1;1,1}}\Bigr|^K\nonumber\\
&=&\epsilon^{-1}\frac{2\sqrt{2\kappa_{1,0}}}{(\kappa_{0,0}-\kappa_{1,0})}
\frac{L}{\sqrt{\Gamma_S\Gamma_P}}\sqrt{\frac{t_1}{t_2}}
\left[1-\exp\left(-\frac{\kappa_{0,0}-\kappa_{1,0}}{2\kappa_{1,0}}
\right)\right]\exp\left[-\frac{\kappa_{1,0}-\kappa_{1,1}}{2\kappa_{1,0}}
\right]\nonumber\\
&=&\epsilon^{-1}\cdot  2.25\cdot 10^4.
\label{6.76}
\ear
This gives:
\be
\label{6.77}
|\vec e_1\cdot\hat{\vec{F}}(K t_2)|\cong \epsilon^{-1}\cdot  5.5\cdot10^{-8}.
\ee

To summarize: With $K\cong 10^7$ iterations, performed in a total time
$Kt_2=\Gamma_S^{-1}=0.12$ s one can get a decent signal of a P-violating
spin rotation for hydrogen $^1_1$H. The total number of atoms required is
$\stackrel {\scriptstyle >}{\sim} 10^{15}$ which does not seem exorbitantly
high. With the help of the parameter $\epsilon$ the signal in absolute terms
but not its statistical significance can be enhanced. Choosing
$\epsilon=10^{-6}$ we can
obtain (on paper) a spin rotation signal of $\approx 5\%$! For hydrogen $^1_1$H
the signal is nearly exclusively due to the \underline{spin-dependent}
weak axial charge $Q^{(2)}_W$ of the proton. This quantity gets a contribution
from $\Delta s_p$ and changes by 15\% if we go from $\Delta s_p=0$ to
$\Delta s_p=-0.19$, the EMC value. Thus, for a 10\% measurement of $\Delta
s_p$,
assuming the EMC value to be correct, one would need $\approx 5\cdot10^{18}$
atoms.
\section{Conclusions}
In this article we have discussed a new type of parity-violating effect
in atoms: polarization rotations in external electric fields. We have
identified 5 types of such rotations in Table 1. An analysis of the
consequences of time reversal invariance showed that certain types of
rotations are not allowed by
this symmetry  for systems with total angular momentum $\frac{1}{2}$.
Previous discussions relevant for our topic concentrated mainly on these
special cases and thus missed a great variety of P-violating effects.

 In this article we have only considered hydrogen-like atoms
and $n=2$ states, where $n$ is the principal quantum number. There, one has
the metastable $2S_{1/2}$ states, which are convenient for
experimental purposes. In addition the $2S_{1/2}$
 and the $2P_{1/2}$ states with opposite
parity are only separated by the small Lamb shift, thus rendering P-violating
state mixing rather large where ``large'' means of order $10^{-12}$
(cf.~(\ref{3.19})).
Correspondingly, the typical order of magnitude for the P-violating
polarization rotations was $10^{-12}$. But we found ways to produce large
enhancement factors in two cases: for rotations of type V we obtained an
enhancement factor of order $(\Gamma_P/\Gamma_S)^{1/2}$ (cf. (\ref{5.46})),
for type III rotations of order $L/(\Gamma_P\Gamma_S)^{1/2}$
(cf. (\ref{6.76})). In both cases this leads numerically to factors of order
$10^4$ for $^1_1$H. The most promising effect for experimental observation
seems for us to be the type III rotation discussed in subsection 6.2 for
$^1_1$H. With a suitable set-up one could obtain polarization rotations of a
few
percent! Here the additional enhancement is produced by interference
of the P-violating amplitude with a small P-conserving one. For a
statistically significant signal $\approx 10^{15}$ polarized
hydrogen atoms in  the $2S$ state would be required, which would have to be
subjected to a sequence of $\approx 10^7$ switches of the electric field
(cf.(\ref{6.40})). This could perhaps be realized with some trap or storage
ring device. For neutral hydrogen this would, of course, require magnetic
storage
similar to a neutron storage ring. In the analogous case of $^3_2\mbox{He}^+$
or higher
$Z$ ions one could use the ion storage rings \cite{10}. Another idea mentioned
to the authors by D.~Habs would be to let an atomic beam pass a suitable
Maser or Laser wave providing the required modulation of the electric field.
Note that the presence of magnetic fields is in principle no hindrance for the
observation of P-violation effects. A constant magnetic field, for instance,
can by itself not induce any P-violating atomic effect.

In our investigation we encountered in various places the Zeldovich dipole
moment of unstable states \cite{23}. We have shown that this dipole
moment is connected with the initial rotation of undecayed atomic states
when they are placed in an electric field (cf. (\ref{5.17})). Of course,
we also confirmed in (\ref{5.22}) the well-known role of this dipole moment
in
giving the angular momentum transfer from the undecayed state to the decay
products (cf. \cite{22}) at large times.

Not all polarization rotation effects we discussed require unstable states,
however. Some rotations can also occur for stable systems. This could be
interesting for atoms with more than one electron where --- contrary to the
hydrogen-like systems --- one can find in some cases a P-violating admixture
of reasonable size already for the ground states.

In this paper we have always assumed sudden changes in the electric field
(cf. (\ref{4.3})). This was done in order to simplify the calculations, i.e.
we could use the ``sudden approximation''. More general changes in the
electric field should produce in essence the same type of polarization
rotations, only the theoretical calculations would be a bit more complicated.
In a list of topics for further study we include this as well as
investigations of other atomic systems with more than one electron and muonic
atoms.

To conclude: We hope to give with this article a new impetus to the study
of P-violation in light atomic systems. There, theoretical calculations
can be pushed to very high accuracy, thus allowing in principle low energy
precision determinations of the electroweak parameters related to the
Z exchange between quarks and leptons. For light atomic systems
the nuclear spin-dependent effects are of comparable size to the
nuclear spin-independent ones. P-violation effects in hydrogen could
allow a determination of the $s$-quark contribution to the total spin of
the proton, thus shedding light on the ``spin crisis'' of the nucleons. For
higher $Z$ mirror nuclei one could combine such measurements with accurate
determinations of the Gamow-Teller transition matrix elements of
$\beta$-decay in order to study the change of the $s$-quark spin contribution
of the nucleons in the nuclear environment. Thus P violation in atoms which
involves atomic, nuclear, and particle physics could make further significant
contributions to all three fields.

\vspace{1cm}
\noindent{\bf Acknowledgements:} The authors would like to thank
J.~Bernab\'eu,  W.~Bernreuther, D.~Dubbers,  U.~Ellwanger,
D.~Habs, R.R.~Lewis,  A.~Sch\"afer, D.~Schwalm, L.M.~Sehgal, B.~Stech,
V.~Telegdi and A.~Wolf  for useful discussions and suggestions.
Special thanks are due to W.~Bernreuther, A.~Sch\"afer
and B.~Stech for a critical reading of the
manuscript.
\newpage
\section*{Appendix A}
\renewcommand{\theequation}{A.\arabic{equation}}
\setcounter{equation}{0}

\subsection*{The axial form factors $G_A^{(q)}(Z,N)$}
Here we discuss the possibility to relate $G^{(u)}_A(Z,N)-G_A^{(d)}(Z,N)$
to experimental results from nuclear $\beta$ decay
and $G^{(s)}_A(Z,N)$ to the EMC-results.

{}From the definition in eq. (\ref{2.6}) we have
\bear
\lefteqn{\bra{} Z,N,T,T_3;I,I_3' | \int d^3 x \, \bar \psi(\vec x) \vec \gamma
\gamma_5 \tau_3 \psi(\vec x) | Z,N,T,T_3;I,I_3 \ket{}} \nn &=& \left[
G_A^{(u)}(Z,N) - G_A^{(d)}(Z,N) \right]
\bra{} I,I_3'|\vec{I} | I,I_3\ket{}.
\label{A.1}
\ear
Here $\psi$ denotes the field operator for up and down quarks
\be\label{A.2}
\psi(\vec x) = \left(
\begin{array}{r}u(\vec x) \\ d(\vec x)\end{array}
\right)\ee
and $T,T_3$ are the quantum numbers of isospin and its third component for
the nuclear state. By $\tau_i$ $ (i=1,2,3)$ we denote as usual the Pauli
matrices
of isospin. From isospin invariance we get immediately
\be\label{A.2a} G_A^{(u)}(Z,N)-G_A^{(d)}(Z,N)=0\ee
for $T=0$ nuclei. For $T\not=0$ we consider nuclear
$\beta^-(\beta^+)$-decay where one  defines the Gamow-Teller
matrix elements as matrix elements of the $d\to u$ $(u\to d)$ axial charged
weak
current (cf. e.g. \cite{A1}, \cite{A2}). For a
$\Delta I=\Delta T=0$ transition this reads:
\bear
&&
\bra{} Z,N,T,T_3;I,I_3' | \int d^3 x \, \bar \psi(\vec x) \vec \gamma \gamma_5
\tau^\pm \psi(\vec x) | Z\mp 1,N \pm1 ,T,T_3\mp 1;I,I_3 \ket{}
\nn
&=& g_A M_{GT}^\mp \frac{1}{\sqrt{I ( I+1)}} \bra{} I, I_3' | \vec I | I,I_3
\ket{}
\label{A.3}\ear
where
\be\label{A.4}
\tau^\pm = \halb(\tau_1 \pm i \tau_2).
\ee

We are here only interested in weak axial charges of light stable nuclei. If
such
a nucleus is to be the daughter in an observable $\beta$-decay process, its
parent
must be a ground state nucleus, too.
For the isospin of the ground state of a light nucleus in general $T=|T_3|$
holds
\cite{A3}. As
$\beta^\mp$-decay implies $|\Delta T_3|=1$, the isospin of relevant nuclei must
then be $T=\frac{1}{2}$. So we are left with mirror nuclei where we  can  get
information about axial charges.
Isospin invariance leads now from (\ref{A.1})
and (\ref{A.3}) to the following relations: If $Z,N$ is the $T_3=\pm 1/2$
isospin partner and $\beta^\mp$  decay daughter of the nucleus $Z\mp1, N\pm1$,
we have (cf.~(\ref{2.8})):
\bear
G_A^{(u)}(Z,N) - G_A^{(d)}(Z,N) &=& \frac{1}{I} (\Delta u(Z,N) - \Delta d(Z,N))
\nn
&=& \pm \frac{g_A M_{GT}^\mp}{\sqrt{I (I+1)}}.
\label{A.5}
\ear

The absolute square of the Gamow-Teller matrix element is obtained from the
$ft_{1/2}$ values. In the simple leading order theory one has
\be\label{A.6}
ft_{1/2} = \tau_0 \left( |M_F|^2 +   |g_AM_{GT}|^2 \right)^{-1}.
\ee
Here $ft_{1/2}$ is the usual notation for the product of half-life $t_{1/2}$
and $f$-function (integrated lepton phase space times Fermi function), $\tau_0$
is a constant and $M_F$ is the Fermi matrix element (cf. \cite{A1}, \cite{A2}).
For superallowed Fermi transitions the theory has been pushed to very high
accuracy by evaluation of a great number of correction terms to the relation
(\ref{A.6}) (cf. \cite{A4} and refs. quoted therein). The resulting value
for $\tau_0$ quoted in \cite{A4} is
\be
\label{A.7}
\eh\tau_0=3070.6\pm 1.6s.
\ee

We are not aware of a calculation similar to the one of \cite{A4} for
transitions between light mirror nuclei. Therefore, to obtain an estimate
of $|M_{GT}|^2$, we will use the lowest order expression (\ref{A.6}) with
$\tau_0$ from (\ref{A.7}). Inserting
\be\label{A.8}
M_F=1\ee
as appropriate for $T=1/2$ we get from the experimental $ft_{1/2}$
values numbers for $|g_AM_{GT}|$ and, using (\ref{A.5}), for $|\Delta u-
\Delta d|$.

The sign of $M_{GT}$ and thus of $\Delta u-\Delta d$ can be obtained
experimentally from the study of the decay of polarized nuclei,
as has been done extensively for instance for the neutron decay. We obtain
a theoretical estimate of the sign of $\Delta u-\Delta d$ for the other
nuclei of our list with the help of a simple shell model calculation,
which we also use to estimate the quantities $\Delta s$.

Consider the axial quark currents of (\ref{2.3}). Expressing them
as effective nucleonic currents we get in nonrelativistic reduction:
\be\label{A.9}
\bar u(\vec x)\vec\gamma\gamma_5u(\vec x)-\bar d(\vec x)\vec\gamma\gamma_5
d(\vec x)\longrightarrow g_A\varphi^\dagger_N(\vec x)\tau_3\vec\sigma\varphi_N
(\vec x),\ee
\be\label{A.10}
\bar s(\vec x)\vec\gamma\gamma_5s(\vec x)\longrightarrow \Delta
s_p\varphi^\dagger_N(\vec x)\vec\sigma\varphi_N(\vec x),\ee
where
\be\label{A.11}
\varphi_N(\vec x)={p(\vec x) \choose n(\vec x)}\ee
is the nonrelativistic Pauli-Schr\"odinger field operator for the nucleons.
The effective nucleonic current in (\ref{A.9}) is completely conventional
whereas the one in (\ref{A.10}) has  -- to our knowledge --
not been discussed in the literature before. The replacement made
in (\ref{A.10}) is dictated by requiring that the quark current
on the l.h.s. and the effective current on the r.h.s. of (\ref{A.10})
give the same result for the matrix elements of free
nonrelativistic nucleons.
We assume  now in our model that the effective currents
of (\ref{A.9}), (\ref{A.10}) with $g_A$ and
$\Delta s_p$ equal to the free nucleon values give the axial charges
for light nuclei. We have then:
\bear\label{A.12}
&&\left[G_A^{(u)}(Z,N)-G_A^{(d)}(Z,N)\right]\langle I,I_3'|\vec I|I,I_3\rangle
\nonumber\\
&& \ \ \ \ \ \
=\langle Z,N,T,T_3;I,I_3'|g_A\int d^3x\varphi_N^\dagger(\vec x)\tau_3\vec
\sigma\varphi_N(\vec x)|Z,N,T,T_3;I,I_3\rangle,\ear
\bear\label{A.13}
&&G_A^{(s)}(Z,N)\langle I,I_3'|\vec I|I,I_3\rangle\nonumber\\
&&\ \ \ \ \ \
=\langle Z,N,T,T_3;I,I_3'|\Delta s_p\int d^3x\varphi_N^\dagger(\vec x)\vec
\sigma\varphi_N(\vec x)|Z,N,T,T_3; I,I_3\rangle.\ear
We evaluated these matrix elements for the nuclei $^2_1\mbox{H},\
^3_2\mbox{He}, \
^6_3\mbox{Li},\ ^{13}_6\mbox{C}$ and $^{17}_8\mbox{O}$ using the simplest
standard shell
model configurations \cite{A3}. The resulting values for $\Delta u-\Delta d$
and $\Delta s$ are listed in Table A1.

In Table A2 we list the values for the weak charges $Q_W^{(1,2)}$ for some
nuclei. $Q_W^{(1)}$ is calculated from (\ref{2.16}) using $\ssw
=0.23$ \cite{16}. The value for $\Delta u-\Delta d$ for the proton equals
$g_A$ from isospin invariance (cf. (\ref{2.9}), (\ref{A.5})). For the other
nuclei we used for $T=0$ (\ref{A.2a}) and for $T=1/2\ |\Delta u-\Delta d|$ from
the
$ft_{1/2}$ values of the parent nuclei in $\beta$ decay and the sign of
$\Delta u-\Delta d$ from Table A1. These $ft_{1/2}$ values were taken from
\cite{A2} except for $^{17}_9\mbox{F}$, where we used \cite{A8}. The column
$\Delta s$, EMC gives for the proton the central value of the EMC result
\cite{8} and for the other nuclei the corresponding shell model
estimate of Table A1. We list finally $Q_W^{(2)}$ for $\Delta s$ equal
to these values and for $\Delta s=0$.

By comparing the columns $\Delta u-\Delta d$ of Tables A1 and A2 we notice the
well-known fact (cf. \cite{seve} and references cited therein) that the
effective axial coupling $g_A$ to be used in (\ref{A.9}) is decreasing
with increasing $A$. It would clearly be of great interest to know if
the effective $\Delta s_p$ to be used in (\ref{A.10}) stays the same or
is also decreasing with $A$.
\section*{Appendix B}
\renewcommand{\theequation}{B.\arabic{equation}}
\setcounter{equation}{0}

In this appendix the non-hermitian mass matrix $\M$ (cf. \ref{3.12}) and its
left and right eigenvectors for $\vec{\E}=0$, the dipole operator
$\underline{\bm{D}}$ (cf. (\ref{3.34})) and the matrix representations
for the T transformation (cf. (\ref{3.30})) and the reflection R
(\ref{3.51b}) are
 given explicitly for the cases $I=0$ and $I=\halb$.

For the Coulomb wave functions of the atomic states the phase conventions of
\cite{18} are used except for an overall sign change in all radial wave
functions. For $I=0$ the basis states, i.e. the eigenstates of $H_0$
(\ref{3.1}),
 are denoted as $|n,l,F,F_3)$ where
$F\equiv j,\ F_3\equiv j_3$;
for $I = \halb$ as $|n,l,j,F,F_3)$ with $l = 0,1,\dots$ indicated in
spectroscopic notation: ($S,P,\dots$). The states are normalized to one. The
zero point of the energy scale is set at the energy center of the
$2P_{\frac{1}{2}}$ states. In the main text
 we have in the notation of the states frequently omitted the quantum numbers
$j$ and/or $F$ if their values are clear from the context.

For the case of nuclear spin $I=0$ the relevant matrix elements of
$H_{PV}$ between $2S$ and $2P$ states are (cf. (\ref{3.14}), (\ref{3.16}) for
$L,
\delta_1$):
\bear\label{B.1}
(2S,\eh ,F_3'|H_{PV}|2P,\eh ,F_3)
&=&(2S,\eh ,F_3'|H_{PV}^{(1)}|2P,\eh ,F_3)\nonumber\\
&=&-i\delta_1(Z,N)L(Z,N)\delta_{F_3'F_3}.\ear
Here we use the Coulomb approximation for the wave functions. The matrix
representations for $\M(\E\vec e_3)$, $\underline{\T}$, $\underline{\R}$ and
$\underline{\bm{D}}$ (cf. (\ref{3.12}), (\ref{3.30}), (\ref{3.51b}),
(\ref{3.34})) are given in Tables B1 -- B3. The
eigenvalues,
right and left eigenvectors (cf. (\ref{3.22})--(\ref{3.24a})) of $\M(0)$ are
given in Table B4, where we are always neglecting terms of order $\delta_1^2$.
For $I=0$ the T-transformation matrix $\underline{\T}$ and the
reflection matrix $\underline{\R}$ satisfy
\be
\label{B.2}
\underline{\T} = - \underline{\T}^\dagger = - \underline{\T}^{-1},
\ee
\be\label{B.3}
\underline{\R}=-\underline{\R}^\dagger=-\underline{\R}^{-1}.\ee
The T transformation of the eigenstates of Table B4 is as follows:
\bear\label{B.4}
\left(\underline{\T}^\dagger|2\hat
S,F,F_3)\right)^T&=&(-1)^{F+F_3}(2\widetilde{
\hat S,F},-F_3|,\nonumber\\
\left(\underline{\T}^\dagger|2\hat
P,F,F_3)\right)^T&=&(-1)^{F-F_3}(2\widetilde{
\hat P,F},-F_3|,\nonumber\\
\left((2\widetilde{\hat S,F},F_3|\underline{\T}\right)^T&=&(-1)^{F+F_3}|2
\hat S,F,-F_3),\nonumber\\
\left((2\widetilde{\hat P,F},F_3|\underline{\T}\right)^T&=&(-1)^{F-F_3}|2
\hat P,F,-F_3),\nonumber\\
&&(F=\eh ,\ F_3=\pm\eh ).\ear

For nuclear spin $I\not=0$ the relevant matrix elements of ${\cal
H}_{PV}^{(1,2)}$ are as follows, where we use again the Coulomb approximation
for the wave functions (cf. (\ref{3.14}), (\ref{3.16}) for $L,\delta_{1,2}$)
\bear\label{B.5}
(2S,\halb,F',F_3'|H^{(1)}_{PV}|2P,\halb,F,F_3)
&&=-i\delta_1(Z,N)L(Z,N)\delta_{F',F}\delta_{F_3',F_3} \ ,\nonumber\\
(2S,\halb,F',F_3'|H^{(2)}_{PV}|2P,\halb,F,F_3)
&&=-i\delta_2(Z,N)L(Z,N) \cdot \nonumber \\
&&\ \ \
   [F(F+1)-I(I+1)-\dv]\delta_{F',F}\delta_{F_3',F_3} \ ,\nonumber\\
(|I-\eh|\leq F,F'\leq I+\eh) &&\ ;
\ear
\bear
\label{B.6}
(2S,\eh,F',F_3'|H^{(2)}_{PV}|2P,\dh,F,F_3)&=&0 \ ,\ \ \ \ \ \ \ \ \ \
       \nonumber\\
(|I-\eh|\leq F'\leq I+\eh&,&\ |I-\dh|\leq F\leq I+\dh) .\ear

For $I=\frac{1}{2}$ the
matrix representations for $\M(\E\vec e_3), \ \underline{\T},\ \underline\R$
and $\underline{\bm{D}}$ are given in Tables B5 -- B7. The eigenvalues
 and  eigenvectors of $\M(0)$ are listed in Table B8. The normalization
conditions (\ref{3.24}), (\ref{3.24a}) are satisfied, neglecting quadratic
terms in $\delta_{1,2}$. For $I=\halb$ the matrices $\underline{\T}$ and
$\underline{\R}$ satisfy
\be\label{B.8}
\underline{\T} =\underline{\T}^\dagger = \underline{\T}^{-1} ,
\ee
\be\label{B.9}
\underline{\R} =\underline{\R}^\dagger = \underline{\R}^{-1} .
\ee
The T  transformation properties of the $2\hat S$ states of Table B8 are as
follows:
\bear\label{B.10}
\left(\underline{\T}^\dagger|2\hat S,\halb ,
F,F_3)\right)^T&=&(-1)^{F+F_3}(2\widetilde{
\hat S,\halb }, F,-F_3|,
\nonumber\\
\left((2\widetilde{\hat S,\halb }, F,F_3|\underline\T\right)^T&=&(-1)^{F+F_3}|2
\hat S,\halb , F,-F_3).
\ear
\section*{Appendix C}
\renewcommand{\theequation}{C.\arabic{equation}}
\setcounter{equation}{0}
In this appendix we collect some useful formulae for eigenvalues and
eigenvectors of non-hermitian matrices.

Let $\calm_0,\calm_1$ be non-hermitean $n\times n$ matrices and let $\lambda$
be
a real parameter. Define
\be\label{C.1}
\calm(\lambda)=\calm_0+\lambda\calm_1.\ee
We assume that the eigenvalue problem for $\calm_0$ is solved and that
all (in general complex) eigenvalues $E_\alpha^0$ $(\alpha=1,...,n)$ are
different from each other:
\be\label{C.2}
E^0_\alpha\not=E^0_\beta\quad{\rm for}\quad \alpha\not=\beta.\ee
There exists then a basis of right and left eigenvectors,
$|\alpha^0\bigl\rangle$ and
$\bigl\langle\tilde\alpha^0|$, such that
\bear\label{C.3}
\calm_0|\alpha^0\bigl\rangle&=&E^0_\alpha|\alpha^0\bigl\rangle,\nonumber\\
\bigl\langle\tilde\alpha^0|\calm_0&=&\bigl\langle\tilde\alpha^0|E^0_\alpha\ear
and we can choose the normalization such that
\bear\label{C.4}
&&\bigl\langle\tilde\alpha^0|\beta^0\bigl\rangle=
\delta_{\alpha\beta},\nonumber\\
&&(1\leq\alpha,\beta\leq n).\ear
The quasiprojectors are defined as (cf. (\ref{3.25}))
\bear\label{C.5}
\Pro^0_\alpha&:=&|\alpha^0\bigl\rangle \,
\bigl\langle\tilde\alpha^0|\nonumber\\
&&(\alpha=1,...,n).\ear

Consider now the eigenvalue problem for $\calm(\lambda)$ (C.1). The eigenvalues
$E_\alpha(\lambda)$ are obtained as solutions of the characteristic equation
\be\label{C.6}
\mbox{det}\left[\calm(\lambda)-E\cdot\eins\right]=0.\ee
The implicit function theorem together with (\ref{C.2}) guarantees
the existence of a neighbourhood $\U$ of $\lambda=0$ such that for
all $\lambda$ with $\lambda\in \U$ the eigenvalues $E_\alpha(\lambda)$
satisfy again
\be\label{C.7}
E_\alpha(\lambda)\not=E_\beta(\lambda)\quad {\rm for}\quad
\alpha\not=\beta.\ee
This in turn implies the existence of right and left eigenvectors and
quasiprojectors for $\lambda\in \U$:
\be\label{C.8}
|\alpha(\lambda)\bigl\rangle,\ \bigl\langle\tilde\alpha(\lambda)|,\
\Pro_\alpha(\lambda).\ee
We want to show that $\Pro_\alpha(\lambda)$ and with
suitable phase conventions also the right and left eigenvectors are
differentiable for $\lambda\in \U'$ where $\U'\subset \U$ is some neighbourhood
of
$\lambda=0$. To see this we construct the resolvent (cf. \cite{mess}):
\be\label{C.9}
\frac{1}{z-\calm(\lambda)}=\sum_\alpha\frac{\Pro_\alpha(\lambda)}
{z-E_\alpha(\lambda)}.\ee
With the help of Cauchy's formula we can write the quasiprojectors as
follows:
\be\label{C.10}
\Pro_\alpha(\lambda)=\frac{1}{2\pi
i}\oint_{C_\alpha}\frac{dz}{z-\calm(\lambda)}
\ee
where the integration path $C_\alpha$ runs in the analyticity domain of
$(z-\calm(\lambda))^{-1}$ and encircles anticlockwise
just the pole at the eigenvalue
$E_\alpha(\lambda)$ and no other pole. This is possible due to (\ref{C.7}) for
sufficiently small $|\lambda|$. But on the closed curve $C_\alpha$ the
matrix elements of $(z-\calm(\lambda))^{-1}$ are rational functions in the
variables $\Re\ z$, $\Im\ z$, $\lambda$ without singularities for $|\lambda|$
sufficiently small: $\lambda\in \U''$, where $\U''$ is a suitable neighbourhood
of $\lambda=0$. Since rational functions are arbitrarily often differentiable,
we can then infer from a well-known theorem of analysis that also $\Pro_\alpha
(\lambda)$ (\ref{C.10}) is arbitrarily often differentiable for $\lambda\in
\U''$. For the correctly normalized eigenvectors we can take
\bear
\label{C.11}
|\alpha(\lambda)\bigl\rangle&=&\Pro_\alpha(\lambda)|\alpha^0\bigl\rangle,
\nonumber\\
\bigl\langle\tilde\alpha(\lambda)|&=&[\Tr(\Pro_\alpha(\lambda)
\Pro_\alpha(0))]^{-1}\,\bigl\langle\tilde\alpha^0|\Pro_\alpha(\lambda),
\ear
where the trace will be $\not=0$ in a suitable neighbourhood $\U'$ of $\lambda
=0$. Thus $\Pro_\alpha(\lambda)$ and, with suitable phase conventions,
also $|\alpha(\lambda)\bigl\rangle$ and
$\bigl\langle\tilde\alpha(\lambda)|$
are differentiable
for $\lambda\in \U'$, q.e.d..

The perturbation expansion in $\lambda$ is obtained from (\ref{C.10}) in a
similar way as for hermitian matrices (cf. \cite{mess}):
\bear\label{C.12}
\Pro_\alpha(\lambda)&=&\frac{1}{2\pi i}\oint_{C_\alpha}dz\,[1-(z-\calm_0)^{-1}
     \lambda\calm_1]^{-1}(z-\calm_0)^{-1}\nonumber\\
&=&\frac{1}{2\pi i}\oint_{C_\alpha}dz\left[(z-\calm_0)^{-1}+(z-\calm_0)^{-1}
\lambda\calm_1(z-\calm_0)^{-1}+...\right].\ear
This leads to
\be\label{C.13}
\Pro_\alpha(\lambda)=\Pro^0_\alpha+\sum_{\beta\not=\alpha}
\frac{1}{E^0_\alpha-E^0_\beta}\cdot\left[\Pro^0_\alpha\lambda
\calm_1\Pro_\beta^0+
\Pro^0_\beta\lambda\calm_1\Pro^0_\alpha\right]+{\cal O}(\lambda^2),
\ee
\be
\label{C.14}
\Tr \, \Pro_\alpha(\lambda)\Pro^0_\alpha=1+{\cal O}(\lambda^2),
\ee
\be
\label{C.15}
|\alpha(\lambda)\bigl\rangle=|\alpha^0\bigl\rangle
+\sum_{\beta\not=\alpha}
\frac{1}{E^0_\alpha-E^0_\beta}\Pro^0_\beta\lambda\calm_1|\alpha^0\bigl\rangle
+{\cal O}(\lambda^2),
\ee
\be
\label{C.16}
\bigl\langle\tilde\alpha(\lambda)|=\bigl\langle\tilde\alpha^0|+
\sum_{\beta\not=\alpha}\frac{1}{E^0_\alpha-E^0_\beta}
\,\bigl\langle\tilde\alpha^0|\lambda\calm_1\Pro_\beta^0+{\cal O}(\lambda^2),
\ee
\bear\label{C.17}
E_\alpha(\lambda)&=&\Tr[(\calm_0+\lambda\calm_1)
\Pro_\alpha(\lambda)]\nonumber\\
&=&E^0_\alpha+\lambda \Tr\,[\calm_1\Pro^0_\alpha]+{\cal O}(\lambda^2).\ear
These expansion formulae (cf. also (4.3) of \cite{12}) have been used
extensively
in our calculations.
\section*{Appendix D}
\renewcommand{\theequation}{D.\arabic{equation}}
\setcounter{equation}{0}
In this appendix we calculate the first few terms in the expansion
of the quantity $\Delta f$ defined in (\ref{4.14}). From (\ref{4.11}) and
(\ref{4.12}) we get
\bear
\Delta f^{r_2,r_1} &=& \Tr \left\{ \M_2^{r_2} \M_1^{r_1} \left(
\pro(2\hat{S},\halb,0) - \pro(2\hat{S},-\halb,0) \right) \right\} \nn
&=& \Tr \left\{ \left[ \M_2^{r_2},\M_1^{r_1}\right] \pro(2\hat{S},\halb,0)
\right\}
\label{D.1}
\ear
where
\bear
\M_j &:=& \M(\E^j \vec e_3)
\label{D.2}
\\
&&(j = 1,2) \nonumber.
\ear
{}From(\ref{D.1}) we deduce immediately $\Delta f^{r_2,r_1} = 0$ for $r_1 = 0$
or $r_2 = 0$. To calculate further terms we write
\bear
\M_j &=& \M_0 - \E^j \underline{D}_3
\label{D.3}
\\
&&(j = 1,2),\nonumber
\ear
where $\M_0$ is the mass matrix for $\vec \E = 0$ and $\underline{D}_3$ the
3-component of the dipole moment operator (cf. Tables B1, B3). We get then:
\bear
\Delta f^{r_2,r_1} &=& ( 2 \tilde{\hat{S}}, \halb,0 | \left[ \M_2^{r_2},
\M_1^{r_1}\right] | 2 \hat{S},\halb,0) \nn
&=&
( 2 \tilde{\hat{S}}, \halb,0 | \left[(\M_0 - \E^2 \underline{D}_3)^{r_2},
(\M_0 - \E^1 \underline{D}_3)^{r_1}\right]
| 2 \hat{S},\halb,0) .
\label{D.4}
\ear
The trick is now to use the fact that $|2 \hat{S},\halb,0)$ and
$( 2 \tilde{\hat{S}}, \halb,0 |$ are right and left eigenvectors, respectively,
of $\M_0$ to the \underline{same} eigenvalue $E(2 \hat{S},\halb,0)$. This leads
to
\bear
\Delta f^{1,1} &=&
( 2 \tilde{\hat{S}}, \halb,0 | \left[(\M_0 - \E^2 \underline{D}_3),
(\M_0 - \E^1 \underline{D}_3)\right]
| 2 \hat{S},\halb,0)  \nn
&=& ( 2 \tilde{\hat{S}}, \halb,0 | \left[ (E(2 \hat{S},\halb,0) - \E^2
\underline{D}_3),
(E(2 \hat{S},\halb,0) - \E^1 \underline{D}_3)\right]
| 2 \hat{S},\halb,0) \nn
&=& 0 \label{D.5}
\ear
and in a similar way to
\be
\label{D.6}
\Delta f^{1,2} = \Delta f^{2,1} = \Delta f^{1,3} = \Delta f^{3,1} = 0.
\ee
The first nonzero value for a coefficient $\Delta f^{r_2,r_1}$ occurs for $r_2
= r_1 = 2$, where we find
\be
\label{D.7}
\Delta f^{2,2} = \E^1 \E^2 (\E^1 - \E^2) ( 2 \tilde{\hat{S}}, \halb,0 | \,
\underline{D}_3 \left[ \underline{D}_3 , \M_0\right] \underline{D}_3 \,
| 2 \hat{S},\halb,0) .
\ee
Explicit evaluation of this matrix element using Tables B1, B3, B4 leads to the
result given in (\ref{4.18}).
\clearpage
\newpage
\section*{Table Captions}
\small
\begin{description}
\item[Table 1]
Types of polarization rotation and angular momentum transfer effects in
electric
fields. It is always assumed that the initial state at time $t=0$ is a
metastable $2\hat{S}$ state for field $\vec \E = 0$. Types I, II, V refer to a
single $\vec \E$-field, constant in space and time for $t>0$, type III to an
$\vec \E$-field constant for a time interval $0 \leq t \leq t_1$, type IV  to
an
$\vec \E$-field piecewise constant in time which is taking on more than one
value.
\item[Table 2]
Energies, inverse lifetimes  and the quantity ${\cal F}$
for $^4_2$He$^+$ and $^{12}_6$C$^{5+}$ (cf.
(\ref{3.14}), (\ref{3.15})).
For $\Gamma_P(2,2),\ \Gamma_P(6,6)$ and $\Gamma_S(6,6)$ we  used the scaling
laws from (\ref{3.20}) together with the numerical values for $\Gamma_P(1,0),\
\Gamma_S(1,0)$ from (\ref{3.18}).
The P-violation parameters $\delta_1(Z,N)$ (\ref{3.16}) are calculated using
$\sin^2\theta_W=0.23$.
\item[Table 3]
Energies, inverse lifetimes, the quantity ${\cal F}$ (\ref{3.15}) and the
P-violation parameters $\delta_{1,2}$ (\ref{3.16}) calculated with $\ssw=0.23$
for  $^1_1$H.
\item[Table 4]
Parametrization of a general density matrix for a $2\hat{S}$ state for an atom
with $I = \halb$ (cf. (\ref{6.6})). $S_0, \vec{S}, S_{ij}$ are real and $S_{ij}
= S_{ji}, S_{ii} = 0$, $\vec K$ is in general complex.
\item[Table 5]
Values for the parameters $\tilde \kappa$ and $\kappa$ of (\ref{6.48}) for
$^1_1$H.
\item[Table A1]
Shell model configurations for the nuclei $^2_1$H, $^3_2$He, $^6_3$Li,
$^{13}_6$C and $^{17}_8$O and the resulting  values for $\Delta u-\Delta d$ and
$\Delta s$. Here $g_A=1.2573(28)$  is the  axial decay constant of the neutron,
$\Delta s_p$ the $s$-quark contribution to the proton spin.
\item[Table A2]
Parameters of some nuclei: $I$ is the spin, $Q_W^{(1,2)}$ are the weak charges
(\ref{2.16}), (\ref{2.17}), $\Delta u,\Delta d, \Delta s$ are the spin
contributions of $u,d,s$ quarks (\ref{2.8}). The numbers are obtained as
explained in the text.
\item[Table B1]
The mass matrix $\M(\vec \E)$ for the $n=2$ states for nuclear spin $I=0$ and
external field $\vec \E = \E \vec e_3;\ {\cal F}=e r_B(Z)\E$ (cf. (\ref
{3.14})-(\ref{3.16})).
\item[Table B2]
The T-transformation matrix $\underline{\T}$ (\ref{3.30})
and the reflection transformation matrix $\underline{\cal R}$ (\ref{3.51b})
 for $I=0$ in the same basis as used in table B1.
\item[Table B3]
The matrix representation of the suitably normalized dipole operator
\newline
 $\underline{\vec D} \, / \, (3 \Qe r_B(Z))$ for $I=0$ in the $n=2$ subspace.
Here $\vec{e}_i \, (i = 1,2,3)$ are the Cartesian unit vectors.
\item[Table B4]
The right and left eigenvectors of $\M(0)$ from Table B1, neglecting terms of
${\cal O} (\delta_1^2)$.
\item[Table B5]
The mass matrix $\M(\vec \E)$ for the $n=2$ states
 for nuclear spin $I=\halb$ and external field $\vec \E =\E\vec e_3;\
{\cal F}=er_B(Z)\E$ (cf. (\ref{3.14})-(\ref{3.16}), (\ref{6.1})).
\item[Table B6]
The T-transformation matrix $\underline{\T}$ (\ref{3.30}) and the reflection
transformation matrix $\underline{\cal R}$ (\ref{3.51b})
for $I=\halb$ in the same basis as used in table B5.
\item[Table B7]
The matrix representation of the suitably normalized dipole operator
$\underline{\vec D} \, / \, (3er_B(Z))$
 for  $I=\halb$ in the $n=2$ subspace. Here $\vec e_i\ (i=1,2,3)$ are the
Cartesian unit vectors.
\item[Table B8] The eigenvalues and right eigenvectors of $\M(0)$ from Table
B5,
neglecting terms of ${\cal O}(\delta_1^2,\delta^2_2,\delta_1\delta_2)$. The
left eigenvectors, written as ket vectors, can be
inferred from the right eigenvectors by replacing $i \Gamma_{S,P}$ by $-i
\Gamma_{S,P}$.
\end{description}
\normalsize
\newpage
\begin{center}
{\bf Table 1}
\end{center}
\renewcommand{\arraystretch}{3.5}
\begin{center}
\begin{tabular}{|c|c|}
\hline
type & rotation of \\
\hline
 I & \parbox{10cm}{complete state vector of undecayed atoms in the external
electric field.
\\}  \\
\hline
 II & \parbox{10cm}{metastable part of the undecayed state vector which
survives for $t \gg \tau_P$ in a weak electric field satisfying $\sqrt{3}
\F(Z,|\vec \E|) / L(Z,N) \ll 1$ (cf. \ref{3.15}), (\ref{3.18}), (\ref{3.20}) .
\\
}\\
\hline
III & \parbox{10cm}{$2 \hat S$ projected state, after switching off $\vec \E$,
relative to initial state.
\\}
\\
\hline
IV & \parbox{10cm}{$2 \hat S$ projected state, after having passed through more
than one $\E$ field, relative to initial state.
\\}
\\
\hline
V & \parbox{10cm}{external field producing device due to angular momentum
transfer from atoms \cite{12}.
\\}\\
\hline
\end{tabular}
\end{center}
\renewcommand{\arraystretch}{1.5}
\begin{center}
\vspace*{1.5cm}
{\bf Table 2}
\end{center}
\nopagebreak[4]
\renewcommand{\arraystretch}{2.0}
\begin{tabular}{|c||p{3.9cm}|p{1.0cm}||p{4.2cm}|p{1.0cm}|}
\hline
  & $^4_2$He$^+$ & Ref. & $^{12}_6$C$^{5+}$ & Ref. \\
\hline \hline
$\Delta(Z,N)/h$ & $1.756 \cdot 10^5 $ MHz
   & \cite{john} & $ 1.424 \cdot 10^7 $ MHz & \cite{john}  \\
 $L(Z,N)/h$ &$ 1.404 \cdot 10^4 $ MHz  & \cite{drak}
   &$7.801 \cdot 10^5$ MHz  &   \cite{kuge} \\
 $\Gamma_P(Z,N)$ &$1.008\cdot 10^{10}$ s$^{-1}$ & &
           $8.165 \cdot 10^{11}$ s$^{-1}$ &
   \\
 $\Gamma_S(Z,N)$ &525.49 s$^{-1}$ & \cite{hind} & $3.840 \cdot 10^5$
             s$^{-1}$  &     \\
 $\sqrt{3}{\cal F} (Z,{\cal E})/L(Z,N)$ &
      ${\cal E}\cdot(12667.9 \, \mbox{V/cm})^{-1}$ &  &
       ${\cal E}\cdot(\mbox{2.112$\cdot 10^6$ V/cm})^{-1}$   &
 \\
$\delta_1(Z,N)$ & $1.36 \cdot 10^{-11} $ & & $5.96 \cdot 10^{-11}$ & \\
\hline
\end{tabular}
\newpage
\begin{center}
{\bf Table 3}
\vspace*{.5cm}
\end{center}
\nopagebreak[4]
\renewcommand{\arraystretch}{2.0}
\renewcommand{\tabcolsep}{0.5cm}
\begin{center}
\begin{tabular}{|c||p{4.4cm}|p{1.0cm}|}
\hline
  & $^1_1$H$^+$ & Ref. \\
\hline \hline
$\Delta(Z,N)/h$ & $ 10969$ MHz
   & \cite{eric}  \\
 $ L(Z,N)/h$ &$ 1058 $ MHz  & \cite{eric}
  \\
 $ A_1(Z,N)/h$ &$ \frac{1}{8}\cdot 1420.4 $ MHz  & \cite{pipk}
   \\
 $ A_2(Z,N)/h$ &$ \frac{1}{24}\cdot 1420.4 $ MHz  & \cite{pipk}
   \\
 $ A_3(Z,N)/h$ &$ \frac{1}{60}\cdot 1420.4 $ MHz  & \cite{pipk}
   \\
 $\Gamma_P(Z,N)$ &$6.3\cdot 10^{8}$ s$^{-1}$ & \cite{beth}
   \\
 $\Gamma_S(Z,N)$ &8.2 s$^{-1}$ & \cite{beth} \\
 $\sqrt{3}{\cal F} (Z,{\cal E})/L(Z,N)$ &
      ${\cal E}\cdot (477.3  \mbox{V/cm})^{-1}$ &
 \\
$\delta_1$ &$ - 4.91 \cdot 10^{-13}$ &
\\
$\delta_2$ & $1.23 \cdot 10^{-12} (1 - g_A^{-1} \Delta s_p) $ &
\\
\hline
\end{tabular}
\end{center}
\vspace*{2 cm}
\begin{center}
{\bf Table 4}
\nopagebreak[4]
\renewcommand{\arraystretch}{1.0}
$$
\begin{array}{|rrr||c|c|}
\hline
&& F & 0 & 1 \\
&& F_3& 0 & F_3 \\
F' & F_3' &  & &  \\
\hline
\hline
\rule[-1.3em]{0cm}{3em}
0 & 0 && 1 - S_0 & \vec{K}^* \cdot \vec{e}_{F_3} \\
\hline
\rule[-1.3em]{0cm}{3em}
1 & F_3' &&  \vec{e}_{F_3'}^* \cdot \vec{K} & \frac{1}{3} S_0 \delta_{F_3',F_3}
+  \vec{S}\cdot  \left(
\vec{e}_{F_3'}^* \times \vec{e}_{F_3} \right)/(2 i) - \vec{e}^*_{F_3'} \!\cdot
\!\vec{e}_i\, S_{ij}\, \vec{e}_j \!\cdot \!\vec{e}_{F_3} \\
\hline
\end{array}
$$
\end{center}
\newpage
\begin{center}
{\bf Table 5}
\end{center}
\renewcommand{\arraystretch}{1.5}
$$
\begin{array}{|rc|c|c|}
\hline
F&F_3&\tilde{\kappa}_{F,F_3}&\kappa_{F,F_3}\\
\hline
0&0&1.02&3.2\\
1&0&0.63&2.1 \\
1&\pm 1&0.70&1.9\\
\hline
\end{array}
$$
\renewcommand{\arraystretch}{1.0}
\renewcommand{\arraystretch}{2.0}
\renewcommand{\tabcolsep}{0.3cm}
\vspace*{1cm}
\begin{center}
{\bf Table A1}
\end{center}
\nopagebreak[4]
$$
\begin{array}{|c|c|c|c|c|}
\hline
\mbox{nucleus}  & \multicolumn{2}{c|}{ \mbox{configuration} } &
 \multicolumn{2}{c|} {}\\
\cline{2-5}
& \mbox{protons} & \mbox{neutrons} & \Delta u - \Delta d & \Delta s \\
\hline
{}_1^2 \mbox{H} & 1 s_{1/2} & 1 s_{1/2} & 0 &2 \Delta s_p \\
{}_2^3 \mbox{He} & (1 s_{1/2})^2 & 1 s_{1/2} & - g_A & \Delta s_p \\
{}_3^6 \mbox{Li} & ( 1 s_{1/2})^2 \,1 p_{3/2} & (1 s_{1/2})^2\, 1 p_{3/2} & 0
&\frac{2}{3} \Delta s_p \\
{}_6^{13}\mbox{C} &( 1 s_{1/2})^2 \,(1 p_{3/2})^4 &
( 1 s_{1/2})^2 \,(1 p_{3/2})^4 \, 1 p_{1/2} & \frac{1}{3} g_A & - \frac{1}{3}
\Delta s_p \\
{}_8^{17}\mbox{O} &( 1 s_{1/2})^2 \,(1 p_{3/2})^4 \, (1 p_{1/2})^2  &
( 1 s_{1/2})^2 \,(1 p_{3/2})^4 \, (1 p_{1/2})^2 \, 1 d_{5/2} & - g_A &
\Delta s_p \\
\hline
\end{array}
$$
\vspace*{0.5cm}
\begin{center}
{\bf Table A2}
\vspace*{0.5cm}

\begin{math}
\renewcommand{\arraystretch}{1.5}
\renewcommand{\tabcolsep}{0.3cm}
\begin{tabular}{|c|c|c|c|c|c|c|
                             c|}
\hline
  &  &  related & & 
         $\Delta u -\Delta d$ & $\Delta s$ &
       \multicolumn{2}{c|}{$Q_W^{(2)}$} \\
\cline{5-8}
nucleus & $I$ & mirror nucleus & $Q_W^{(1)}$ & GT & 
           EMC & $\Delta s_{EMC}$ & $\Delta s = 0 $ \\
\hline
\hline
p & \eh &  n & 0.08 & 1.2573(28)& $-0.19$ &$-0.23$ & $-0.20$\\
$_1^2$H & 1 &   & $-0.92$ &0 &$ -0.38$ & $-0.030 $ &0 \\
$_2^3$He & \eh  &  $_1^3$H & $-0.84$ &$ -1.21$&$-0.19$ & 0.16& 0.19\\
$_2^4$He & 0 &    &$ -1.84$  &0 &0 &0 & 0\\
$_3^6$Li & 1 &    & $-2.76$  & 0&$ -0.13$ & $-0.010$ &0 \\
$_6^{12}$C & 0 &    & $-5.52$ & 0&0 & 0& 0\\
$_6^{13}$C & \eh  & $_7^{13}$N  & $-6.52$ &0.32  &0.06 & $-0.041$&$-0.051$ \\
$_8^{17}$O & \fh &  $_9^{17}$F  & $-8.36$ &$-1.09$  &$ -0.19$&0.029 &0.035 \\
  \hline
\end{tabular}
\end{math}
\end{center}
\raisebox{-19cm}{

\begin{rotate}{-90}
\raisebox{-13cm}{
\begin{minipage}[b]{0.80\textheight}
\renewcommand{\arraystretch}{2.5}
\renewcommand{\tabcolsep}{0.1cm}
\begin{center}
{\bf Table B1}
\vspace*{.5cm}

\begin{tabular}{|c||c|ccc|ccc|
                             c|}
\hline
 & $2P, \dh, \ \dh$ & $2P, \dh, \ \eh $ & $2P,\eh, \ \eh $ & $2S,\eh, \ \eh $
    & $2P,\dh, -\eh $ & $2P,\eh, -\eh $ & $2S, \eh, -\eh $ & $2P, \dh, -\dh $\\
\hline
\hline
 $2P, \dh, \ \dh$ & $\Delta - \frac{i}{2} \Gamma_P$  & 0
   & 0& 0& 0 & 0  & 0 & 0\\
\hline
 $2P, \dh, \ \eh $ & 0& $\Delta - \frac{i}{2} \Gamma_P $ & 0 &
                         $-\sqrt{6}{\cal F}$& 0
   & 0 & 0& 0\\
 $2P,\eh, \ \eh $ & 0&0& $  - \frac{i}{2} \Gamma_P  $& $i \delta_1
L+\sqrt{3}{\cal F}$
             & 0 &  0 &  0& 0\\
 $2S,\eh, \ \eh $  & 0& $-\sqrt{6}{\cal F}$ &$-i \delta_1 L+\sqrt{3}{\cal F}$
    & $L-\ih \Gamma_S$& 0& 0&0& 0\\
\hline
 $2P,\dh, -\eh $ & 0& 0& 0 & 0
   & $\Delta - \frac{i}{2} \Gamma_P $ & 0& $-\sqrt{6}{\cal F}$& 0\\
 $2P,\eh, -\eh $ & 0 &
        0&0 & 0
      & 0 & $  - \frac{i}{2} \Gamma_P  $& $i \delta_1 L-\sqrt{3}{\cal F}$ & 0\\
 $2S, \eh, -\eh $  & 0 & 0 &
       0 &
        0& $-\sqrt{6}{\cal F}$ & $-i \delta_1 L-\sqrt{3}{\cal F}$&$L-\ih
\Gamma_S$ &  0\\
\hline
 $2P, \dh, -\dh $& 0&0&0&0&0&0&0&$\Delta - \frac{i}{2} \Gamma_P$ \\
\hline
\end{tabular}
\end{center}
\end{minipage}
}     
\end{rotate}
}
\newpage
\begin{center}
{\bf Table B2}
\end{center}
\begin{eqnarray*}
\underline{\cal T} (I=0) = - \underline{\cal R} (I=0)  &=&    \left(
\begin{array}{c|ccc|ccc|c}
                  0 & 0 & 0 & 0 & 0 & 0 & 0 & 1  \\
\hline
                  0 & 0 & 0 & 0 & -1 & 0 & 0 & 0  \\
                  0 & 0 & 0 & 0 & 0 & 1 & 0 & 0  \\
                  0 & 0 & 0 & 0 & 0 & 0 & -1 & 0  \\
 \hline
                 0 & 1 & 0 & 0 & 0 & 0 & 0 & 0  \\
                  0 & 0 & -1 & 0 & 0 & 0 &0 & 0  \\
                  0 & 0 & 0 & 1 & 0 & 0 & 0 & 0  \\
\hline
                  -1 & 0 & 0 & 0 & 0 & 0 & 0 & 0  \\
                   \end{array} \right)   \nonumber \\
         & & \nonumber \\
\end{eqnarray*}
\raisebox{-21cm}{
\begin{rotate}{-90}
\raisebox{-13cm}{
\begin{minipage}[b]{0.80\textheight}
\renewcommand{\arraystretch}{2.5}
\renewcommand{\tabcolsep}{0.05cm}
\begin{center}
{\bf Table B3}
\vspace*{.5cm}

\begin{tabular}{|c||c|ccc|ccc|
                             c|}
\hline
 & $2P, \dh,\ \dh$ & $2P, \dh, \ \eh $ & $2P,\eh, \ \eh $ & $2S,\eh, \ \eh $
    & $2P,\dh, -\eh $ & $2P,\eh, -\eh $ & $2S, \eh, -\eh $ & $2P, \dh, -\dh $\\
\hline
\hline
 $2P, \dh, \ \dh$ & 0  & 0
   & 0& $
-\sqrt{\frac{1}{2}}(\mbox{\boldmath$e$}_1\!\!-\!i\mbox{\boldmath$e$}_2)$& 0 & 0
 & 0 & 0\\
\hline
 $2P, \dh, \ \eh $ & 0& 0 & 0 &
                         $\sqrt{\frac{2}{3}}\mbox{\boldmath$e$}_3$& 0
   & 0 &
$-\sqrt{\frac{1}{6}}(\mbox{\boldmath$e$}_1\!\!-\!i\mbox{\boldmath$e$}_2)$& 0\\
 $2P,\eh, \ \eh $ & 0&0& 0& $-\sqrt{\frac{1}{3}}\mbox{\boldmath$e$}_3$
             & 0 &  0 &
$-\sqrt{\frac{1}{3}}(\mbox{\boldmath$e$}_1\!\!-\!i\mbox{\boldmath$e$}_2)$ & 0\\
 $2S,\eh, \ \eh $  &
$-\sqrt{\frac{1}{2}}(\mbox{\boldmath$e$}_1\!\!+\!i\mbox{\boldmath$e$}_2)$&
          $\sqrt{\frac{2}{3}}\mbox{\boldmath$e$}_3$
&$-\sqrt{\frac{1}{3}}\mbox{\boldmath$e$}_3$
    & 0&
$\sqrt{\frac{1}{6}}(\mbox{\boldmath$e$}_1\!\!-\!i\mbox{\boldmath$e$}_2)$&

$-\sqrt{\frac{1}{3}}(\mbox{\boldmath$e$}_1\!\!-\!i\mbox{\boldmath$e$}_2)$&0&
0\\
\hline
 $2P,\dh, -\eh $ & 0& 0& 0 &
$\sqrt{\frac{1}{6}}(\mbox{\boldmath$e$}_1\!\!+\!i\mbox{\boldmath$e$}_2)$
   & 0 & 0&$\sqrt{\frac{2}{3}}\mbox{\boldmath$e$}_3$ & 0\\
 $2P,\eh, -\eh $ & 0 &
        0&0 &
$-\sqrt{\frac{1}{3}}(\mbox{\boldmath$e$}_1\!\!+\!i\mbox{\boldmath$e$}_2)$
      & 0 & 0& $\sqrt{\frac{1}{3}}\mbox{\boldmath$e$}_3$ & 0\\
 $2S, \eh, -\eh $  & 0 &
$-\sqrt{\frac{1}{6}}(\mbox{\boldmath$e$}_1\!\!+\!i\mbox{\boldmath$e$}_2)$ &

$-\sqrt{\frac{1}{3}}(\mbox{\boldmath$e$}_1\!\!+\!i\mbox{\boldmath$e$}_2)$ &
        0& $\sqrt{\frac{2}{3}}\mbox{\boldmath$e$}_3$ &
$\sqrt{\frac{1}{3}}\mbox{\boldmath$e$}_3$&0
        &
$\sqrt{\frac{1}{2}}(\mbox{\boldmath$e$}_1\!\!-\!i\mbox{\boldmath$e$}_2)$\\
\hline
 $2P, \dh, -\dh $&
0&0&0&0&0&0&$\sqrt{\frac{1}{2}}
(\mbox{\boldmath$e$}_1\!\!+\!i\mbox{\boldmath$e$}_2)$&       0 \\
\hline
\end{tabular}
\end{center}
\end{minipage}
}     
\end{rotate}
}
\newpage
\renewcommand{\arraystretch}{1.6}
\begin{center}
{\bf Table B4}
\end{center}
\begin{table}[h]
$$
\begin{array}{|c|l|} \hline
\mbox{eigenvalue}&\mbox{right and left eigenvectors as ket-vectors}\\
\hline\hline
\Delta - \frac{i}{2} \Gamma_P & \vert 2 \hat{P},\frac{3}{2},\pm\frac{3}{2} ) =
\vert 2 P,\frac{3}{2},\pm\frac{3}{2}  ) \\
- \frac{i}{2} \Gamma_P & \vert 2 \hat{P} ,\frac{1}{2},\pm\frac{1}{2} ) =
\vert 2 P ,\frac{1}{2},\pm\frac{1}{2} )
+ i \delta_1 \lambda \vert 2 S,\frac{1}{2},\pm\frac{1}{2}  ) \\
L - \frac{i}{2} \Gamma_S & \vert 2 \hat{S} ,\frac{1}{2},\pm\frac{1}{2} ) =
\vert 2 S ,\frac{1}{2},\pm\frac{1}{2} )
+ i \delta_1 \lambda \vert 2 P,\frac{1}{2},\pm\frac{1}{2}  ) \\
\hline
\mbox{\rule{0em}{1.3em}}
\Delta - \frac{i}{2} \Gamma_P & \vert 2\widetilde{
\hat{P},\frac{3}{2}},\pm\frac{3}{2}  )
= \vert 2 P,\frac{3}{2},\pm\frac{3}{2}) )\\
- \frac{i}{2} \Gamma_P & \vert 2\widetilde{ \hat{P}
,\frac{1}{2}},\pm\frac{1}{2} ) =
\vert 2 P, \frac{1}{2},\pm\frac{1}{2}  )
+ i \delta_1 \lambda^* \vert 2 S,\frac{1}{2},\pm\frac{1}{2}  ) \\
L - \frac{i}{2} \Gamma_S & \vert 2\widetilde{\hat{
S},\frac{1}{2}},\pm\frac{1}{2}  ) =
\vert 2 S ,\frac{1}{2},\pm\frac{1}{2}  )
+ i \delta_1 \lambda^* \vert 2 P,\frac{1}{2},\pm\frac{1}{2} ) \\
\hline
\end{array}
$$
\end{table}
$$
 \lambda := \frac{L(Z,N)}{L(Z,N) - \frac{i}{2} (\Gamma_S(Z,N) - \Gamma_P(Z,N))}
$$
\renewcommand{\arraystretch}{1.0}
\renewcommand{\arraystretch}{2.5}
\renewcommand{\tabcolsep}{0.5cm}
\begin{center}
\vspace*{0.5cm}
{\bf Table B5 a}
\vspace*{.5cm}

\begin{tabular}{|c||p{1.2cm}|p{1.2cm}p{1.2cm}p{1.2cm}p{1.2cm}|}
\hline
 &\zpdhzz & \zpdhze & \zpdhee & \zpehee & \zsehee \\
\hline
\hline
\zpdhzz & $\Delta - \frac{i}{2} \Gamma_P +\frac{3}{8} A_3$
   &0 & 0& 0& 0 \\
\hline
 \zpdhze  & 0& $\Delta - \frac{i}{2} \Gamma_P +\frac{3}{8} A_3$
   & 0& 0& $-\frac{3}{\sqrt{2}} {\cal F}$  \\
 \zpdhee &0 &0 & $\Delta - \frac{i}{2} \Gamma_P -\frac{5}{8} A_3$ &
           $-\frac{5}{16 \sqrt{2}} A_3$&
    $\sqrt{\frac{3}{2}} {\cal F} $\\
 \zpehee &0 & 0& $-\frac{5}{16 \sqrt{2}} A_3$& $
               - \frac{i}{2} \Gamma_P +\frac{1}{4} A_2 $&
   $ \sqrt{3} {\cal F}+i \delta_1 L + \ih \delta_2 L $  \\
 \zsehee &0 & $ -\frac{3}{\sqrt{2}}{\cal F}$ & $\sqrt{\frac{3}{2}} {\cal F}$ &
$\sqrt{3} {\cal F}
    -i \delta_1 L - \ih \delta_2 L$ & $L - \frac{i}{2} \Gamma_S+\frac{1}{4}
  A_1$
 \\
\hline
\end{tabular}
\end{center}
\newpage
\renewcommand{\arraystretch}{2.5}
\renewcommand{\tabcolsep}{0.5cm}
\begin{center}
{\bf Table B5 b}
\vspace*{.5cm}

\begin{tabular}{|c||p{1.2cm}p{1.2cm}p{1.2cm}p{1.2cm}p{1.2cm}p{1.2cm}|}
\hline
 & \zpdhzn & \zpdhen & \zpehen & \zsehen
    & \zpehnn & \zsehnn \\
\hline
\hline
 \zpdhzn & $\Delta - \frac{i}{2} \Gamma_P +\frac{3}{8} A_3$
   & 0& 0& $-\sqrt{6} {\cal F}$ & 0 & 0 \\
 \zpdhen & 0& $\Delta - \frac{i}{2} \Gamma_P -\frac{5}{8} A_3$ &
                         $-\frac{5}{16 \sqrt{2}} A_3$& 0
   & 0& $-\sqrt{6} {\cal F} $\\
 \zpehen & 0& $-\frac{5}{16 \sqrt{2}} A_3$& $
                       - \frac{i}{2} \Gamma_P +\frac{1}{4} A_2 $&
   $i \delta_1 L + \ih \delta_2 L $& 0&  $\sqrt{3} {\cal F} $\\
 \zsehen & $-\sqrt{6} {\cal F}$ & 0& $-i \delta_1 L - \ih \delta_2 L$ &
   $L - \frac{i}{2} \Gamma_S+\frac{1}{4} A_1$ & $\sqrt{3} {\cal F} $& 0\\
 \zpehnn & 0& 0&0 & $\sqrt{3} {\cal F}$ & $- \frac{i}{2} \Gamma_P -\frac{3}{4}
A_2$
   & $i \delta_1 L -\dh i \delta_2 L$\\
 \zsehnn & 0& $-\sqrt{6} {\cal F}$ & $\sqrt{3} {\cal F} $&0 & $-i \delta_1 L
+\dh
    i \delta_2 L$
   & $L - \frac{i}{2} \Gamma_S-\frac{3}{4} A_1 $\\
\hline
\end{tabular}
\end{center}
\newpage
\renewcommand{\arraystretch}{2.5}
\renewcommand{\tabcolsep}{0.5cm}
\begin{center}
{\bf Table B5 c}
\vspace*{.5cm}

\begin{tabular}{|c||p{1.2cm}p{1.2cm}p{1.2cm}p{1.5cm}|p{1.5cm}|}
\hline
 & \zpdhzme & \zpdheme & \zpeheme & \zseheme & \zpdhzmz  \\
\hline
\hline
 \zpdhzme & $\Delta - \frac{i}{2} \Gamma_P +\frac{3}{8} A_3$
   &0 &0 & $-\frac{3}{\sqrt{2}} {\cal F}$ & 0\\
 \zpdheme & & $\Delta - \frac{i}{2} \Gamma_P -\frac{5}{8} A_3$ &
                                     $-\frac{5}{16 \sqrt{2}} A_3$&
    $-\sqrt{\frac{3}{2}} {\cal F} $&0\\
 \zpeheme &0 & $-\frac{5}{16 \sqrt{2}} A_3$& $
                       - \frac{i}{2} \Gamma_P +\frac{1}{4} A_2 $&
   $ -\sqrt{3} {\cal F}+i \delta_1 L + \ih \delta_2 L $ &0 \\
 \zseheme & $ -\frac{3}{\sqrt{2}}{\cal F}$ & $-\sqrt{\frac{3}{2}} {\cal F}$ &
$-\sqrt{3} {\cal F}
    -i \delta_1 L - \ih
   \delta_2 L$ & $L - \frac{i}{2} \Gamma_S+\frac{1}{4} A_1$
  &0 \\
\hline
 \zpdhzmz &0 &0 & 0& 0&
   $  \Delta -\frac{i}{2} \Gamma_P + \frac{3}{8} A_3$  \\
\hline
\end{tabular}
\end{center}
\begin{center}
\vspace*{0.5cm}
{\bf Table B6}
\vspace*{.5cm}
\end{center}
\renewcommand{\arraystretch}{1.0}
\begin{eqnarray*}
\hspace*{-2cm}
\underline{\cal T} (I=\eh) = \underline{\cal R} (I=\eh)
 =     \left( \begin{array}{r|rrrr|rrrrrr|rrrr|r}
     0 & 0 & 0 & 0 & 0 & 0 & 0 & 0 & 0 & 0 & 0 & 0 & 0 & 0 & 0 &-1  \\
\hline
     0 & 0 & 0 & 0 & 0 & 0 & 0 & 0 & 0 & 0 & 0 & 1 & 0 & 0 & 0 & 0  \\
     0 & 0 & 0 & 0 & 0 & 0 & 0 & 0 & 0 & 0 & 0 & 0 &-1 & 0 & 0 & 0  \\
     0 & 0 & 0 & 0 & 0 & 0 & 0 & 0 & 0 & 0 & 0 & 0 & 0 &-1 & 0 & 0  \\
     0 & 0 & 0 & 0 & 0 & 0 & 0 & 0 & 0 & 0 & 0 & 0 & 0 & 0 & 1 & 0  \\
\hline
     0 & 0 & 0 & 0 & 0 &-1 & 0 & 0 & 0 & 0 & 0 & 0 & 0 & 0 & 0 & 0  \\
     0 & 0 & 0 & 0 & 0 & 0 & 1 & 0 & 0 & 0 & 0 & 0 & 0 & 0 & 0 & 0  \\
     0 & 0 & 0 & 0 & 0 & 0 & 0 & 1 & 0 & 0 & 0 & 0 & 0 & 0 & 0 & 0  \\
     0 & 0 & 0 & 0 & 0 & 0 & 0 & 0 &-1 & 0 & 0 & 0 & 0 & 0 & 0 & 0  \\
     0 & 0 & 0 & 0 & 0 & 0 & 0 & 0 & 0 &-1 & 0 & 0 & 0 & 0 & 0 & 0  \\
     0 & 0 & 0 & 0 & 0 & 0 & 0 & 0 & 0 & 0 & 1 & 0 & 0 & 0 & 0 & 0  \\
\hline
     0 & 1 & 0 & 0 & 0 & 0 & 0 & 0 & 0 & 0 & 0 & 0 & 0 & 0 & 0 & 0  \\
     0 & 0 &-1 & 0 & 0 & 0 & 0 & 0 & 0 & 0 & 0 & 0 & 0 & 0 & 0 & 0  \\
     0 & 0 & 0 &-1 & 0 & 0 & 0 & 0 & 0 & 0 & 0 & 0 & 0 & 0 & 0 & 0  \\
     0 & 0 & 0 & 0 & 1 & 0 & 0 & 0 & 0 & 0 & 0 & 0 & 0 & 0 & 0 & 0  \\
\hline
    -1 & 0 & 0 & 0 & 0 & 0 & 0 & 0 & 0 & 0 & 0 & 0 & 0 & 0 & 0 & 0  \\
 \end{array} \right)
\end{eqnarray*}
\begin{center}
{\bf Table B7 a}
\vspace*{.5cm}
\end{center}
\hspace*{-2.5cm}
\begin{math}
\renewcommand{\arraystretch}{2.5}
\renewcommand{\tabcolsep}{0.5cm}
\begin{tabular}{|c||ccccc|}
\hline
 &\zpdhzz & \zpdhze & \zpdhee & \zpehee & \zsehee \\
\hline
\hline
\zpdhzz & $0$
   &0 & 0& 0& $-\frac{1}{\sqrt{2}} (\ex -i \ey )$ \\
 \zpdhze  & 0& 0
   & 0& 0& $\frac{1}{\sqrt{2}} \ez $  \\
 \zpdhee &0 &0 & 0 &
          0&
    $ -\frac{1}{\sqrt{6}} \ez $\\
 \zpehee &0 & 0& 0& 0&
   $ -\frac{1}{\sqrt{3}} \ez $  \\
 \zsehee &$-\frac{1}{\sqrt{2}} (\ex +i \ey )$ & $ \frac{1}{\sqrt{2}} \ez$ &
    $ -\frac{1}{\sqrt{6}} \ez $ &  $ -\frac{1}{\sqrt{3}} \ez $  & 0
 \\
\hline
\end{tabular}
\end{math}
\newline
\newline \ .
\begin{center}
{\bf Table B7 b}
\vspace*{.5cm}
\end{center}
\hspace*{-2.0cm}
\begin{math}
\renewcommand{\arraystretch}{2.5}
\renewcommand{\tabcolsep}{0.1cm}
\hspace{-0.8cm}
\begin{tabular}{|c||cccccc|}
\hline
 & \zpdhzn & \zpdhen & \zpehen & \zsehen
    & \zpehnn & \zsehnn \\
\hline
\hline
 \zpdhze & 0
   & 0& 0& $-\frac{1}{2} (\ex -i \ey )$ & 0 & 0 \\
 \zpdhee & 0& 0 &
                         0& $-\frac{1}{2\sqrt{3}} (\ex -i \ey )$
   & 0& $-\frac{1}{\sqrt{3}} (\ex -i \ey )$\\
 \zpehee  & 0& 0 &
                         0& $-\frac{1}{\sqrt{6}} (\ex -i \ey )$
   & 0& $\frac{1}{\sqrt{6}} (\ex -i \ey )$\\
  \zsehee &  $\frac{1}{2\sqrt{3}} (\ex -i \ey )$& $-\frac{1}{2\sqrt{3}}
     (\ex -i \ey )$& $-\frac{1}{\sqrt{6}} (\ex -i \ey )$ &
   0 & $\frac{1}{\sqrt{6}} (\ex -i \ey )$& 0\\
\hline
\end{tabular}
\end{math}
\newpage
\begin{center}
{\bf Table B7 c}
\vspace*{.5cm}
\end{center}
\hspace*{-2.0cm}
\begin{math}
\renewcommand{\arraystretch}{2.5}
\renewcommand{\tabcolsep}{0.5cm}
\begin{tabular}{|c||cccc|}
\hline
  & \zpdhze & \zpdhee & \zpehee & \zsehee \\
\hline
\hline
 \zpdhzn & $0$
    & 0& 0&  $\frac{1}{2\sqrt{3}} (\ex +i \ey )$\\
 \zpdhen  & 0& 0
   & 0&  $-\frac{1}{2\sqrt{3}} (\ex +i \ey )$ \\
 \zpehen  &0 & 0 &
          0& $-\frac{1}{\sqrt{6}} (\ex +i \ey )$
    \\
 \zsehen  & $-\frac{1}{2} (\ex +i \ey )$& $-\frac{1}{2\sqrt{3}} (\ex +i \ey )$
            & $-\frac{1}{\sqrt{6}} (\ex +i \ey ) $ &
   $ 0 $  \\
 \zpehnn &0 & 0 &
    0 &     $\frac{1}{\sqrt{6}} (\ex +i \ey )$
 \\
  \zsehnn & 0 & $-\frac{1}{\sqrt{3}} (\ex +i \ey )$ &
        $\frac{1}{\sqrt{6}} (\ex +i \ey )$ & 0 \\
\hline
\end{tabular}
\end{math}
\newline
 \vspace{2cm}
\begin{center}
{\bf Table B7 d}
\vspace*{.5cm}
\end{center}
\begin{math}
\renewcommand{\arraystretch}{2.5}
\renewcommand{\tabcolsep}{0.3cm}
\begin{tabular}{|c||cccccc|}
\hline
 & \zpdhzn & \zpdhen & \zpehen & \zsehen
    & \zpehnn & \zsehnn \\
\hline
\hline
 \zpdhzn  & 0 & 0 & 0 & $ \sqrt{\frac{2}{3}} \ez $ &
        0 & 0 \\
 \zpdhen  & 0 & 0 & 0 & 0 &
        0 & $ \sqrt{\frac{2}{3}} \ez $ \\
 \zpehen & 0 & 0 & 0 & 0 &
        0 & $ -\frac{1}{\sqrt{3}} \ez $ \\
 \zsehen & $ \sqrt{\frac{2}{3}} \ez $ & 0 & 0 & 0 &
        $ -\frac{1}{\sqrt{3}} \ez $ & 0 \\
 \zpehnn  & 0 & 0 & 0 & $ -\frac{1}{\sqrt{3}} \ez $ &
        0 & 0 \\
  \zsehnn & 0 & $ \sqrt{\frac{2}{3}} \ez $ & $ -\frac{1}{\sqrt{3}} \ez $ & 0 &
        0 & 0 \\
\hline
\end{tabular}
\end{math}
\newpage
\begin{center}
{\bf Table B7 e}
\vspace*{.5cm}
\end{center}
\hspace*{-1.0cm}
\begin{math}
\renewcommand{\arraystretch}{2.5}
\renewcommand{\tabcolsep}{0.5cm}
\begin{tabular}{|c||cccc|}
\hline
 & \zpdhzme & \zpdheme & \zpeheme & \zseheme   \\
\hline
\hline
 \zpdhzn & $0$
    & 0& 0&  $-\frac{1}{2\sqrt{3}} (\ex -i \ey )$\\
 \zpdhen  & 0& 0
   & 0&  $-\frac{1}{2\sqrt{3}} (\ex -i \ey )$ \\
 \zpehen  &0 & 0 &
          0& $-\frac{1}{\sqrt{6}} (\ex -i \ey )$
    \\
 \zsehen  & $\frac{1}{2} (\ex -i \ey )$& $-\frac{1}{2\sqrt{3}} (\ex -i \ey )$
            & $-\frac{1}{\sqrt{6}} (\ex -i \ey ) $ &
   $ 0 $  \\
 \zpehnn &0 & 0 &
    0 &     $-\frac{1}{\sqrt{6}} (\ex -i \ey )$
 \\
  \zsehnn & 0 & $\frac{1}{\sqrt{3}} (\ex -i \ey )$ &
        $-\frac{1}{\sqrt{6}} (\ex -i \ey )$ & 0 \\
\hline
\end{tabular}
\end{math}
\newline
 \vspace{2cm}
\begin{center}
{\bf Table B7 f}
\vspace*{.5cm}
\end{center}
\hspace*{-1.5cm}
\begin{math}
\renewcommand{\arraystretch}{2.5}
\renewcommand{\tabcolsep}{0.1cm}
\hspace{-1.5cm}
\begin{tabular}{|c||cccccc|}
\hline
 & \zpdhzn & \zpdhen & \zpehen & \zsehen
    & \zpehnn & \zsehnn \\
\hline
\hline
 \zpdhzme & 0
   & 0& 0& $\frac{1}{2} (\ex +i \ey )$ & 0 & 0 \\
 \zpdheme & 0& 0 &
                         0& $-\frac{1}{2\sqrt{3}} (\ex +i \ey )$
   & 0& $\frac{1}{\sqrt{3}} (\ex +i \ey )$\\
  \zpeheme & 0& 0 &
                         0& $-\frac{1}{\sqrt{6}} (\ex +i \ey )$
   & 0& $-\frac{1}{\sqrt{6}} (\ex +i \ey )$\\
  \zseheme &  $-\frac{1}{2\sqrt{3}} (\ex +i \ey )$& $-\frac{1}{2\sqrt{3}}
     (\ex +i \ey )$& $-\frac{1}{\sqrt{6}} (\ex +i \ey )$ &
   0 & $-\frac{1}{\sqrt{6}} (\ex +i \ey )$& 0\\
\hline
\end{tabular}
\end{math}
\newpage
\begin{center}
{\bf Table B7 g}
\vspace*{.5cm}
\end{center}
\hspace*{-1.0cm}
\vspace*{1cm}
\begin{math}
\renewcommand{\arraystretch}{2.5}
\renewcommand{\tabcolsep}{0.5cm}
\hspace{-1.5cm}
\begin{tabular}{|c||ccccc|}
\hline
 & \zpdhzme & \zpdheme & \zpeheme & \zseheme & \zpdhzmz \\
\hline
\hline
 \zpdhzme  & 0
   & 0& 0& $\frac{1}{\sqrt{2}} \ez $ & 0\\
 \zpdheme   &0 & 0 &
          0&
    $ \frac{1}{\sqrt{6}} \ez $ &0 \\
 \zpeheme  & 0& 0& 0&
   $ \frac{1}{\sqrt{3}} \ez $ &0 \\
 \zseheme   & $ \frac{1}{\sqrt{2}} \ez$ &
    $ \frac{1}{\sqrt{6}} \ez $ &  $ \frac{1}{\sqrt{3}} \ez $  & 0
        &$\frac{1}{\sqrt{2}} (\ex -i \ey )$\\
 \zpdhzmz
   &0 & 0& 0& $\frac{1}{\sqrt{2}} (\ex +i \ey )$ & $0$
 \\
\hline
\end{tabular}
\end{math}
\newpage
\begin{center}
{\bf Table B8}
\end{center}
$\ $
\hspace{-3.2cm}
\begin{math}
\renewcommand{\arraystretch}{2.5}
\renewcommand{\tabcolsep}{0.3cm}
\begin{tabular}{l|l}
   eigenvalue
& \hspace{2cm}  
   right eigenvector
\\
\hline
\hline
$\Delta -\ih \Gamma_P +\da A_3$ &  $|\mbox{$ 2P, \frac{3}{2} ,2,F_3 $} )$ \\
 & $\{ F_3 =\pm 2,\pm 1,0 \}$ \\
\hline
$\Delta -\ih \Gamma_P -\mbox{$\frac{5}{8} $}A_3
      $
   &      $  | 2\hat{P},\dh ,  1,F_3) \, =  | 2P,\dh ,  1,F_3)-
                \left( \mbox{$\frac{5A_3}{16 \sqrt{2}}$} \right)\cdot
           \mbox{$\frac{1}{(\Delta -\mbox{$\frac{5}{8} $}A_3 -
              \mbox{$\ev $}A_2) }$}  | 2P,\eh ,  1,F_3)     $ \\
 $\ +\left( \mbox{$\frac{5A_3}{16 \sqrt{2}}$} \right)^2 \cdot
 \mbox{$\frac{1}{\Delta -\mbox{$\frac{5}{8} $}A_3 -
              \mbox{$\ev $}A_2 }$}$
&  $\ \ \ \ \ \ \ \ +i(\delta_1 + \eh \delta_2 ) \mbox{$ \frac{L
        \left( \mbox{$\frac{5A_3}{16 \sqrt{2}}$} \right)}
          { (\Delta -\mbox{$\frac{5}{8} $}A_3 -
              \mbox{$\ev $}A_2)( \Delta  -L-\ih (\Gamma_P-\Gamma_S)
        -\ev A_1 - \mbox{$\frac{5}{8} $}A_3)
          } $}
          | 2S,\eh ,  1,F_3)   $ \\
&  $\{ F_3 =\pm 1,0 \}$ \\
\hline
$-\ih \Gamma_P+\ev A_2 $ &
  $  | 2\hat{P},\eh ,  1,F_3) \, =| 2P,\eh ,  1,F_3)+
     i(\delta_1 + \eh \delta_2 )\mbox{$\frac{L
        }{
       L- \frac{i}{2} (\Gamma_S -\Gamma_P)+\frac{1}{4} (A_1-A_2)
        }$}
           | 2S,\eh ,  1,F_3)
       $
       \\
$\ -\left( \mbox{$\frac{5A_3}{16 \sqrt{2}}$} \right)^2 \cdot
 \mbox{$\frac{1}{\Delta -\frac{5}{8} A_3 -
            \frac{1}{4}  A_2 }$}$
& $ \ \  \ \ \ \ \ \ \ \ \ \ \ \ \ \ \ \ \ \ \ \ \ \
           +\mbox{$ \frac{
        \left( \mbox{$\frac{5A_3}{16 \sqrt{2}}$} \right)}
      { (\Delta -\mbox{$\frac{5}{8} $}A_3 -
              \mbox{$\ev $}A_2)} $}  | 2P,\dh ,  1,F_3)
         $ \\
&  $\{ F_3 =\pm 1,0 \}$ \\
\hline
$L-\ih \Gamma_S +\ev A_1 $& $| 2\hat{S},\eh ,  1,F_3)\, =
 |2S,\eh ,  1,F_3) + i (\delta_1 +\eh \delta_2)
    \frac{L}{L- \frac{i}{2} (\Gamma_S -\Gamma_P)+\frac{1}{4} (A_1-A_2) }\,
    |2P,\eh ,  1,F_3) $ \\
  & $ \ \  \ \ \ \ \ \ \
           +i(\delta_1 +\eh \delta_2)\mbox{$\frac{L
        \left( \mbox{$\frac{5A_3}{16 \sqrt{2}}$} \right)}{
     ( \Delta  -L-\ih (\Gamma_P-\Gamma_S)
        -\ev A_1 - \mbox{$\frac{5}{8} $}A_3)
       (L- \frac{i}{2} (\Gamma_S -\Gamma_P)+\frac{1}{4} (A_1-A_2))
        }$}
           | 2P,\dh ,  1,F_3) $ \\
&  $\{ F_3 =\pm 1,0 \}$ \\
\hline
$L-\ih \Gamma_S -\dv A_1 $& $| 2\hat{S},\eh , 0,0)\, =
   |2S,\eh , 0,0) + i (\delta_1 -\dh  \delta_2)
    \frac{L}{L- \frac{i}{2} (\Gamma_S -\Gamma_P)-\frac{3}{4} (A_1-A_2) }\,
    |2P,\eh , 0,0)   $ \\
\hline
$-\ih \Gamma_P -\dv A_2 $ &  $| 2\hat{P},\eh , 0,0)\, =
   |2P,\eh , 0,0) + i (\delta_1 -\dh  \delta_2)
    \frac{L}{L- \frac{i}{2} (\Gamma_S -\Gamma_P)-\frac{3}{4} (A_1-A_2) }\,
    |2S,\eh , 0,0)   $ \\
\hline
\end{tabular}
\end{math}
\newpage
\section*{Figure Captions}
\begin{description}
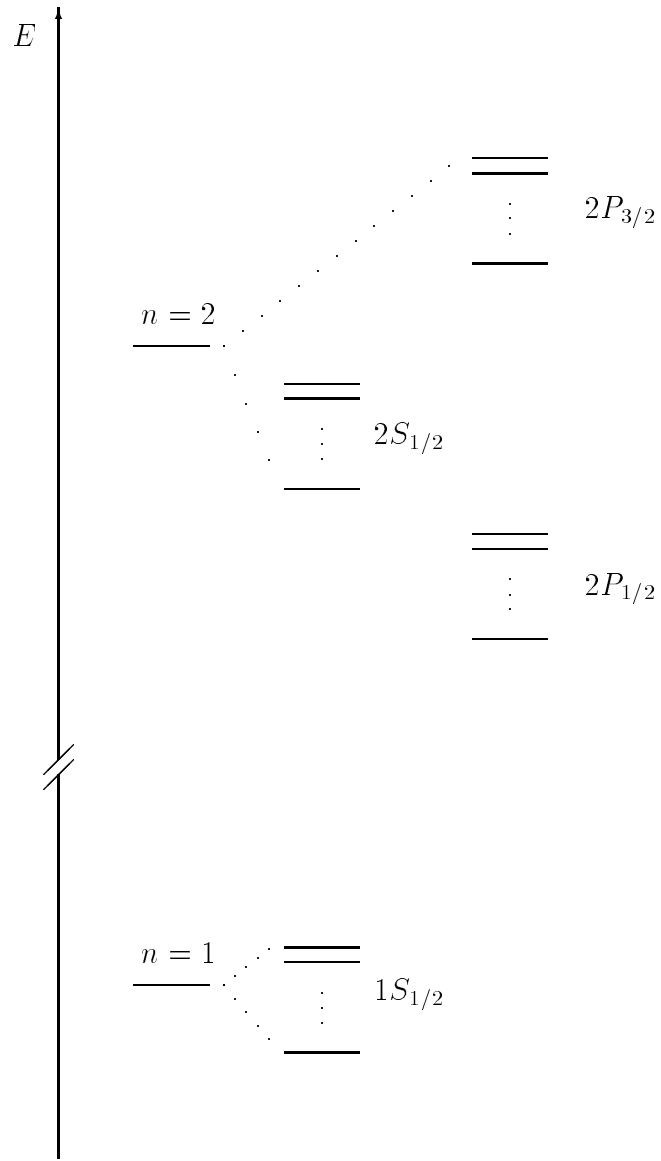
\item[Figure 1]
Schematic diagram of the $n=1$ and $n=2$ energy levels of a hydrogen like
system described by the Hamiltonian $H_0$ of (\ref{3.1}). Indicated are the
Coulomb approximation, the fine and hyperfine structure.

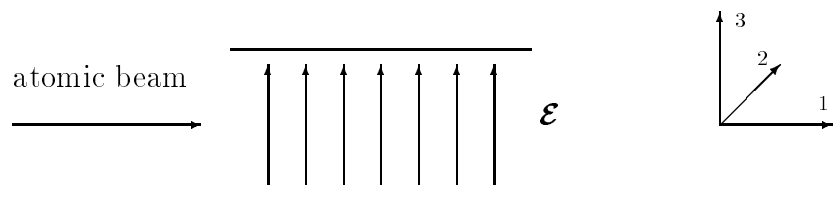
\item[Figure 2]
An atomic beam traversing the electric field of a plane plate capacitor.

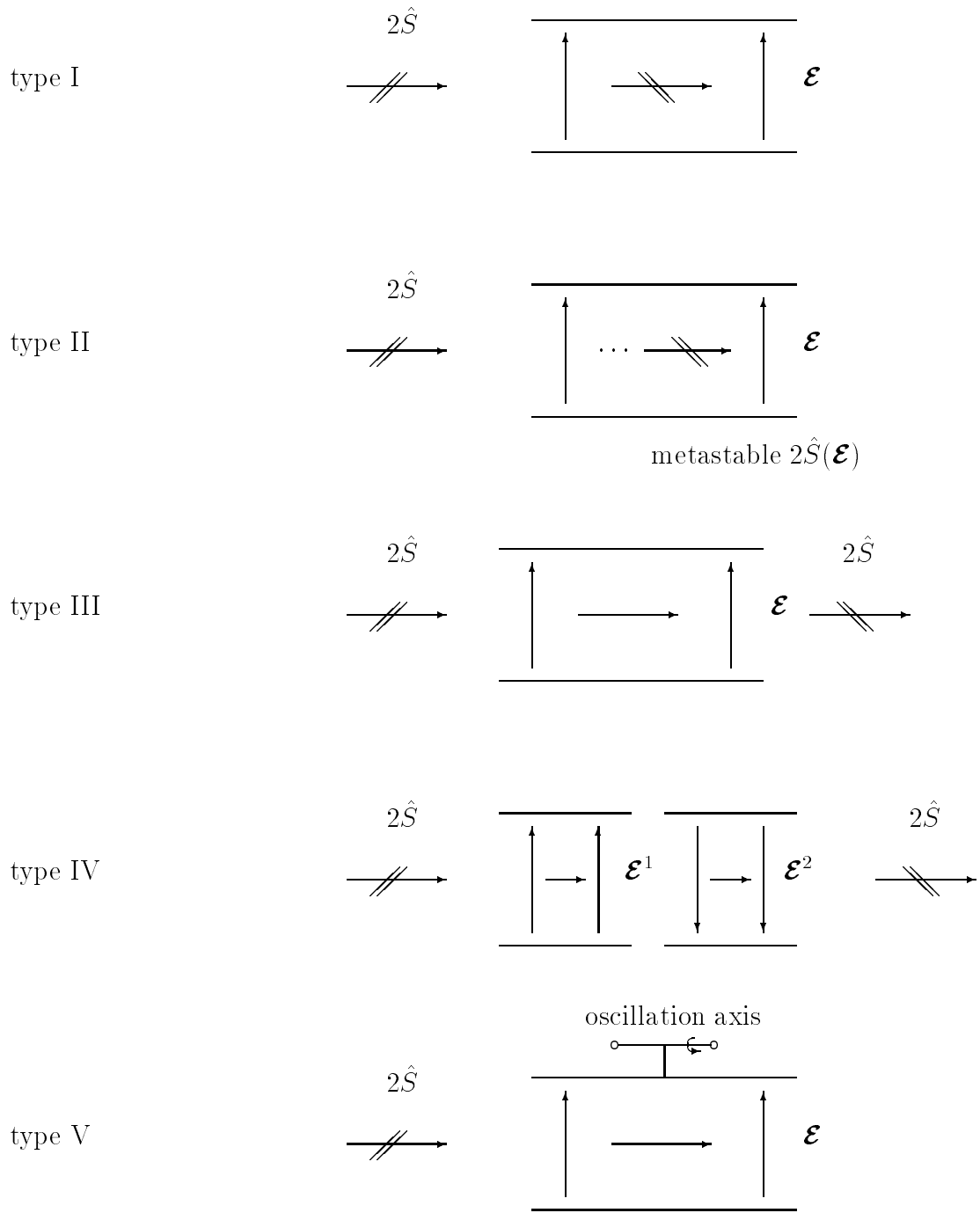
\item[Figure 3]
Schematic arrangements for type I -- V rotations defined in the text and in
table 1 with an atomic beam passing through capacitors. For the type V
rotation we draw a capacitor which is suspended such that it can oscillate
around one axis as indicated.

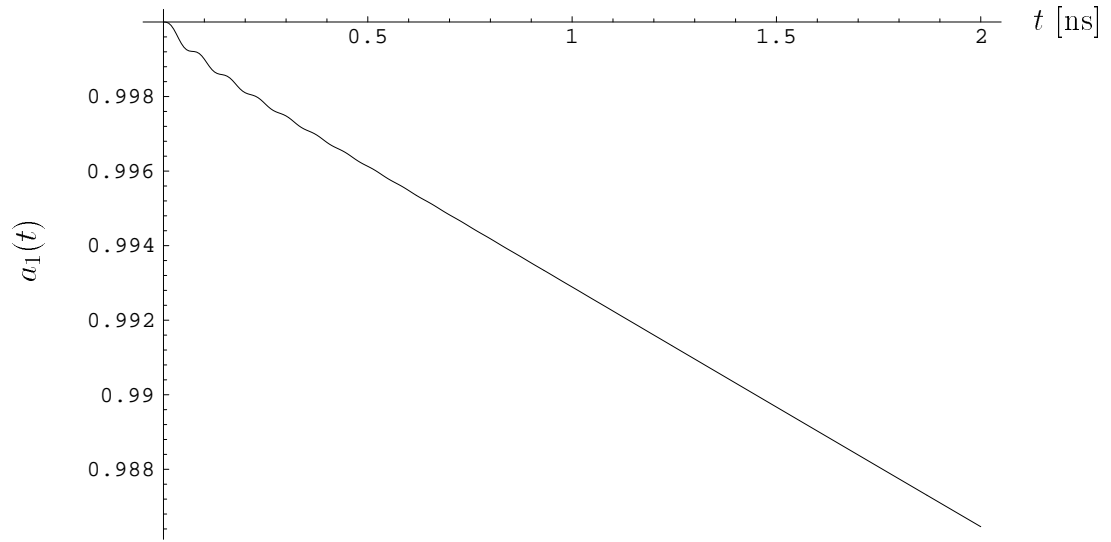
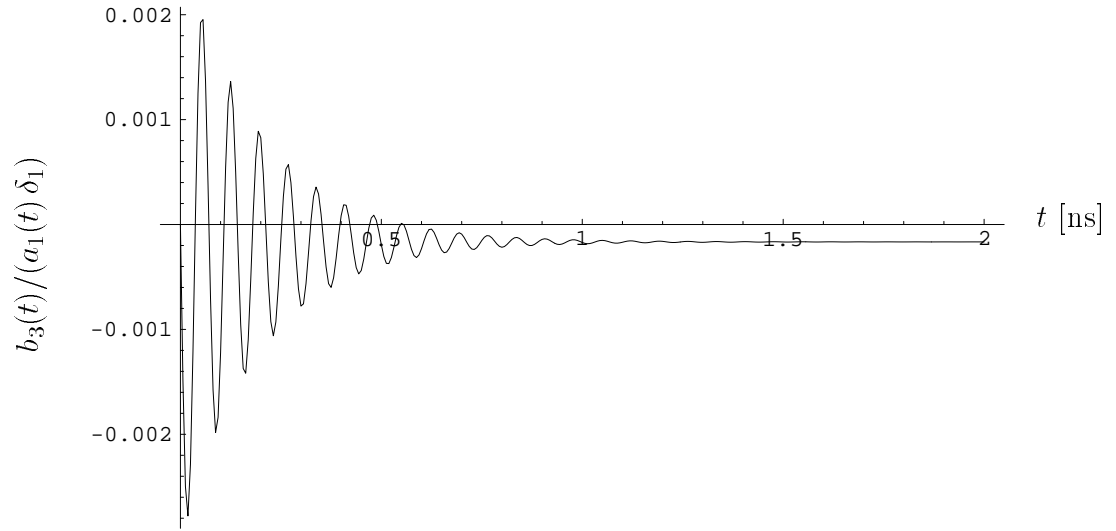
\item[Figure 4]
The functions $a_1$ and the ratios $b_1/a_1, b_3/(a_1 \delta_1(2,2)), b_4/(a_1
\delta_1(2,2))$ (cf. (\ref{5.8}), (\ref{5.9})) for $^4_2$He$^+$ for $|\vec\E| =
320$ V/cm as function of time $t$.

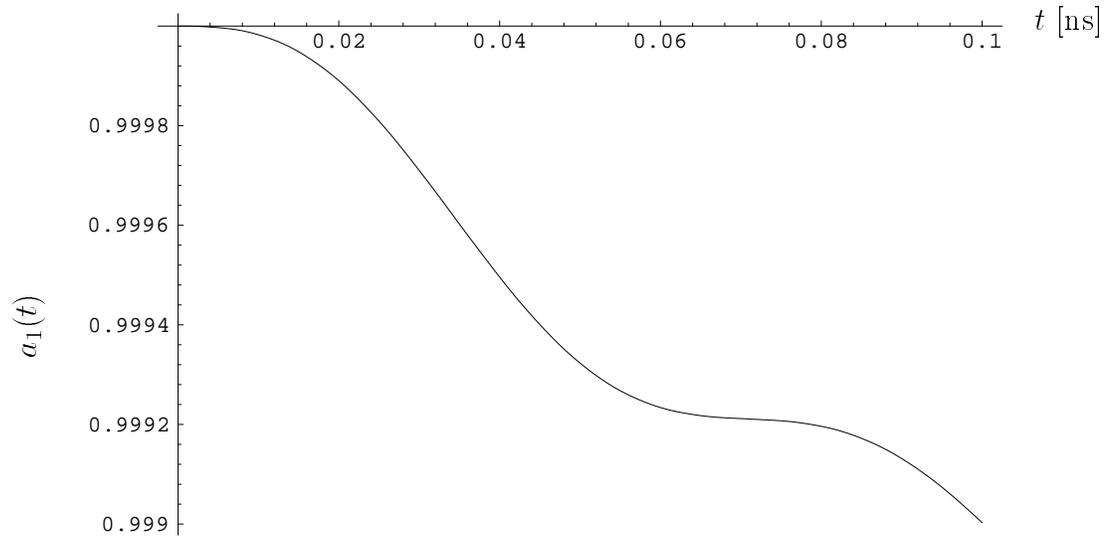
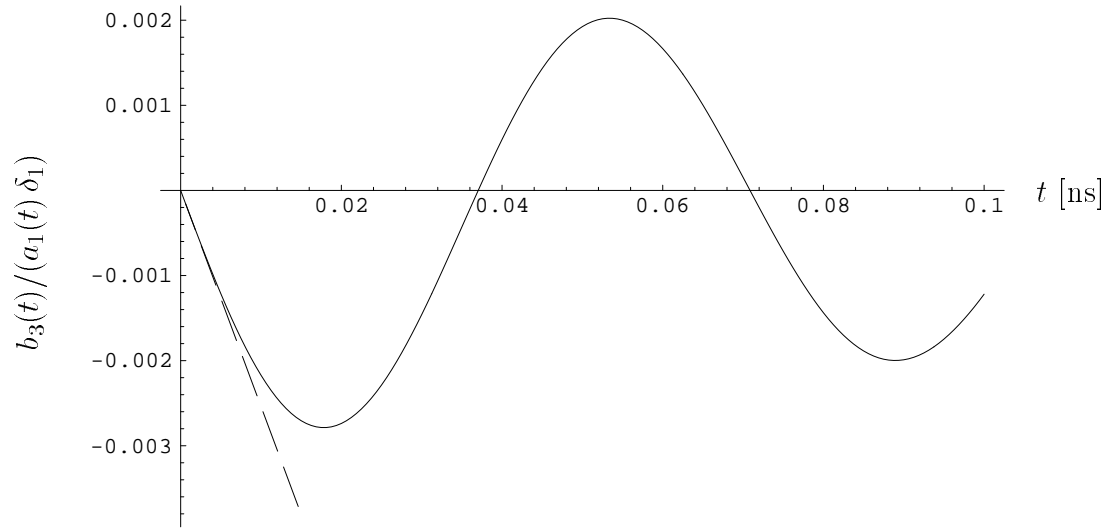
\item[Figure 5]
Same as Figure 4 but for very short times. Also indicated is the limiting
behaviour for $b_3/(a_1 \delta_1(2,2))$ from (\ref{5.22b}) as dashed line.

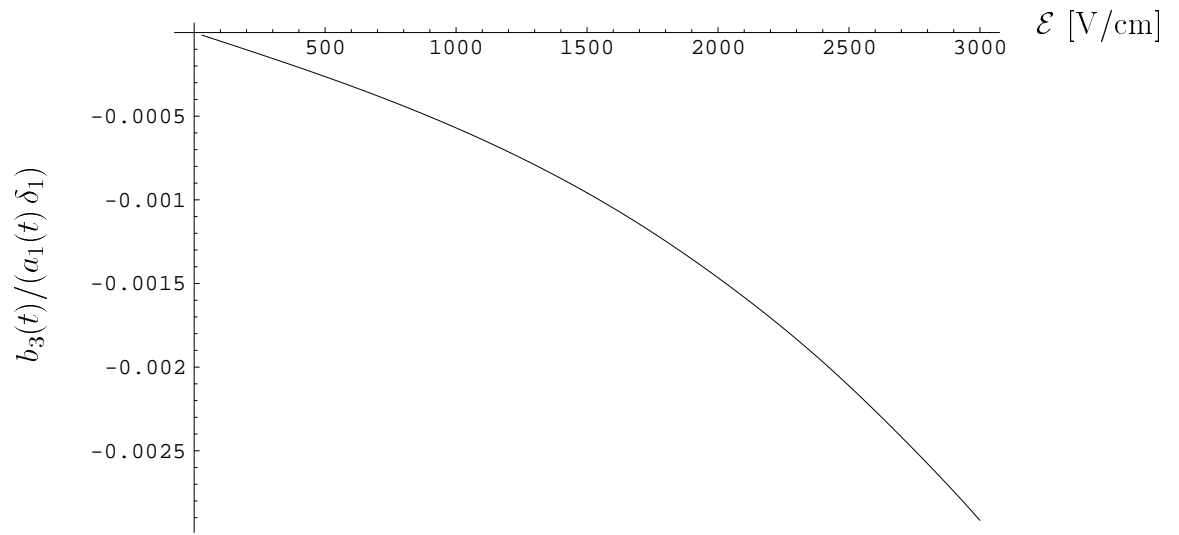
\item[Figure 6]
The behaviour of the ratios $b_i(t,|\vec\E|^2)/(a_1(t,|\vec\E|^2)
\delta_1(2,2))$ $(i = 3,4)$ for \mbox{$t = 2$ ns}, i.~e. in the asymptotic
plateau region (cf. Figure 4), as function of $\E=|\vec\E|$.

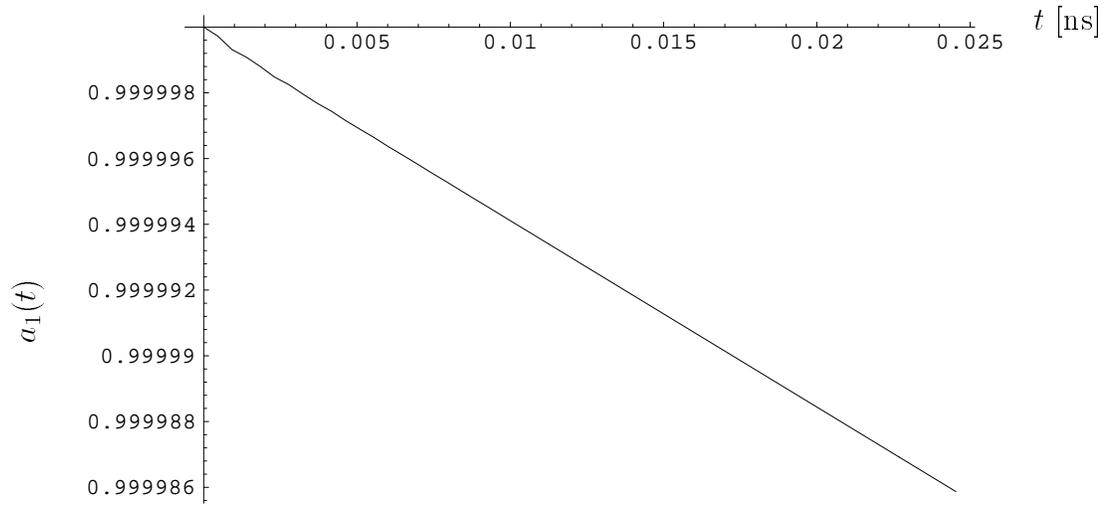
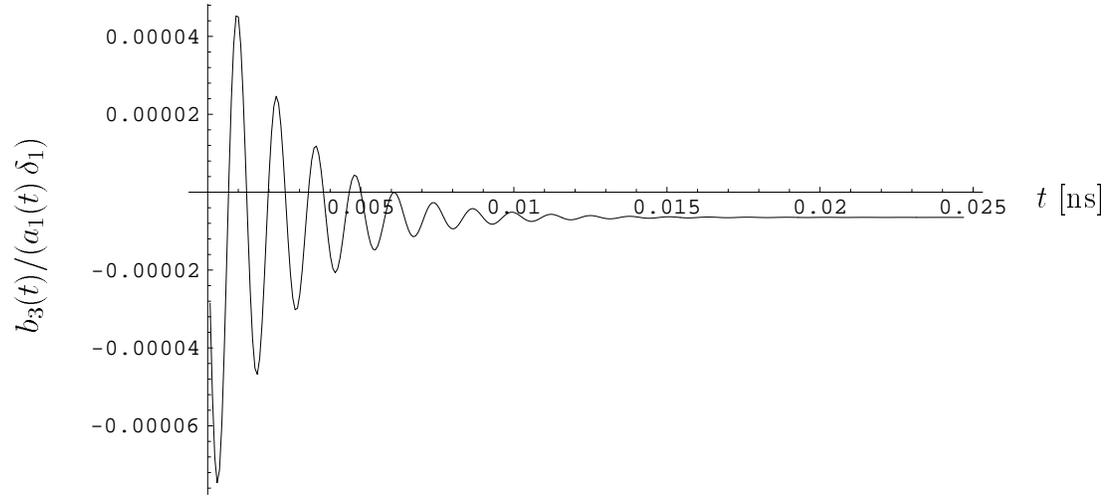
\item[Figure 7]Same as Figure 4 but for the ion $^{12}_6$C$^{5+}$. The electric
field is 1 kV/cm.

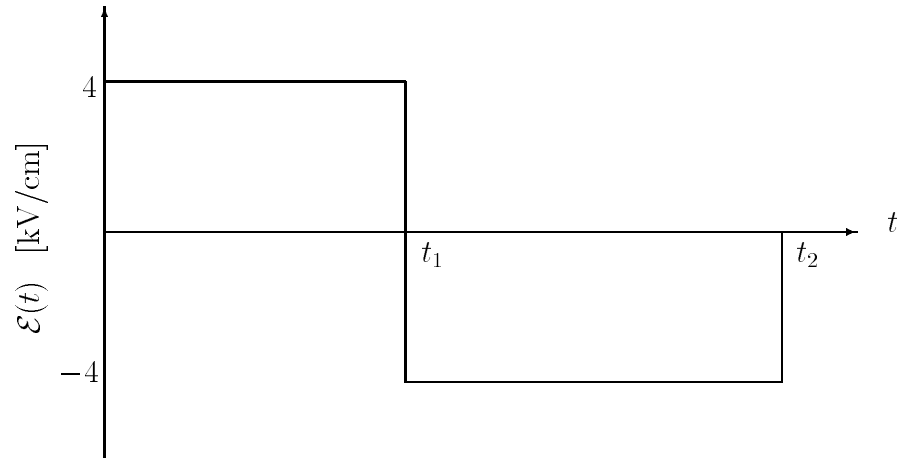
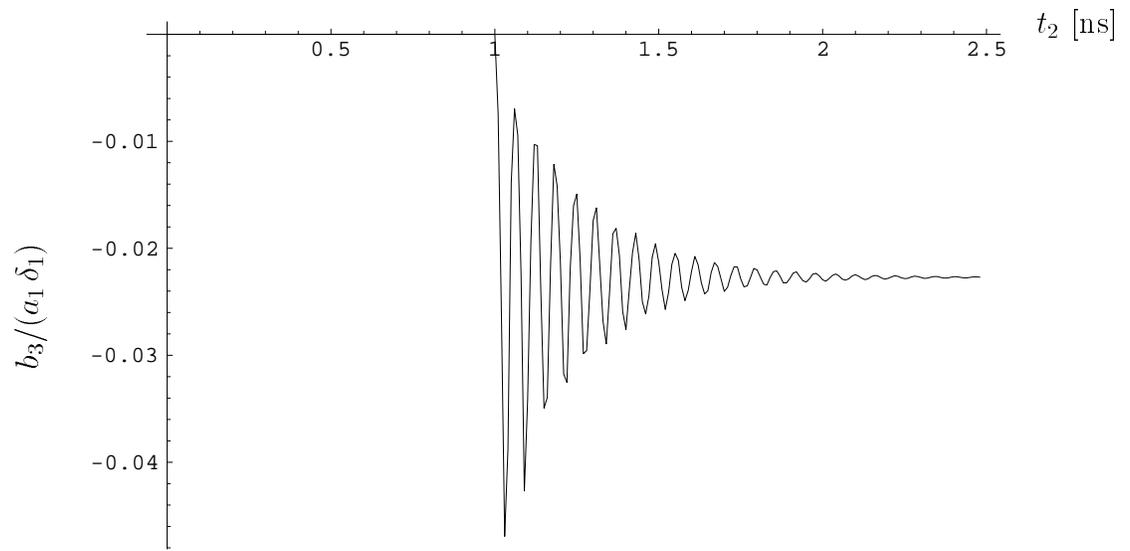
\item[Figure 8]
Rotations of type IV for $^4_2$He$^+$. Plotted are the  variation of the
electric field with time (a), the function $a_1(t_2,t_1;\E^2,\E^1)$ (b), and
the ratios
$b_{3,4}(t_2,t_1;\E^2,\E^1)/$ $[a_1(t_2,t_1;\E^2,\E^1) \delta_1(2,2)]$ (c,d)
(cf. (\ref{5.24}), (\ref{5.25})).
The parameters are given in (\ref{5.33}). For $t_2<t_1=1$ ns  the results
correspond to one electric field  $\E^1$ only which is switched off at $t_2$.

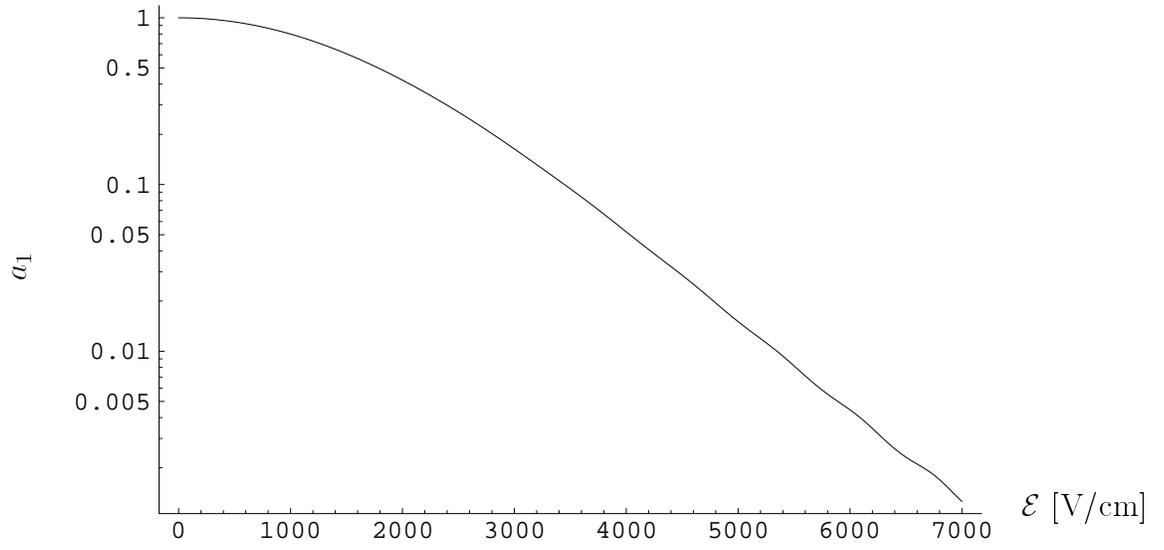
\item[Figure 9]
The dependence of $a_1(t_2,t_1;\E^2,\E^1)$ and
$b_{3,4}(t_2,t_1;\E^2,\E^2)/[a_1(t_2,t_1;\E^2,\E^1) \delta_1(2,2)]$ on $\E^1$
for $\E^2 = - \E^1$, $t_1 = 1$ ns, $t_2 = 3$ ns.

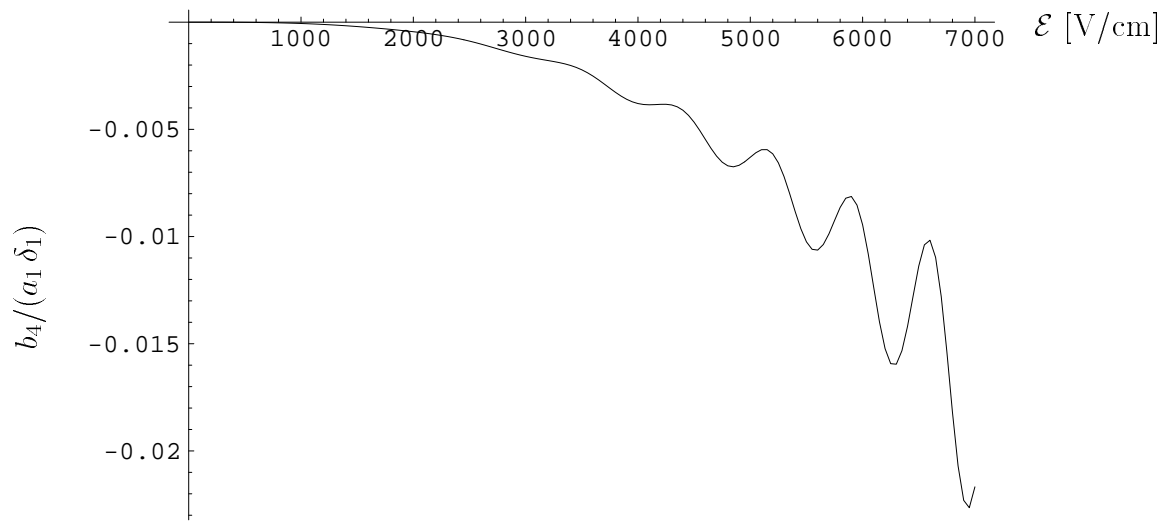
\item[Figure 10]
An interference device for observing the type IV rotation. The time $t$ is the
proper time of the atoms which traverse an electric field $\E^1 \vec{e}_3$ for
time $t_1$, are then split to traverse fields $\E^1 \vec{e}_3$ and
$\E^2\vec{e}_3$ respectively for a time $t_2 - t_1$ and are let to interfere,
split in beams $\pm$ and analysed for their $2 \hat{S}$ content at time $t_2$.

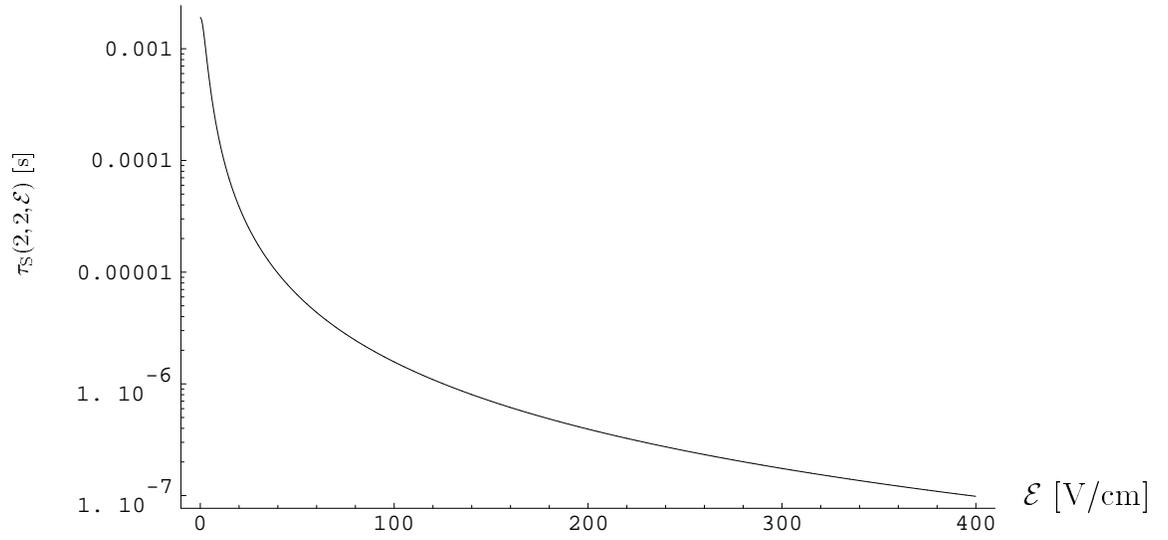
\item[Figure 11]
The lifetime $\tau_S(2,2,\E)$ for $^4_2$He$^+$ in an external field $\vec\E$ as
function of $\E=|\vec\E|$ (a). The inverse lifetime as function of $[\sqrt{3}
{\cal F}(2,\E)/L(2,2)]^2$ (b).

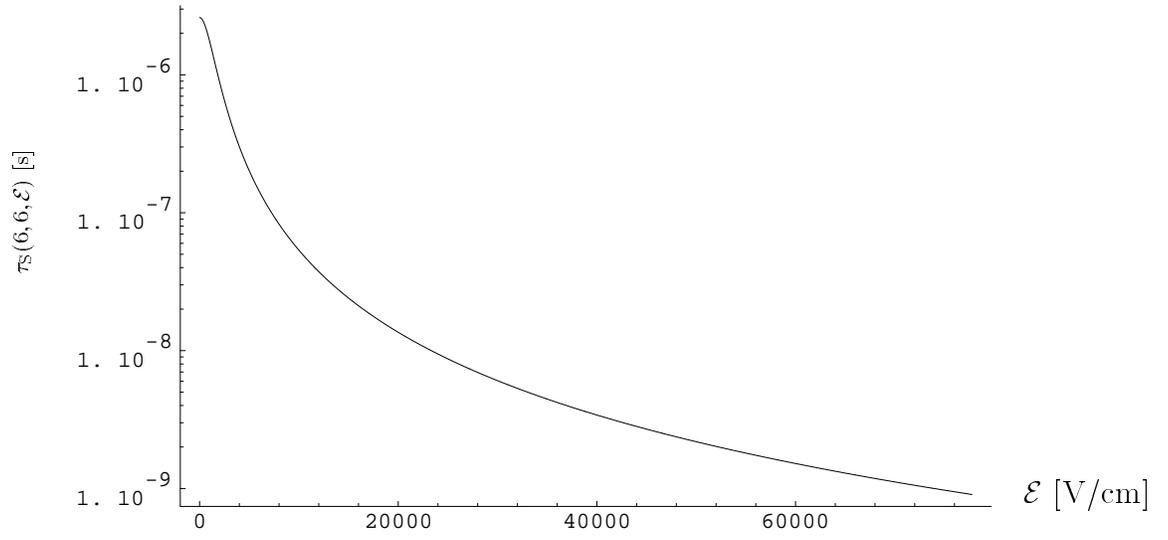
\item[Figure 12]
Same as Figure 11 but for $^{12}_6$C$^{5+}$.

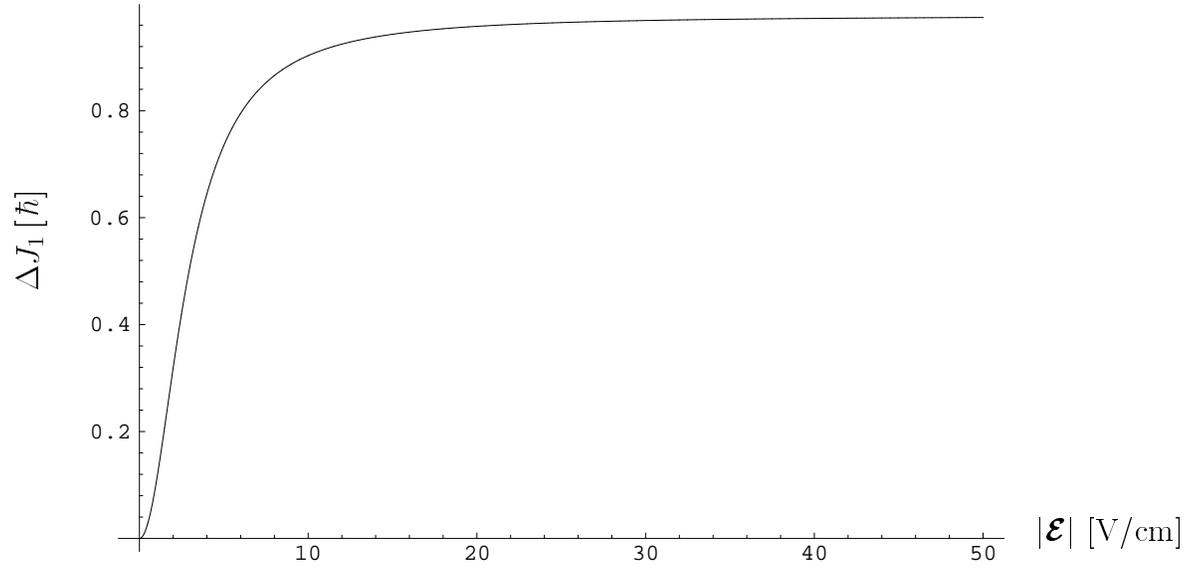
\item[Figure 13]
Results for the angular momentum transfers $\Delta \, J_1$ and $\Delta J_2$ for
$^4_2$He$^+$ as function of the electric field $|\vec\E|$ (cf. \ref{5.38}).

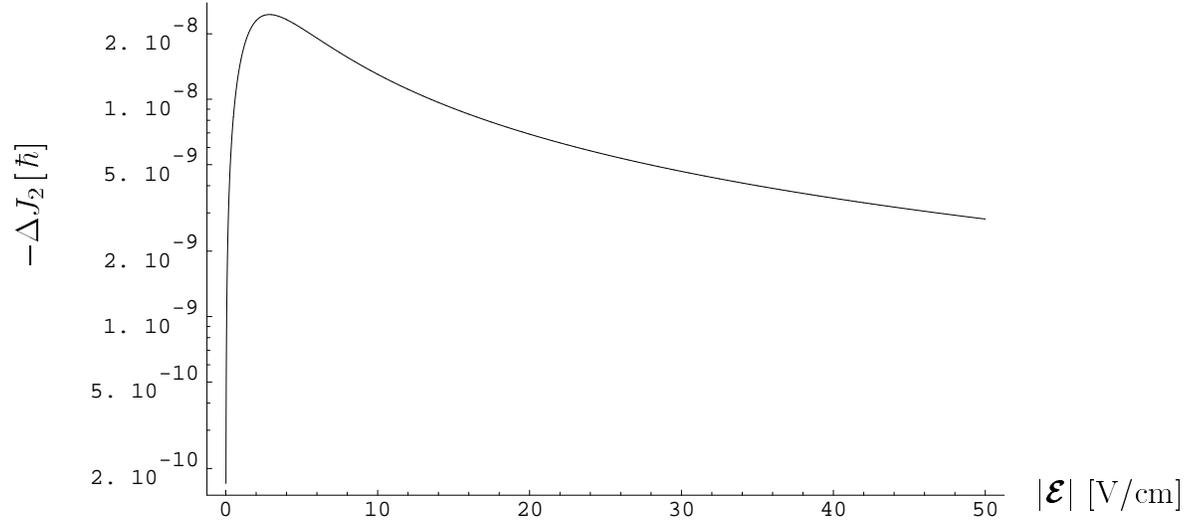
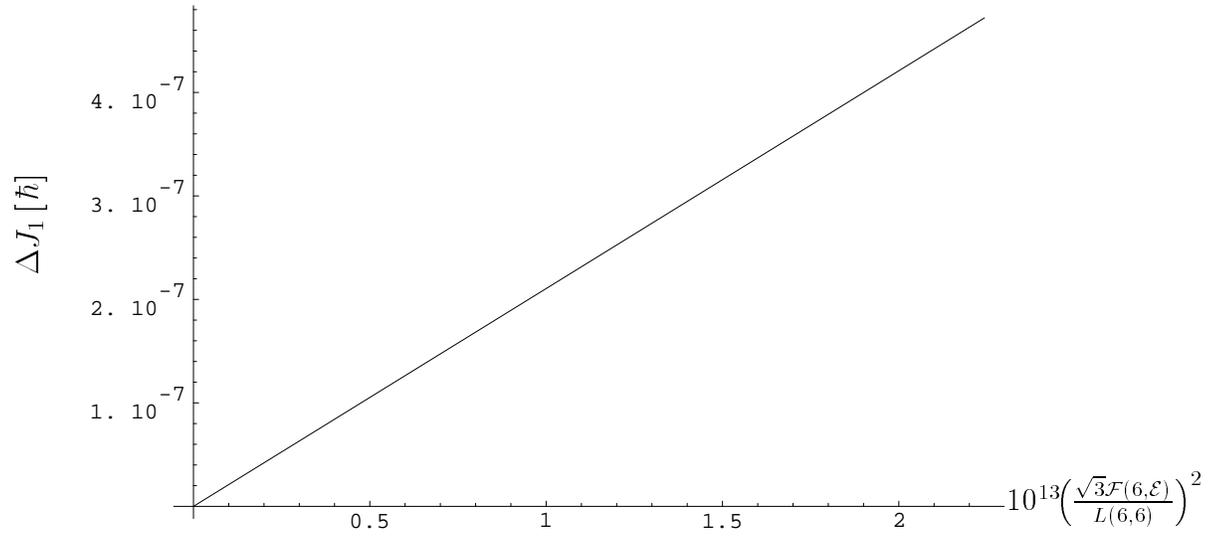
\item[Figure 14]
Same as Figure 13 but for $^{12}_6$C$^{5+}$.

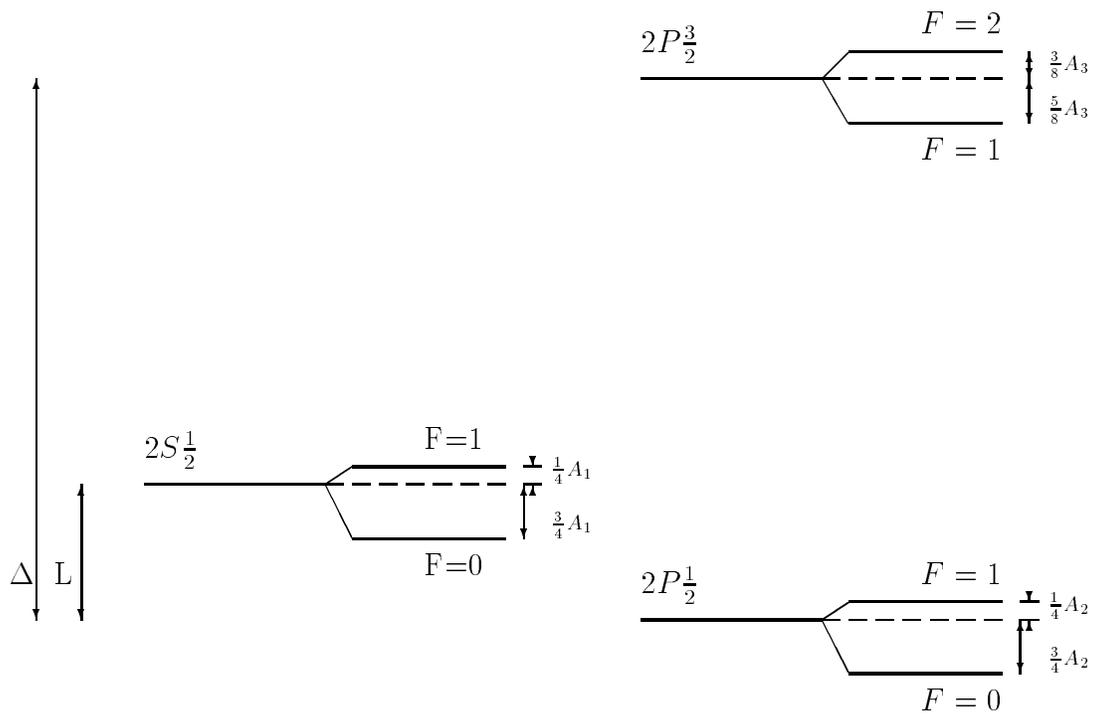
\item[Figure 15]
Scheme of the energy levels of the $n=2$ states for an atom with $I=\halb$ at
zero external field.

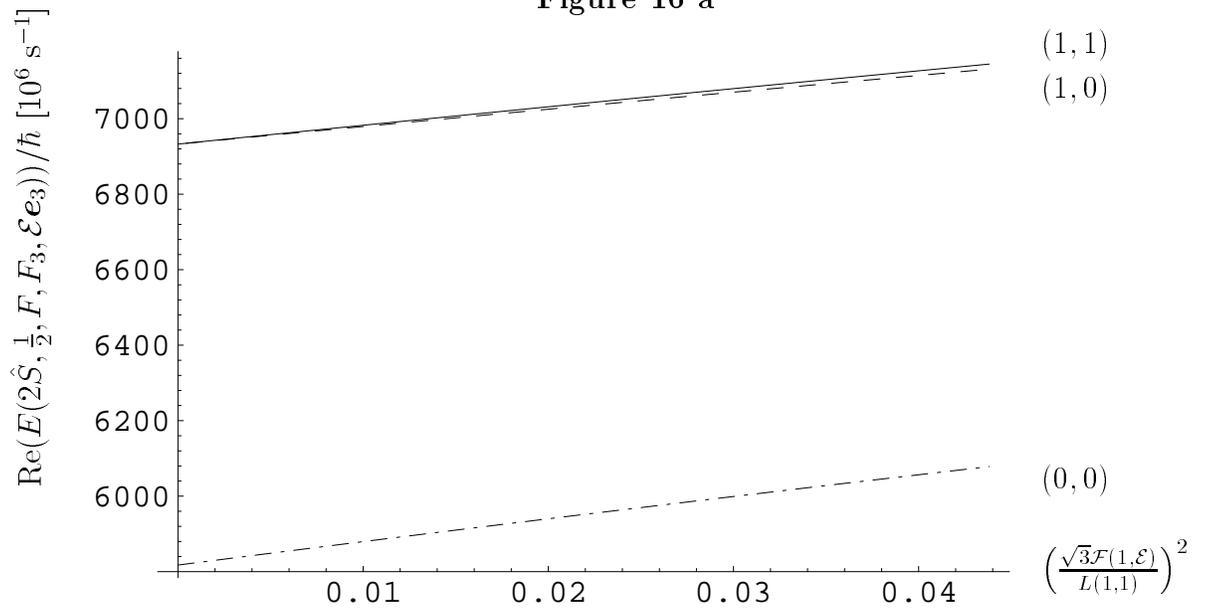
\item[Figure 16]
The real  parts of the complex energies (\ref{6.10}) and the decay rates
(\ref{6.10a})
 of the $2\hat S$ states in an electric field $\E\vec e_3$ for $^1_1$H as
function of $(\sqrt3{\cal F}(1,\E)/L(1,1))^2=(\E/477 \ \mbox{V\ cm}^{-1})^2$.
The curves are marked by the values $(F,F_3)$.

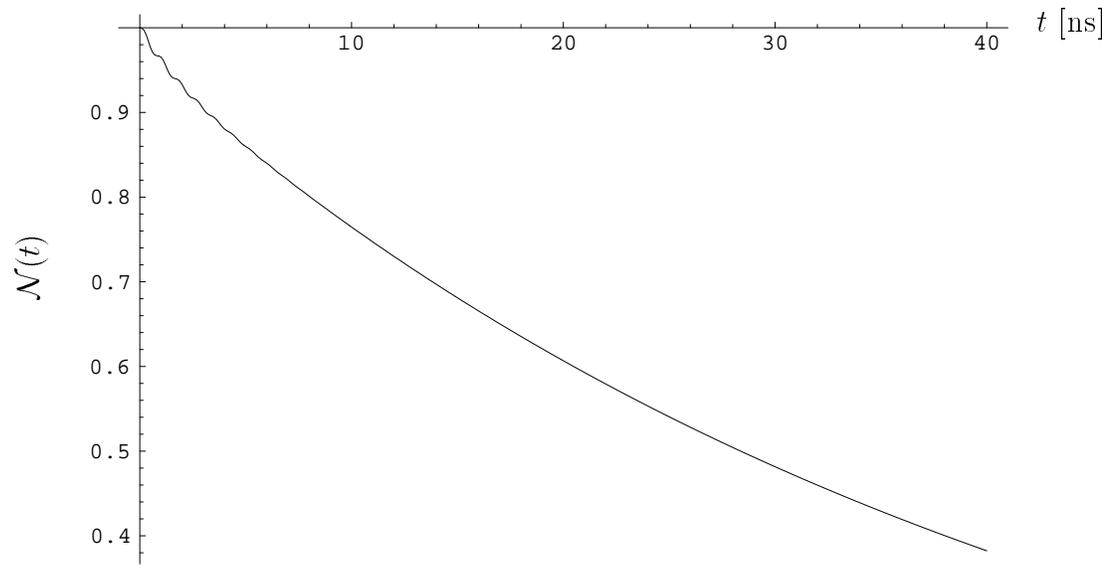
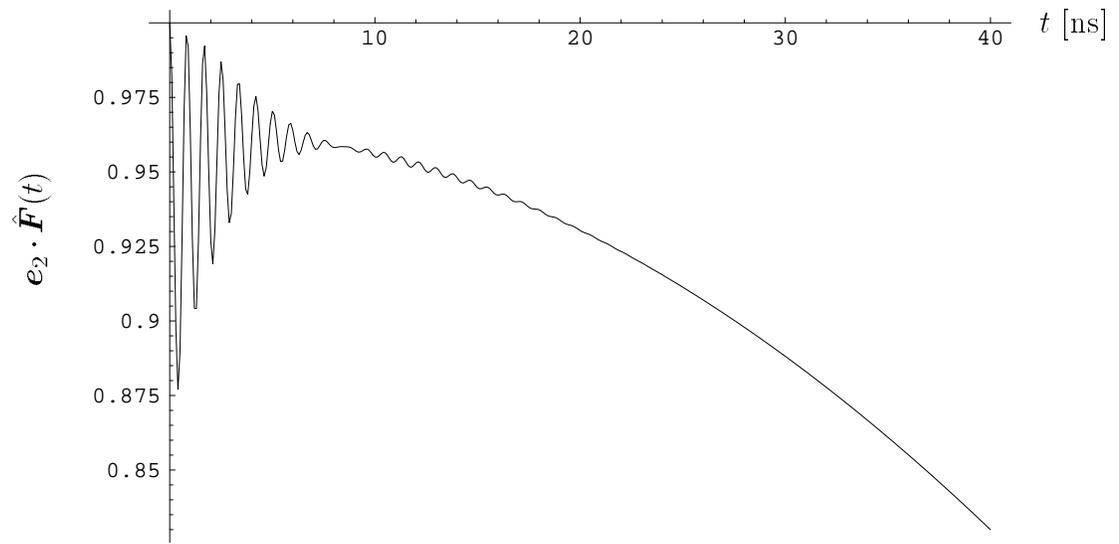
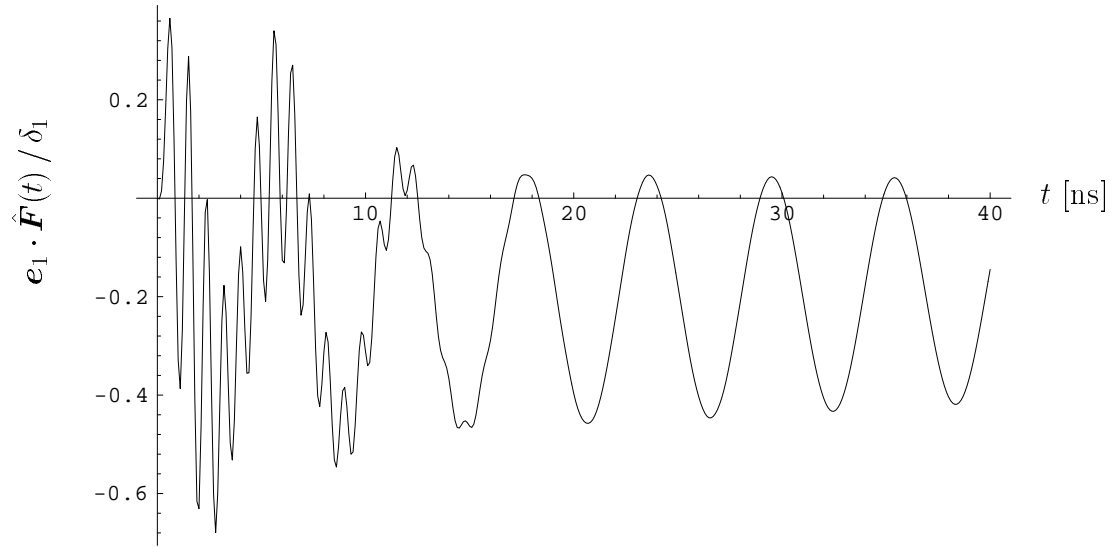
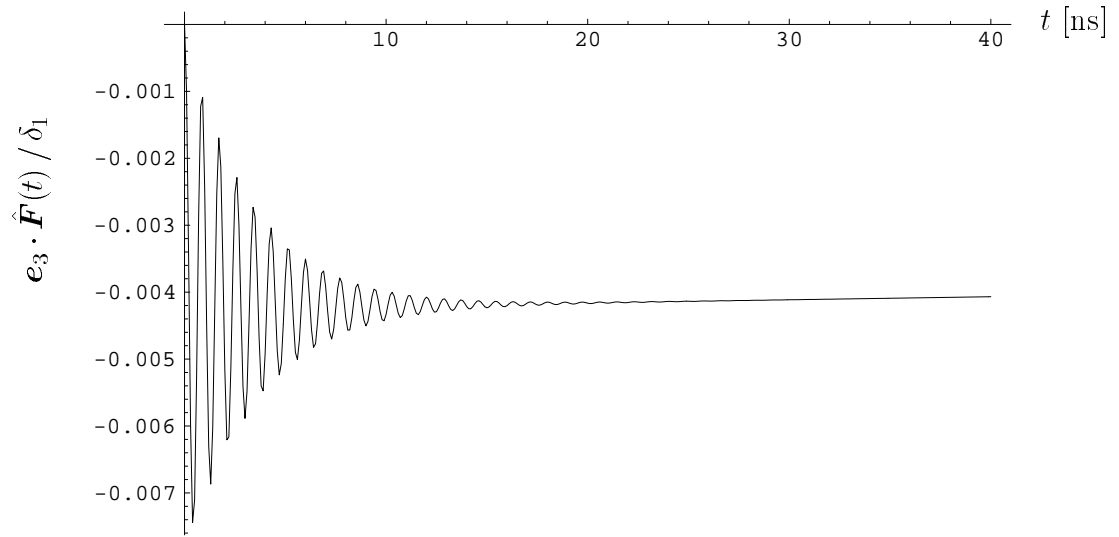
\item[Figure 17]
Results for angular momentum rotation of ordinary $2S$-hydrogen $^1_1$H in a
weak electric field $\vec \E = \E \vec{e}_3$, $\E = 100$ V/cm.
Shown are the time evolution of the norm ${\cal N}(t)$ and of the components of
the angular momentum per unit norm $\hat{F}(t)$.

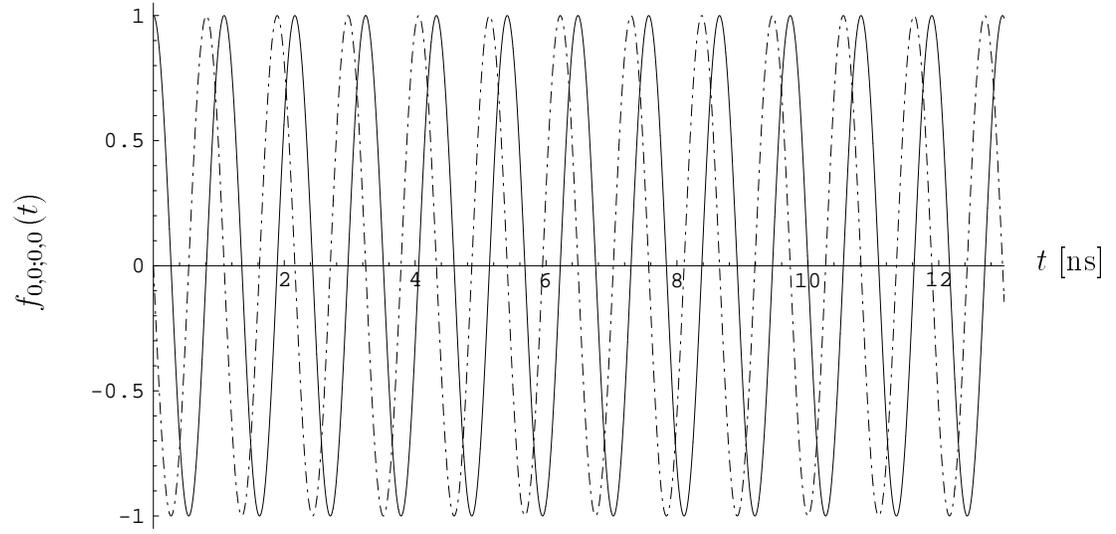
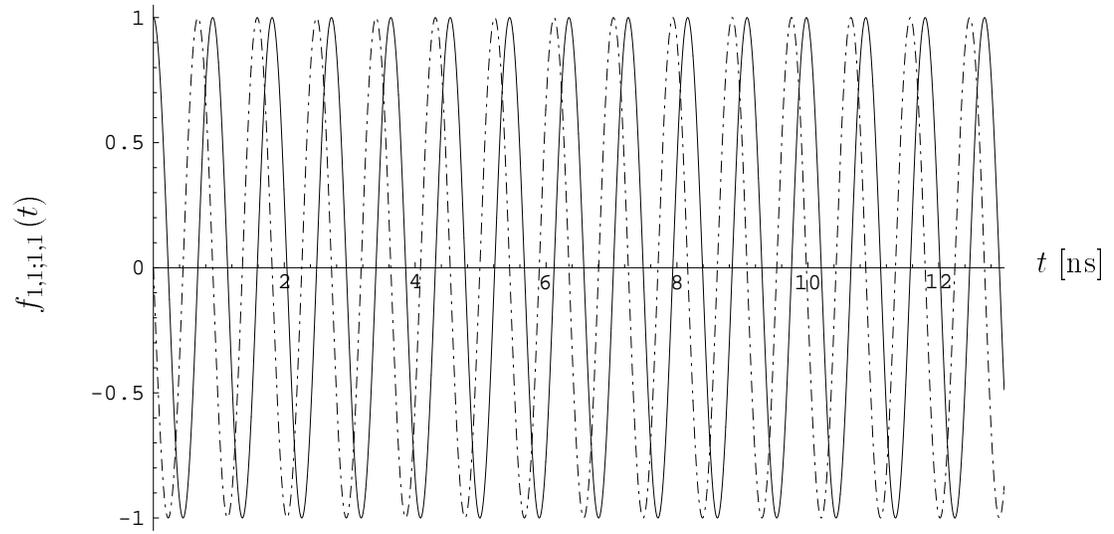
\item[Figure 18]
The functions $f_{F',F'_3;F,F_3}(t)$ (\ref{6.28}) for $^1_1$H and $\E=100\
$V/cm with $\Delta s_p=0$. The P-conserving functions are shown in $a,b,c$, the
P-violating function  $f_{1,0;0,0}$ in $d$. Full lines represent the real,
dashed lines the imaginary parts.
\end{description}
\end{document}